 \documentclass[twocolumn]{aastex61}
\hypersetup{linkcolor=red,citecolor=blue,filecolor=cyan,urlcolor=magenta}

\usepackage{natbib}

\newcommand\aastex{AAS\TeX}

\usepackage{gensymb}
\usepackage{longtable}
\usepackage{rotating}

\usepackage{xcolor}
\accepted{December 7, 2018}
\shorttitle{\aastex\ \textit{HST/WFC3} NIR Library of Substellar Objects}
\shortauthors{Manjavacas et al.}
\begin{document}

\title{Cloud Atlas: Hubble Space Telescope Near-Infrared Spectral Library of Brown Dwarfs, Planetary-mass companions, and hot Jupiters}
\shorttitle{HST/WFC3 Infrared Spectral Library of substellar objects}

\correspondingauthor{Elena Manjavacas}
\email{elenamanjavacas@email.arizona.edu}

\author{Elena Manjavacas}
\affil{Department of Astronomy/Steward Observatory, The University of Arizona, 933 N. Cherry Avenue, Tucson, AZ 85721, USA}

\author{D\'aniel Apai}
\affiliation{Department of Astronomy/Steward Observatory, The University of Arizona, 933 N. Cherry Avenue, Tucson, AZ 85721, USA}
\affiliation{Department of Planetary Science/Lunar and Planetary Laboratory, The University of Arizona, 1629 E. University Boulevard, Tucson, AZ 85718, USA}
\affiliation{Earths in Other Solar Systems Team, NASA Nexus for Exoplanet System Science.}

\author{Yifan Zhou}
\affiliation{Department of Astronomy/Steward Observatory, The University of Arizona, 933 N. Cherry Avenue, Tucson, AZ 85721, USA}

\author{Ben W. P. Lew}
\affiliation{Department of Planetary Science/Lunar and Planetary Laboratory, The University of Arizona,
1640 E. University Boulevard, Tucson, AZ 85718, USA}

\author{Glenn Schneider}
\affiliation{Department of Astronomy/Steward Observatory, The University of Arizona, 933 N. Cherry Avenue, Tucson, AZ 85721, USA}

\author{Stan Metchev}
\affiliation{The University of Western Ontario, Department of Physics and Astronomy, 1151 Richmond Avenue, London, ON N6A 3K7, Canada}

\author{Paulo A. Miles-P\'aez}
\affiliation{The University of Western Ontario, Department of Physics and Astronomy, 1151 Richmond Avenue, London, ON N6A 3K7, Canada}
\affiliation{Department of Astronomy/Steward Observatory, The University of Arizona, 933 N. Cherry Avenue, Tucson, AZ 85721, USA}

\author{Jacqueline Radigan}
\affiliation{Utah Valley University, 800 West University Parkway, Orem, UT 84058, USA}

\author{Mark S. Marley}
\affiliation{NASA Ames Research Center, Mail Stop 245-3, Moffett Field, CA 94035, USA}

\author{Nicolas Cowan}
\affiliation{Department of Earth \& Planetary Sciences, 3450 University St. Montreal, Quebec H3A 0E8, Canada}

\author{Theodora Karalidi}
\affiliation{Department of Astronomy and Astrophysics, University of California, Santa Cruz, California, USA}

\author{Adam J. Burgasser}
\affiliation{Center for Astrophysics and Space Science, University of California San Diego, La Jolla, CA 92093, USA}

\author{Luigi R. Bedin}
\affiliation{INAF — Osservatorio Astronomico di Padova, Vicolo Osservatorio 5, I-35122 Padova, Italy}

\author{Patrick J. Lowrance}
\affiliation{IPAC-Spitzer, MC 314-6, California Institute of Technology, Pasadena, CA 91125, USA}

\author{Parker Kauffmann}
\affiliation{Department of Astronomy/Steward Observatory, The University of Arizona, 933 N. Cherry Avenue, Tucson, AZ 85721, USA}





\begin{abstract}
{Bayesian atmospheric retrieval tools can place  constraints on the properties of brown dwarfs and hot Jupiters atmospheres. To fully exploit these methods, high signal-to-noise spectral libraries with well-understood uncertainties are essential.}

We present  a  high signal-to-noise spectral library (1.10-1.69~$\mu$m) of the thermal emission of 76 brown dwarfs and  hot Jupiters. All our spectra have been acquired with the Hubble Space Telescope's Wide Field Camera 3 instrument and its G141 grism.

The near-infrared spectral types of these objects range from L4 to Y1. Eight of our targets have estimated masses below the deuterium-burning limit. {We analyze the database to identify peculiar objects and/or multiple systems, concluding that this sample  includes two very-low-surface-gravity objects and five intermediate-surface-gravity objects.  In addition, spectral indices designed to search for composite atmosphere brown dwarfs, indicate that eight objects in our sample are strong candidates to have such atmospheres. None of these objects are overluminous, thus their composite atmospheres are unlikely a companion-induced artifact. }Five of the eight confirmed candidates have been reported as photometrically variable, suggesting that composite atmospheric indices are useful in identifying  brown dwarfs with strongly heterogeneous cloud covers.


We compare hot Jupiters and brown dwarfs in a near-infrared color-magnitude diagram. {We confirm that 
 the coldest hot Jupiters in our sample have  spectra similar to mid-L dwarfs, and the hottest hot Jupiters have spectra similar to those of M-dwarfs.}
Our sample provides a uniform dataset of a broad range of ultracool atmospheres, allowing large-scale, comparative studies, and providing a HST legacy spectral library.
\end{abstract}




\section{Introduction} \label{sec:intro}


 
{Over the past decade, increasingly detailed observations are available on a wide range of objects:  spectroscopic information is now available on hot Jupiters} \citep[e.g.,][]{Stevenson2014, Ranjan2014, Line2016, Evans2017, Sheppard2017}, directly imaged exoplanets \citep[e.g.,][]{Samland2017, Rajan2017}, and over a thousand brown dwarfs \citep[e.g.,][]{Cushing2005, Kirkpatrick, 2006ApJ...637.1067B, Apai2013, Buenzli2014,Schneider2015}. {These  datasets have enabled major steps in the complexity and quantitative evaluation of atmospheric models}. A particularly significant advancement has been the adaptation of Bayesian modeling framework first for hot Jupiters \citep[e.g.,][]{Madhusudhan2009,Line2013,Lee2014,Gandhi2017,Pinhas2018,FisherHeng2018} and smaller transiting planets \citep[][]{Benneke2012}, then for directly-imaged exoplanets \citep[][]{Todorov2016,Lavie2017}, and, most recently, for brown dwarfs \citep[][]{Line2015,Line2017,Madhusudhan2016}. 

{The Bayesian modeling framework -- although often less detailed than forward models -- has two key advantages: first, it provides a probabilistic assessment of the fitted parameters and degeneracies, even if the parameter space is highly complex. Second, it enables systematic, comprehensive, unbiased modeling of large number of atmospheres, allowing for detailed comparative studies of the posterior distributions of the model parameters (e.g., C/O ratios, molecular abundances, surface gravities). Although the information provided by posterior distributions is very powerful, it must be remembered that the probabilities derived for individual model components are only correct {under the assumption that the data and uncertainties are correctly represented by the priors, and that the modeling framework itself is complete and correct}. For example, data with hidden biases (resulting in incorrect priors)  will yield systematically incorrect posteriors. In this sense, due to typical observational biases, it is particularly challenging to compare objects over a broad range of parameters (e.g., very different temperatures or surface gravities). In fact, no  spectral library with well understood systematics exists for ultracool atmospheres (hot Jupiters, directly imaged exoplanets, brown dwarfs).
{\em In short, to exploit the potential of atmospheric retrievals and enable rigorous comparative studies of atmospheres, homogeneous spectral datasets with well-understood systematics are required.} }

{Comprehensive spectral libraries exist for brown dwarfs} \citep[the SpeX spectral library\footnote{{The SpeX Prism Library is composed by low-resolution, near-infrared spectra, primarily of low-mass stars and brown dwarfs, obtained with the SpeX spectrograph mounted on the 3~m NASA Infrared Telescope Facility on Mauna Kea, Hawaii. The data provided here have been obtained using the prism-dispersed mode of SpeX with an average resolution of $\sim$120 and spectra spanning 0.90-2.50 $\mu$m: \url{http://pono.ucsd.edu/~adam/browndwarfs/spexprism/library.html}} }, the Montreal spectral library\footnote{https://jgagneastro.wordpress.com/the-montreal-spectral-library/}, and references therein]{Kirkpatrick1999,Kirkpatrick_2000,Leggett2000,Burgasser2002,McLean,Cushing2005,Kirkpatrick,Burgasser2006} built from ground-based spectroscopy of hundreds of brown dwarfs in dozens of studies. These libraries have played and continue to play essential roles in a broad range of brown dwarf studies. {However, existing ground-based spectral libraries were built from data that are non-uniform in terms of instruments, observing conditions,  setups, and usually reduced slightly differently by different groups}.

\begin{figure*}
\centering
\includegraphics[width=0.7\textwidth]{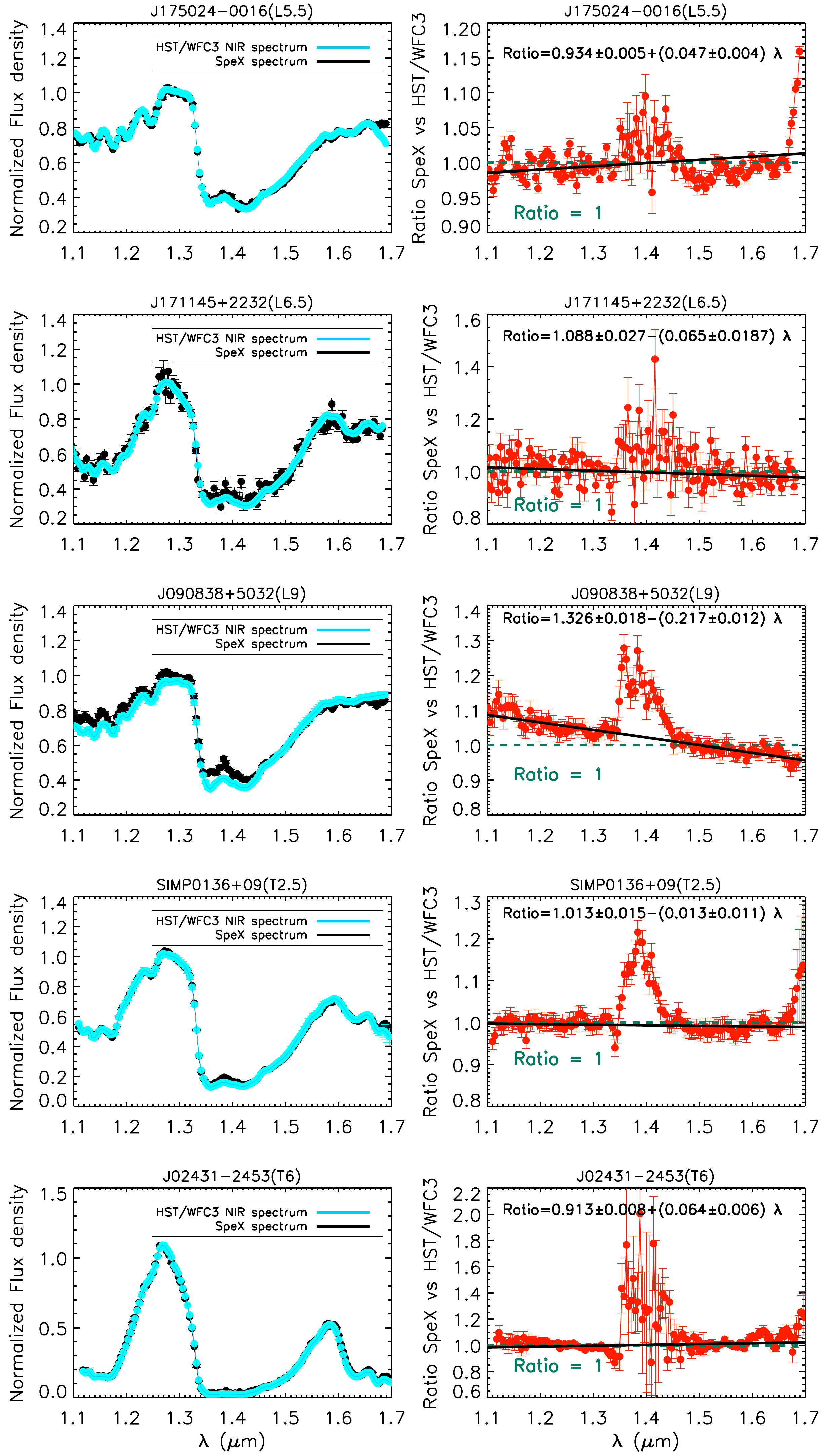}

\caption{\label{SpeX_HST} {In the left column, we show a direct comparison between the SpeX and the HST/WFC3 near infrared spectra. In the right column we show the ratio between the SpeX and the HST/WFC3 near infrared spectra. These plots reveal clear differences in the water band at 1.4~$\mu$m between both spectra, and color trends shown by non-zero slopes of the linear fits.}}
\end{figure*}

While {these spectral libraries remain powerful}, these datasets have several limitations for atmospheric retrieval studies: first, it is not possible to reliably capture the variety of differences in data acquisition, quality, and reduction with priors due to the unknowns involved. Second, ground-based observations unavoidably are influenced by telluric absorption, most notably by water bands. Although it is possible to correct for these to some extent, their time-varying nature and the optical depth in the bands lead to limited reliability in these bands. In fact, quantitative comparisons (Apai, priv. comm.) of some brown dwarfs with SpeX spectra and Hubble Space Telescope Wide Field Camera 3 near-infrared grism spectra, revealed  mismatches in water band shape, and overall color (wavelength-dependent slope). {We show these differences in Fig. \ref{SpeX_HST}. In the left column, we show the direct comparison of the near infrared SpeX (black) and HST/WFC3 spectra (blue) for randomly selected brown dwarfs with spectra in the SpeX and in our HST/WFC3 near infrared library, and with spectral types between L5.5 and T6.  In the right column, we show the ratio of the SpeX and the HST/WFC3 near infrared spectra for the object in the left column. We show a green line indicating where a perfect match between the SpeX and the HST/WFC3 spectra should be (ratio Spex vs HST/WFC3 equal 1). In addition, we fit a line to the slope of the ratio between the two spectra, avoiding the water band at 1.4~$\mu$m, (see black line) showing that in most of the cases, the slope is non-zero, indicating color trends on the SpeX spectra. In these plots a common mismatch between the SpeX and the HST/WFC3 near infrared  spectra at the 1.4~$\mu$m  water band is also evident, due to imperfect telluric correction}. Given that the photometric precision and instrumental systematics of the HST/WFC3 instrument are very well understood, and that the HST/WFC3 near infrared spectra are not affected by  tellurics, the comparison reveals that low-level biases exist in the ground-based spectral libraries.  While {these} corrections are well-suited for forward-modeling and object-to-object comparisons, they are often limiting for retrieval studies.

{For example, \citet[][]{Line2015} applied Bayesian atmospheric retrieval tools to SpeX spectra to derive thermal structures and molecular abundances of some brown dwarfs. Nevertheless, they could only reach convergence in their Markov Chain Monte Carlo retrievals if they assumed that the SpeX spectral uncertainties were underestimated. Thus, they artificially increased the SpeX spectral uncertainties that could reach a maximum of a factor of 100. 
In this case, it is impossible to disentangle if \citet[][]{Line2015} models were incomplete, or the uncertainties of the datasets were not accurately estimated and understood. Therefore, to properly test retrieval models, there is an obvious need for a uniform, space-based spectral library with well-understood spectral uncertainties.}



{In this paper, {we are presenting a high signal-to-noise  spectral library with 76 near-infrared WFC3/HST spectra of brown dwarfs}, low-mass companions to stars, and hot Jupiters. Our study supplements the HST Cloud Atlas Treasury program data (GO 14241, PI: D. Apai) with other published datasets (see Section \ref{sec:data_red}), carefully analyzing and correcting for (typically very small) data reduction differences. {The advantage of the HST/WFC3 instrument is that it provides  near-infrared spectroscopy (1.10--1.69~$\mu$m, S/N up to {3,000} in the $J$-band) where the SED (Spectral Energy Distribution) of these objects peak, and we observe the dominant absorbing species for brown dwarfs}.}



{Besides the presentation of our HST/WFC3 near infrared spectral library, the goal of this study is to provide a comprehensive characterization of the objects of our sample. This spectral characterization is important to validate if the results provided by the retrieval models match with the expectations for a given object. In fact, \cite{Line2015} used two well-characterized T dwarfs, Gl 570D and HD 3651B, to test their retrievals, confirming that the effective temperatures, surface gravities, masses, radii, etc., were consistent with the expected values for those T-dwarfs. Therefore, we aim to provide a basic characterization of our sample, and also to identify peculiar objects: extremely red or blue brown dwarfs, revealing low surface gravity or low metallicity objects, and overluminous brown dwarfs, potentially revealing  multiple systems. In addition, we use spectral indices \citep{Burgasser2006,Burgasser2010,Bardalez_Gagliuffi} to search for  spectral binaries. As a by-product of this analysis, we found that these spectral indices can also be useful to search for variable brown dwarfs. 
In addition, we show the potential of our spectral library by performing a novel direct photometric and spectroscopic comparison between hot Jupiters and brown dwarfs, that can be only be accomplished at this level of accuracy with data acquired from the space. This comparison confirms that some hot Jupiters share effective temperatures and spectra with some M and L dwarfs.}

{In Section \ref{targets}, we describe the targets that we include in the near infrared HST/WFC3 spectral library.  In Section \ref{SpT} we derive spectral types for our sample using the SpeX Prism Spectral Library. In Section \ref{cmd} {we compare the L and T dwarfs with other brown dwarfs from \cite{Dupuy_Liu2012} using a color-magnitude diagram}. In Section \ref{lowg_objects} we search for low surface gravity objects in our sample using low surface gravity spectral indices. In Section \ref{composite}, we search for composite atmosphere objects within our sample. In Section \ref{classification} we measure the water and methane bands on objects of our sample, to trace the change on the depth of those bands with the near infrared spectral type. In Section \ref{hot_jupiters} we directly compare colors and spectra of brown dwarfs and hot Jupiter in a color-magnitude diagram and finding the best matching brown dwarf to every of the hot Jupiters in our sample. Finally, in Section \ref{conclusions} we present our conclusions.}

\section{Targets}\label{targets}

\begin{figure*}
\centering
\includegraphics[height=1.2\textwidth]{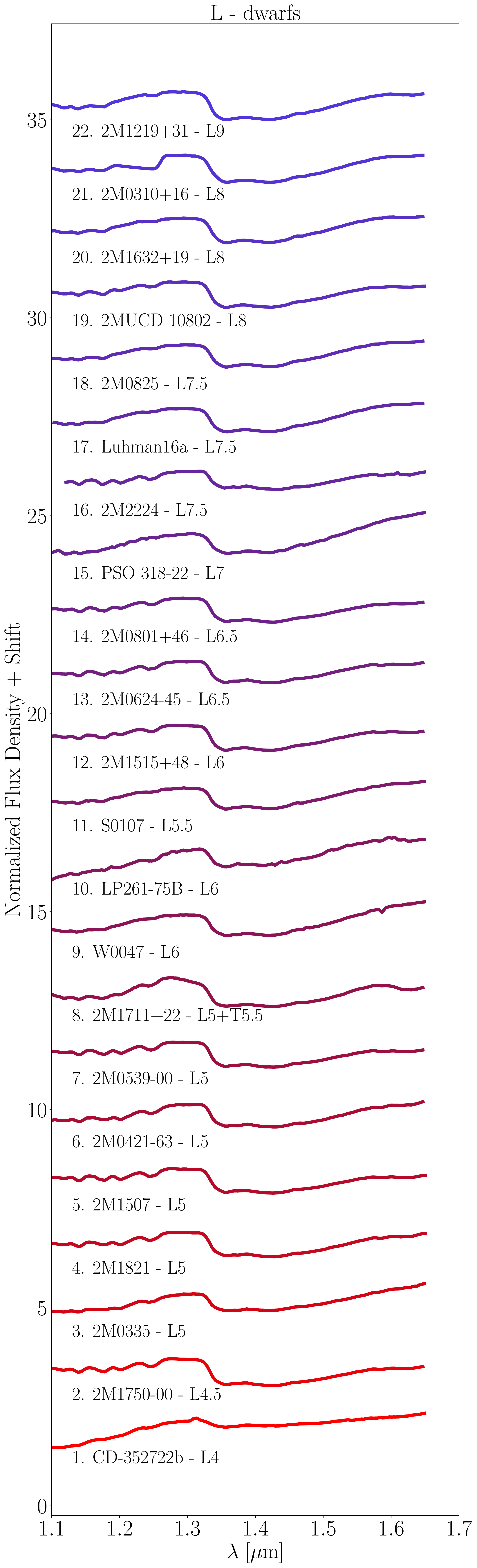}
\includegraphics[height=1.2\textwidth]{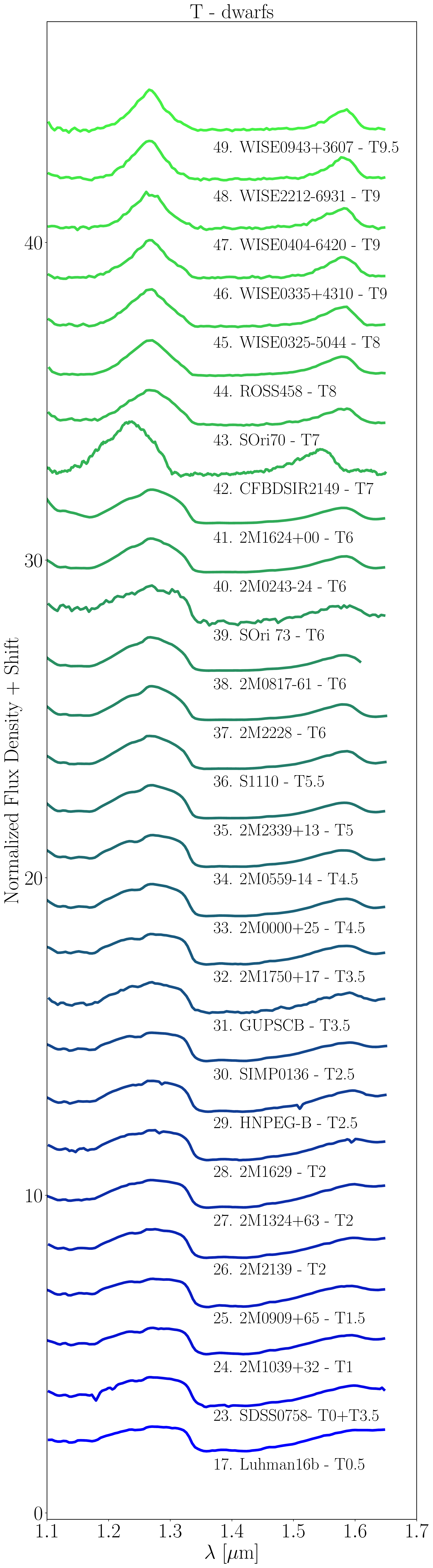}

\caption{\label{plot_L} L and T dwarfs with HST/WFC3 spectra used as a part of this study. Flux is normalized at 1.25~$\mu$m.}
\end{figure*}

\begin{figure*}
\centering
\includegraphics[height=1.2\textwidth]{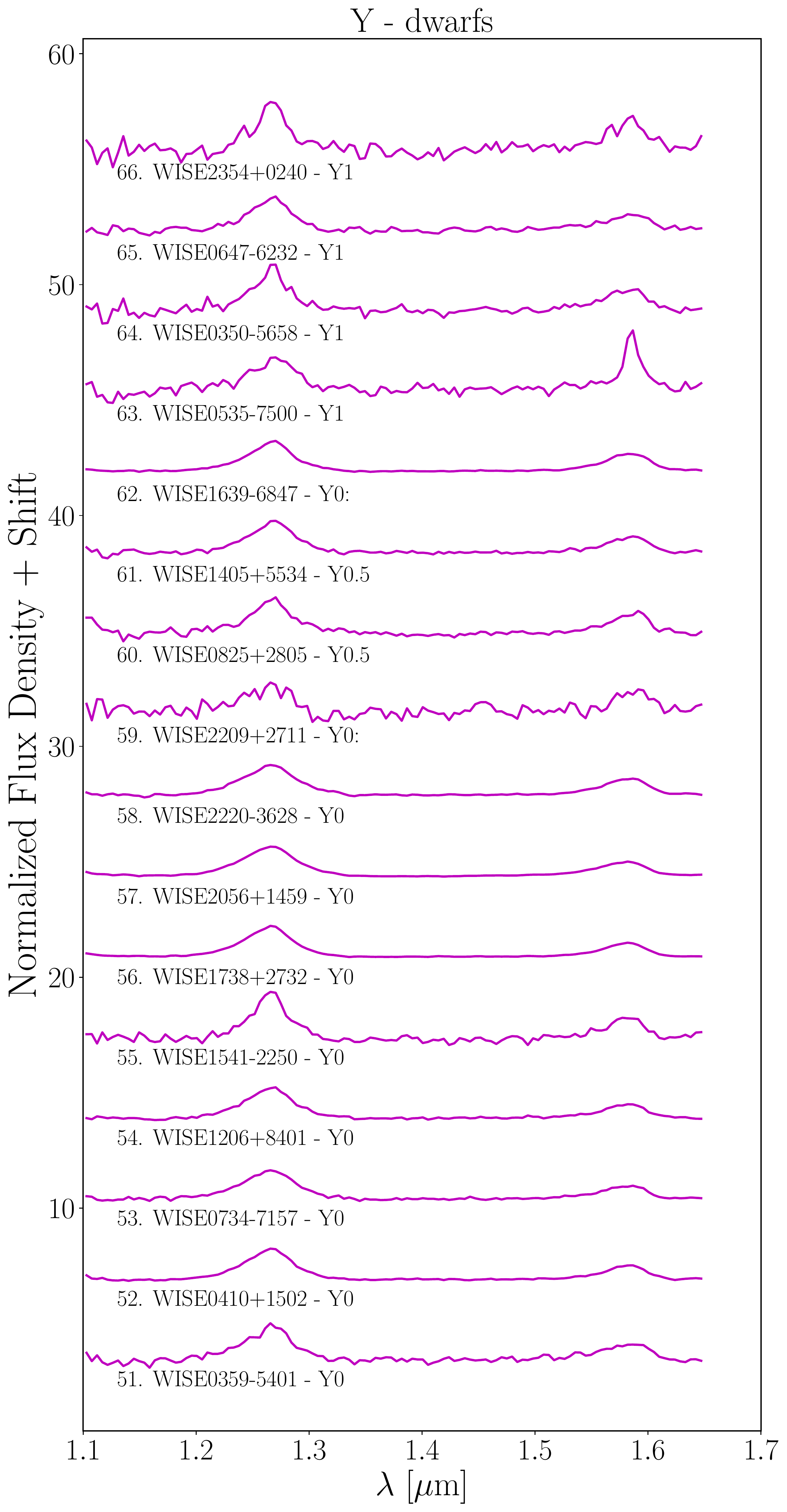}

\caption{\label{plot_Y} Y dwarfs with HST/WFC3 spectra used as a part of this study.  Flux is normalized at 1.25~$\mu$m.}
\end{figure*}

\begin{figure*}
\centering
\includegraphics[height=1.\textwidth]{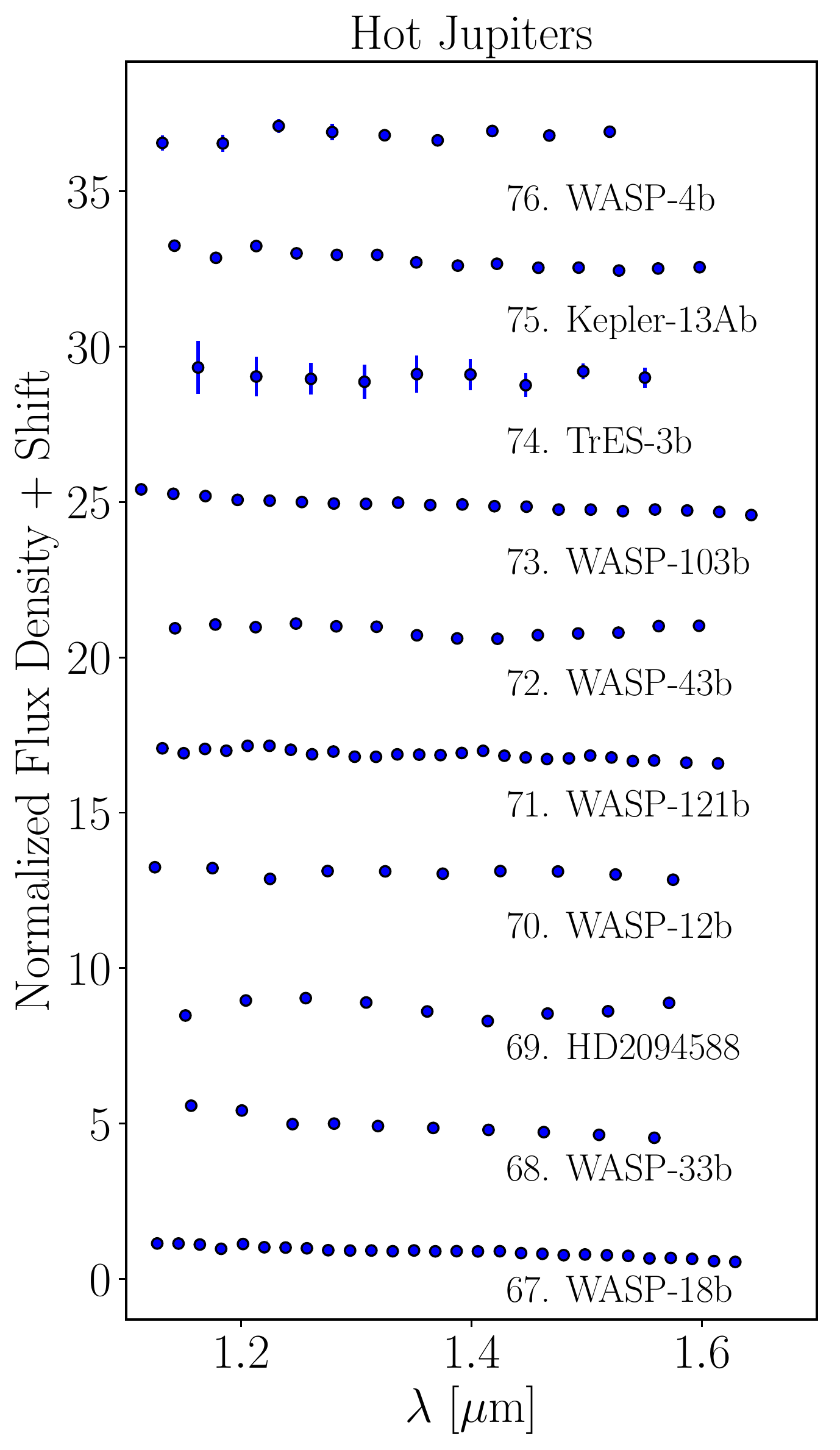}

\caption{\label{plot_Hot_Jupiters} Hot Jupiters with HST/WFC3 emission spectra used as a part of this study. Some error bars are smaller that the symbol for the measurement.}
\end{figure*}

We compiled {all} spectra of brown dwarfs  (including planetary-mass brown dwarfs), brown dwarf companions to stars, and hot Jupiter with emission spectra with 
published data from HST/WFC3 and its G141 grism \citep{MacKenty2010}. {In addition, we present  seven unpublished spectra observed as part of the \textit{Cloud Atlas} treasury program (GO 14241, PI Apai), and  other two brown dwarf spectra that belong to the HST program GO 13299 and 14051 (PI Radigan)}.
These spectra cover the wavelength range between $\sim$1.10 and 1.69~$\mu$m, with a spectral resolving power $R={\lambda \over \Delta \lambda}=130$ at 1.4 $\mu$m. The image scale of WFC3/IR is 0.13 arcsec/pixel.  In Tables \ref{all_data_hst} and \ref{all_hot_jupiters} we provide the list of objects with names, celestial coordinates, and HST program identifiers in which the objects were observed, as well as the most relevant references in which this spectra were first published {(\citealt{Buenzli2012}, \citealt{Buenzli2014}, \citealt{Buenzli2015}, \citealt{Yang2015}, \citealt{Yang2016}, \citealt{Lew2016},  \citealt{Manjavacas2018},     \citealt{Pena_ramirez2015},    \citealt{Zhou2018},  \citealt{Biller2018},  \citealt{Apai2013},  \citealt{Schneider2015}, \citealt{Line2016},  \citealt{Haynes2015}, \citealt{Cartier2017},  \citealt{Stevenson2014},  \citealt{Ranjan2014},  \citealt{Beatty2017},  \citealt{Stevenson2014_Nat}, \citealt{Evans2017},  \citealt{Sheppard2017}, and this work)}. In Table \ref{all_data_hst}, for brown dwarfs and substellar companions, we  specify their spectral types, {Two Micron All Sky Survey} (2MASS, \citealt[][]{Cutri2003})  photometry, and trigonometric parallaxes, {the signal-to-noise as measured at 1.25~$\mu$m from the corresponding HST programs, the HST program for which each object was observed, and the references in which these spectra were published}. For hot Jupiters (Table~\ref{all_hot_jupiters}) we also list the spectral types of the host stars, the star--planet separations, and the radii of the planets. In Figures~\ref{plot_L}, \ref{plot_Y}, and \ref{plot_Hot_Jupiters} we show the spectra of the objects used in our study, {76 in total {(22 L dwarfs, 28 T dwarfs, 16 Y dwarfs and 10 hot Jupiters)}, from which nine are presented here for the first time}. {Some of the objects in our sample in Table~\ref{all_data_hst} have been classified as planetary-mass objects (M $ < 13\ \mathrm{M_{Jup}}$). We list these objects in Table~\ref{planetary_mass} along with their estimated masses, ages, young moving group membership (if known), and key references.} 

{The observing log, including the observing dates, number of orbits per visit, exposure time of each single exposure, and the number of single exposures per orbit is compiled in Table~\ref{quality_spectra} in the Appendix}.

{The datasets and the data reduction  by the  authors that  published  the spectra presented in Tables \ref{all_data_hst} and \ref{all_hot_jupiters} are described in Appendix \ref{sec:data_red}}.
 
\begin{longrotatetable}
\begin{deluxetable*}{llcclrlllllll}
\tablecaption{Sample of L, T and Y dwarfs with HST/WFC3 spectroscopy \label{all_data_hst}}
\tablewidth{600pt}
\tabletypesize{\scriptsize}
\tablehead{
\colhead{Num.} & \colhead{Name} & 
\colhead{SpT} & \colhead{$\mathrm{SpT_{SpeX}^{a}}$} &\colhead{R. A. (J2000)} & 
\colhead{Dec. (J2000)} & \colhead{$J$ [mag]} & 
\colhead{$H$  [mag]} & \colhead{$K_{s}$ [mag]} & \colhead{$\pi_{T}$ [mas] } & \colhead{$\mathrm{SNR^{b}}$} & \colhead{$\mathrm{HST\, \,GO^{c}}$} & \colhead{Ref.} 
} 
\startdata
	1 & CD-352722b              & L4 & L7.5 & 06 09 19.21 & -35 49 31.20 &  &   &  & 44.63$\pm$0.03  & 45 & 14241 &  TS\\ 
	2 & 2MASS J17502484--0016151 & L4.5 & L5 & 17 50 24.84 & -00 16 15.11 & 13.29$\pm$0.02 & 12.41$\pm$0.02 & 11.84$\pm$0.02 & 45.16$\pm$0.16 & 2000 & 12550 &  1 ,13  \\ 
	3 & 2MASS J03552337+1133437 & L5 & L5.5 &  03 55 23.37 & +11 33 43.70 & 14.05$\pm$0.02 & 12.53$\pm$0.03 & 11.53$\pm$0.02& 108.70$\pm$2.36 & 322 &14241 & TS, 14   \\     
	4 & 2MASS J18212815+1414010 & L5 & L5.5 & 18 21 28.15 & +14 14 01.04 & 13.43$\pm$0.02 & 12.39$\pm$0.02 & 11.65$\pm$0.02 & 122.00$\pm$13.00 & 370 &13176 & 2, 15 \\    
	5 & 2MASSW J1507476--162738  & L5 & L5.5 & 15 07 47.69 & -16 27 38.62 & 12.83$\pm$0.03 & 11.89$\pm$0.02 & 11.31$\pm$0.02 & 106.50$\pm$0.19 & 526 &13176 & 2 \\    
	6 & 2MASSI J0421072--630602  & L5 & L5   & 04 21 07.19 & -63 06 02.25 & 15.56$\pm$0.05 & 14.28$\pm$0.04 & 13.44$\pm$0.04 & & 1250 & 12550 & 1\\ 
	7 & 2MASS J05395200--0059019 & L5 & L5   & 05 39 52.00 & -00 59 01.90 & 14.03$\pm$0.03 & 13.10$\pm$0.03 & 12.52$\pm$0.02 & 81.97$\pm$2.69  & 2000 & 12550 & 1, 13 \\ 
	8 & 2MASSI J1711457+223204&L5.0+T5.5&   & 17 11 45.73 & +22 32 04.41 & 17.08$\pm$0.18 & 15.79$\pm$0.11 & 14.72$\pm$0.09 & 33.11$\pm$4.71 & 1111 & 12550 & 1, 16 \\     
	9 & 2MASS J00470038+6803543 & L6 & L5.5 & 00 47 01.06 & +68 03 52.10 & 15.60$\pm$0.07 & 13.97$\pm$0.04 & 13.05$\pm$0.03 & 82.00$\pm$3.00 & 370 &14241 & 3, 17  \\ 
	10 & LP261--75B              & L6 & L4.5 & 09 51 04.60 & +35 58 09.80 & 17.22$\pm$0.21 & 15.89$\pm$0.14 & 15.14$\pm$0.13 & 29.60$\pm$2.80 & 333 & 14241 & 4, 18 \\ 
	11 & 2MASS J01075242+0041563& L6 & L5.5 & 01 07 52.42 & +00 41 56.40 & 15.82$\pm$0.06 & 14.51$\pm$0.04& 13.71$\pm$0.04 & 64.13$\pm$4.51 & 208 & 14241 &  TS, 19 \\     
	12 & 2MASSW J1515008+484742 & L6 & L6   & 15 15 00.83 & +48 47 41.69 & 14.11$\pm$0.03 & 13.09$\pm$0.03 & 12.50$\pm$0.02 &  & 1666 & 12550 & 1, 20\\ 
	13 & 2MASS J06244595--4521548& L6.5 & L5 & 06 24 45.95 & -45 21 54.89 & 14.48$\pm$0.03 & 13.33$\pm$0.03 & 12.59$\pm$0.03 & 86.21$\pm$4.46 & 1428 & 12550 & 1, 19 \\ 
	14 & 2MASSW J0801405+462850 & L6.5 & L6 & 08 01 40.56 & +46 28 49.84 & 16.27$\pm$0.13 & 15.45$\pm$0.14 & 14.53$\pm$0.10 & & 1428 & 12550 & 1  \\ 
    15 & PSO J318.5--22          & L7 & L7.5 & 21 14 08.03 & -22 51 35.84 & 16.71$\pm$0.19 &	15.72$\pm$0.17 &	14.74$\pm$0.12 & 45.10$\pm$1.70 & 285 & 14188 & 10, 21  \\
  	16 & 2MASSW J2224438--015852 & L7.5 &L4.5& 22 24 43.82 & -01 58 52.14 & 14.07$\pm$0.03 & 12.81$\pm$0.03 & 12.02$\pm$0.02 & 86.20$\pm$1.10 & 47 & 14241 & 2, 22 \\ 
	17 & Luh 16AB              & L7.5+T0.5& & 10 49 18.92 & -53 19 10.08 & 11.5$\pm$0.04 & 10.37$\pm$0.04 & 9.44$\pm$0.07 & 501.40$\pm$0.09  & 500 & 13280 & 5, 23   \\ 
	18 & 2MASSI J0825196+211552 & L7.5 & L6 & 08 25 19.69 & +21 15 52.12 & 15.10$\pm$0.03 & 13.79$\pm$0.03 & 13.03$\pm$0.03 & 93.46$\pm$0.87 & 2000 &12550 & 1, 24  \\  
	19 &2MUCD 10802             & L8 & L7.5 & 09 08 38.04 & +50 32 08.82 & 14.55$\pm$0.02 & 13.48$\pm$0.03 & 12.95$\pm$0.03 & & 967 & 14241 & 1  \\ 
	20 &2MASS J16322911+1904407 & L8 & L8   & 16 32 29.11 & +19 04 40.71 & 15.86$\pm$0.07 & 14.61$\pm$0.04 & 14.00$\pm$0.05 & 65.79$\pm$2.16 & 1428 & 12550 & 1, 24  \\ 
	21 &2MASSW J0310599+164816  & L8 & T2   & 03 10 59.87 & +16 48 15.60 & 16.02$\pm$0.08 & 14.93$\pm$0.07 & 14.31$\pm$0.07 & 36.90$\pm$3.40 & 1428 & 12550 & 1, 16  \\ 
	22 &2MASS J12195156+3128497 & L9 & L8   & 12 19 51.56 & +31 28 49.71 & 15.91$\pm$0.08 & 14.91$\pm$0.07 & 14.31$\pm$0.07 & & 1428 &12550 & 1 \\
	23 &SDSS J075840.33+324723.4& T0+T3.5 & & 07 58 40.03 & +32 47 18.39 & 14.95$\pm$0.04 & 14.11$\pm$0.04 & 13.88$\pm$0.06 & & 3333 &13299 & TS   \\ 
	24 &2MASS J10393137+3256263 & T1 & T2   & 10 39 31.38 & +32 56 26.40 & 16.41$\pm$0.15 & 15.34$\pm$0.11 & 15.15$\pm$0.16 & & 1250 & 12550 &  1\\ 
	25 &2MASS J09090085+6525275 & T1.5 & T1 & 09 09 00.86 & +65 25 27.57 & 16.03$\pm$0.09 & 15.21$\pm$0.09 & 15.17$\pm$0.15 & &  1250 &12550 & 1  \\ 
	26 &2MASS J21392676+0220226 & T2 & T1.5 & 21 39 26.77 & +02 20 22.70 & 14.71$\pm$0.01 & 14.16$\pm$0.05 & 13.58$\pm$0.04 & 101.50$\pm$2.00 & 172 & 12314 & 11, 6, 25 \\     
	27 &2MASS J13243553+6358281 & T2 & T0.5 & 13 24 35.54 & +63 58 28.15 & 15.59$\pm$0.07 & 14.57$\pm$0.06 & 14.05$\pm$0.06 & & 1111  & 12550 &  1, 6 \\ 
	28 &2MASS J16291840+0335371 & T2 & T2   & 16 29 18.62 & +03 35 35.01 & 15.29$\pm$0.04 & 14.48$\pm$0.03 & 14.04$\pm$0.03 & & 3333 &14051 & TS  \\ 
	29 &HN PEG B                & T2.5&T2.5 & 21 44 28.47 & +14 46 07.80 & 15.86$\pm$0.03 & 15.40$\pm$0.03 & 15.12$\pm$0.03 & 54.37$\pm$0.85 & 2222 & 14241 & 9 \\ 
	30 &SIMP J013656.5+093347.3 & T2.5&T2.5 & 01 36 56.62 & +09 33 47.30 & 13.46$\pm$0.03 & 12.77$\pm$0.03 & 12.56$\pm$0.02 & 162.87$\pm$1.06 & 370 &  12314 & 6, 11, 26 \\ 
	31 &GU PSC B                & T3.5&T1.5 & 01 12 36.48 & +17 04 31.80 &  &  &   & 21.00$\pm$0.07  & 100 & 14241 & TS \\    
	32 &2MASS J17503293+1759042 &T3.5 &T3.5 & 17 50 32.94 & +17 59 04.30 & 16.34$\pm$0.10 & 15.95$\pm$0.13& 15.48$\pm$0.189 & 36.23$\pm$4.46 & 909 & 12500 & 1, 16  \\ 
	33 &2MASS J00001354+2554180 &T4.5 &T4.5 & 00 00 13.54 & +25 54 18.10 & 15.06$\pm$0.04 & 14.73$\pm$0.07 & & 14.84$\pm$0.12 & 667  & 12550 & 1 \\     
	34 &2MASS J05591914--1404488 &T4.5 &T4.5 & 05 59 19.14 & -14 04 48.88 & 13.80$\pm$0.02 & 13.67$\pm$0.04 & 13.57$\pm$0.06 &96.15$\pm$0.96  & 833 & 12500 &1, 22 \\ 
	35 &2MASS J2339101+135230   & T5.4 & T5 & 23 39 10.25 & +13 52 28.50 & 16.24$\pm$0.11 & 15.82$\pm$0.15 & 16.14$\pm$0.31 & & 909 & 12500 &1 \\ 
	36 &2MASS J1110100+0116130  &T5.5 &T5.5 & 11 10 09.99 & +01 16 13.09 & 16.34$\pm$0.12 & 15.92$\pm$ 0.14 & 52.10$\pm$1.20 &52.10$\pm$1.20 & 147 &14241 &  TS, 27 \\     
	37 &2MASS J2228288--4310262  & T6 & T6   & 22 28 28.89 & -43 10 26.27 & 15.66$\pm$0.07 & 15.36$\pm$0.12 & 15.29$\pm$0.21 & 92.10$\pm$2.60 &  243 & 12314 & 12, 6, 25 \\ 
	38 &2MASS J0817300--6155158  & T6 & T6   & 08 17 30.01 & -61 55 15.82 & 13.61$\pm$0.02 & 13.53$\pm$0.03 & 13.52$\pm$0.04 & 204.08$\pm$12.49 & 1000 & 12550 & 1, 19  \\ 
	39 &S Ori 73                & T4.5&T7.5 & 05 38 10.10 & -02 36 26.00 &  &  &    & & 5 & 12217 & 7 \\ 
	40 &2MASS J0243137--245329   & T6 & T6   & 02 43 13.72 & -24 53 29.80 & 15.38$\pm$0.05 & 15.14$\pm$0.11 & 15.22$\pm$0.17 & 93.46$\pm$3.49 & 667 & 12550 & 1, 22  \\ 
	41 &2MASS J1624143+0029158  & T6 & T6   & 16 24 14.37 & +00 29 15.82 & 15.49$\pm$0.05 & 15.52$\pm$0.10 & 11.00$\pm$0.20 & 90.91$\pm$1.65 & 909 & 12550 & 1, 28 \\ 
	42 &CFBDSIR2149--0403        & T7 &      & 21 49 47.20 & -04 03 08.90 &  &    &   &18.30$\pm$1.80 & 6.7 &14241 & TS, 29 \\    
    43 &S Ori 70                & T7 & T7.5 & 05 38 14.19 & -02 45 11.80 & 20.91$\pm$0.07 & 20.83$\pm$0.12 & 20.91$\pm$0.15  & & 5&12217 & 7  \\ 
	44 &ROSS458C                & T8 & T7.5 & 13 00 41.94 & +12 21 14.72 & 16.68$\pm$0.01 & 17.01$\pm$0.04 & 16.89$\pm$0.06 & 85.54$\pm$1.53& 65 & 14241 & TS  \\ 
	45 &WISEA J032504.5--504403.0& T8 &      & 03 25 04.52 & -50 44 03.00 & 18.43$\pm$0.26 & 16.21$\pm$0.15 & $>$12.918 & & 8 & 13178 & 8  \\ 
	46 &WISEA J033515.1+431044.7& T9 &      & 03 35 15.07 & +43 10 44.70 & $>$18.652 &  & & 14.52$\pm$0.06 & 5 &12970 & 8  \\ 
	47 &WISEA J040443.5--642030.0& T9 &      & 04 04 43.50 & -64 20 30.00  & 18.44$\pm$0.18 & 15.73$\pm$0.06 & $>$12.297 & & 23 &13178 & 8 \\    
	48 &WISEA J221216.3--693121.6& T9 &      & 22 12 16.27 & -69 31 21.60 & 17.26$\pm$0.12 & 14.87$\pm$0.06 & $>$12.278 &  & 4 & 12970 & 8  \\ 
	49 &WISEA J094306.0+360723.3& T9.5 &    & 09 43 05.99 & +36 07 23.57 & 19.74$\pm$0.05 & 20.37$\pm$0.20 &  & & 2 & 12970 &  8\\     
	50 &WISEA J154214.0+223005.2& T9.5 &    & 15 42 14.00 & +22 30 05.20 & 20.25$\pm$0.13 & 21.80$\pm$0.80 & 96.00$\pm$41.00  & & 2 & 12230 & 8, 27 \\ 
	51 &WISEA J035934.1--540154.8& Y0 &      & 03 59 34.07 & -54 01 54.80 & $>$19.031 &  & 15.38$\pm$0.05 & 75.40$\pm$6.62 & 2 & 12970 & 8, 34\\ 
	52 &WISEA J041022.8+150247.9& Y0 &      & 04 10 22.75 & +15 02 47.90  & $>$18.170 &  & 14.11$\pm$0.05 & 153.40$\pm$4.00 &2 & 12970 & 8, 34\\     
	53 &WISEA J073444.0--715743.8& Y0 &      & 07 34 44.03 & -71 57 43.80 & 20.13$\pm$0.08 &  & & 67.60$\pm$8.70 & 2 & 12970 & 8, 34 \\ 
	54 &WISEA J120604.3+840110.5& Y0 &      & 12 06 04.38 & +84 01 10.60 & $>$18.734 &  & 15.06$\pm$0.05 & 85.10$\pm$9.30 & 2 & 13178 &8, 34  \\ 
	55 &WISE J154151.7--225024.9 & Y0 &      & 15 41 51.66 & -22 50 24.99 & 20.99$\pm$0.03 & 20.99$\pm$0.52 & & 167.10$\pm$4.20 &2 & 12970 & 8, 34 \\ 
	56 &WISEA J173835.5+273258.8& Y0 &      & 17 38 35.53 & +27 32 59.10 & 19.47$\pm$0.08 & 20.39$\pm$0.33 &  & 128.50$\pm$6.30 & 2 & 12230 & 8, 34\\     
	57 &WISEA J205628.9+145953.6& Y0 &      & 20 56 28.92 & +14 59 53.22 & 16.48$\pm$0.07 & 13.84$\pm$0.04 & 11.73$\pm$0.25 & 138.30$\pm$3.90 &5  &12230 & 8, 34  \\ 
	58 &WISEA J222055.3--362817.5& Y0 &      & 22 20 55.32 & -36 28 17.50 & $>$18.772 &  & 14.71$\pm$0.06 & 84.10$\pm$5.90 & 3 &12970 & 8, 34\\ 
	59 &WISEA J220905.8+271143.6& Y0: &     & 22 09 05.73 & +27 11 44.00 & 22.58$\pm$0.14 & 22.98$\pm$0.31 &  154.40$\pm$5.70 & & 1 &  12970 &8, 34 \\ 
	60 &WISEA J082507.4+280548.2& Y0.5 &    & 08 25 07.36 & +28 05 48.56 & $>$18.444 &  & 14.58$\pm$0.06 & 139.00$\pm$4.30 & 1 & 12970 & 8, 34 \\ 
	61 &WISEA J140518.3+553421.3& Y0.5 &    & 14 05 18.39 & +55 34 21.40 & 21.06$\pm$0.06 & 21.45$\pm$0.41 &  & 144.30$\pm$8.60 & 1 &12230 & 8, 34  \\     
	62 &WISEA J163940.8--684739.4& Y0pec &   & 16 39 40.83 & -68 47 38.60 & 20.57$\pm$0.05 &   &  &228.10$\pm$8.90 & 5 & 12970 & 8, 34 \\ 
	63 &WISEA J053516.9--750024.6& Y1 &      & 05 35 16.87 & -75 00 24.60 & 17.94$\pm$0.14 & 14.90$\pm$0.05 & $>$12.349 & 79.50$\pm$8.80 & 1 & 12970 &8, 34 \\ 
	64 &WISEA J035000.3--565830.5& Y1 &      & 03 50 00.31 & -56 58 30.50  & $>$18.699 &  & 14.74$\pm$0.04 &168.80$\pm$8.50 & 1 & 12230 &  8, 34 \\ 
	65 &WISEA J064723.2--623235.4& Y1 &      & 06 47 23.24 & -62 32 35.40 & 22.45$\pm$0.07 &   &   & 83.70$\pm$5.70 & 1 & 12970 &8, 34 \\ 
	66 &WISEA J235402.8+024014.1& Y1 &      & 23 54 02.79 & +02 40 14.10 & $>$18.263 &  & 15.01$\pm$0.09 & & 1 &13178 & 8\\ 
\enddata
\tablecomments{{References for first publication of HST/WFC3 near infrared spectra:} TS - This study, [1] - \cite{Buenzli2014}, [2] - \cite{Yang2015}, [3] - \cite{Lew2016}, [4] - \cite{Manjavacas2018}, [5] - \cite{Buenzli2015}, [6] - \cite{Yang2016}, [7] - \cite{Pena_ramirez2015}, [8] - \cite{Schneider2015}, [9] - \cite{Zhou2018}, [10] - \cite{Biller2018}, [11] - \cite{Apai2013}, [12] - \cite{Buenzli2012}.\\
References for trigonometric parallax measurements: [13] - \cite{Andrei2011}, [14] - \cite{Faherty2013}, [15] - \cite{Sahlmann2016}, [16] - \cite{Vrba2004}, [17] - \cite{Gizis2015}, [18] - \cite{Liu_Dupuy_Allers2016}, [19] - \cite{Faherty2012}, [20] - \cite{Wang2018},  [21] - \cite{Liu2013}, [22] - \cite{Dupuy_Liu2012}, [23] - \citep{Bedin2017}, [24] - \cite{Dahn2002}, [25] - \cite{Marocco2013}, [26] - \cite{Weinberger}, [27] - \cite{Tinney2014}, [28] - \cite{Tinney2003}, [29] - \cite{Delorme2017}, [30] - \cite{Marsh2013}, [31] - \cite{Luhman_Esplin2016}, [32] - \cite{Kirkpatrick2011}, [33] - \cite{Leggett2017}, [34] - \cite{Martin2018} }
\tablenotetext{a}{{$\mathrm{SpT_{SpeX}^{a}}$ is the spectral type of each  the  spectra compiled in this work derived by comparison to the SpeX Prism Spectral Library}}
\tablenotetext{b}{SNR measured at 1.25~$\mu$m}
\tablenotetext{c}{Further details of the program-specific observation plan can be found at:  {http://www.stsci.edu/cgi-bin/get-proposal-info?id=\#\#\#\#\#\&submit=Go\&observatory=HST},  where \#\#\#\#\# should be replaced by the given GO program number.}
\end{deluxetable*}
\end{longrotatetable}

\begin{table*}
\begin{center}
\scriptsize
\caption{Sample of hot Jupiters with HST/WFC3 spectroscopic data \label{all_hot_jupiters}}
\begin{tabular}{l l l l l c c c c c c}
\hline
\hline
	Num. & Name & RA & DEC & SpT host & $\pi_{T}^{a}$ [mas] & Separation [au] & Planet radius [$\mathrm{R_{Jup}}$] & $SNR^{b}$ & HST Prog. & Ref.   \\ 
    \hline
    67 & WASP-18b &		01 37 25.03	&	-45 40 40.39 &	F6V &	7.91$\pm$0.30& 0.02014$\pm$0.00034 & 1.165$\pm$0.077  & 37.8 &13467	& 9  \\
    68 &WASP-33b   & 02 26 51.06 & +37 33 01.73  &  A5 & 8.51$\pm$0.24 & 0.02555$\pm$0.00043 & 1.497$\pm$0.045  &38.9	&  12495 & 2 \\ 
	69 & HD 209458B   & 22 03 10.77 & +18 53 03.54 & G0V & 20.47$\pm$0.02 & 0.04723$\pm$0.00079 & 1.359$\pm$0.015   & 6.4	& 13467 & 1 \\ 
	70 &WASP-12b   & 06 30 32.79 & +29 40 20.29 & G0V  & 2.57$\pm$0.27& 0.02253$\pm$0.00038 &  1.790$\pm$0.090    & 12.3	&13467 & 4  \\ 
    71 &WASP-121b    & 07 10 24.06 & -39 05 50.55  & F6V &	3.82$\pm$0.25 & 0.02544$\pm$0.00050 & 1.865$\pm$0.044   &13.3 &14767 &  8 \\ 
    72 &WASP-43b    & 10 19 38.01 & -09 48 22.59 & K7V & $11.49\pm0.04$& 0.01424$\pm$0.00041&  0.930$\pm$0.070 &12.5	&13467 & 7 \\ 
	73 &WASP-103b   & 16 37 15.57 & +07 11 00.07 & F8V & 2.13$\pm$0.16 & 0.01987$\pm$0.00033 & 1.528$\pm$0.073    & 8.3	& 14050 &  3\\ 
	74 &TrES-3b   & 17 52 07.02 & +37 32 46.18 & K0V  &	$4.29\pm0.02$ & 0.02272$\pm$0.00038 & 1.336$\pm$0.031   & 2.5  &12181 & 5 \\ 
    75 &Kepler-13Ab    & 19 07 53.15 & +46 52 05.91 & A0 &$1.91\pm0.01$ & 0.04171$\pm$0.00078&  1.406$\pm$0.038 &10.8	 &13308 & 6 \\ 
	76 &WASP-4b   & 23 34 15.08 & -42 03 41.14 & G7V  &	3.63$\pm$0.70 & 0.02304$\pm$0.00042 &  1.341$\pm$0.023    & 4.4  &12181 & 5 \\

\hline
\end{tabular}
\end{center}
\tablecomments{[1] - \cite{Line2016}, [2] - \cite{Haynes2015}, [3] - \cite{Cartier2017}, [4] - \cite{Stevenson2014}, [5] - \cite{Ranjan2014}, [6] - \cite{Beatty2017}, [7] - \cite{Stevenson2014_Nat}, [8] - \cite{Evans2017}, [9] - \cite{Sheppard2017}}
\tablenotetext{a}{Trigonometric parallaxes delivered by \cite{Gaia2018}}
\tablenotetext{b}{{The SNR is measured between 1.05 and 1.65 $\mu$m}}
\end{table*}

\begin{deluxetable*}{llccccc}
\tablecaption{{Age and mass estimates for the planetary-mass objects of our sample}. \label{planetary_mass}}
\tablewidth{700pt}
\tabletypesize{\small}
\tablehead{
\colhead{Num.} & \colhead{Name} & SpT & \colhead{Mass ($\mathrm{M_{Jup}}$)} & \colhead{Age (Myr)} & \colhead{YMG} & Ref. 
} 
\startdata			
    \hline
	3 & 2MASS J03552337+1133437 & L5.0 & 13--30 &  $\sim$120 & AB-Doradus & 1  \\       
	9 & 2MASS J00470038+6803543 & L6.0 & $\mathrm{20^{+3}_{-7}}$ & $\sim$120 &AB-Doradus & 2  \\     
    15 & PSO J318.5-22 & L7.0 & 8.3$\pm$0.5 & 23$\pm$3 & $\beta$-Pic & 3, 4  \\ 
	30 &SIMP J013656.5+093347.3 & T2.5 & 12.7$\pm$1.0 & $\sim$120 &AB-Doradus & 5  \\ 
	31 &GU PSC B & T3.5 & $11.9^{+2}_{-1.5}$ & $\sim$120 & AB-Doradus & 2\\    
	36 &2MASS J11101001+0116130 & T5.5 & 10--12 & $\sim$120 & AB-Doradus  & 6\\     
	42 &CFBDSIR2149-0403 &  T7.0 & 2--13 & $<$500 & None & 7\\    
	44 &ROSS458C & T8.0 & 5--20 & $<$1000 & None & 8\\ 
\enddata
\tablecomments{References: [1] - \cite{Faherty2013}, [2] - \cite{Aller2016}, [3] - \cite{Liu2013}, [4] - \cite{Allers2016} [5] - \cite{Gagne2017}, [6] - \cite{Gagne2015_s1110}, [7] - \cite{Delorme2017}, [8] - \cite{Burningham2011}.}
\end{deluxetable*}




\section{Spectral Types}\label{SpT}

   The spectral types provided in column 3 of Table~\ref{all_data_hst} for brown dwarfs and substellar companions are those given in the literature from the different sources. To provide a homogeneous spectral type classification, {we compared our HST/WFC3 spectra to the spectra in the  SpeX Prism Spectral Library}. We compared our HST/WFC3 spectra using a modified $\chi^{2}$ metric as presented in \cite{Cushing2008}:

\begin{equation}\label{eq_cushing}
    G = \sum_{\lambda}\left[\displaystyle\frac{C(\lambda)-\alpha T(\lambda)}{\sigma_{c}(\lambda)}\right]^{2},
    \label{chi}
    \end{equation}
    where $C(\lambda)$ is the spectrum of our object, $T(\lambda)$ is the {comparison} spectrum,  $\alpha$ is a scaling factor that minimizes $G$, and $\sigma_{c}(\lambda)$ are the uncertainties of the spectrum.  We  additionally checked the best spectral matches by visual inspection. 
    
In column 4 of Table~\ref{all_data_hst} we show the resulting spectral types for each object. 
We found that the spectral types derived using the SpeX  library are consistent with those published
for each object in the literature, matching typically to within +/- 1.5 spectral sub-types. The only exceptions are some of the known intermediate- or low-surface gravity objects in our sample. For these objects {(CD-352722, LP 261-75b, 2MASS J2224438-015852, 2MASS J0310559+164816, S Ori 73)} the difference in spectral types with respect to the literature values can be up to $\pm$3 spectral sub-types. These differences are expected, as the SpeX spectral library is mostly composed of field, i.e., {high-surface gravity brown dwarfs}.

\section{Color-magnitude diagrams}\label{cmd}

\begin{figure}
\centering
\includegraphics[width=0.45\textwidth]{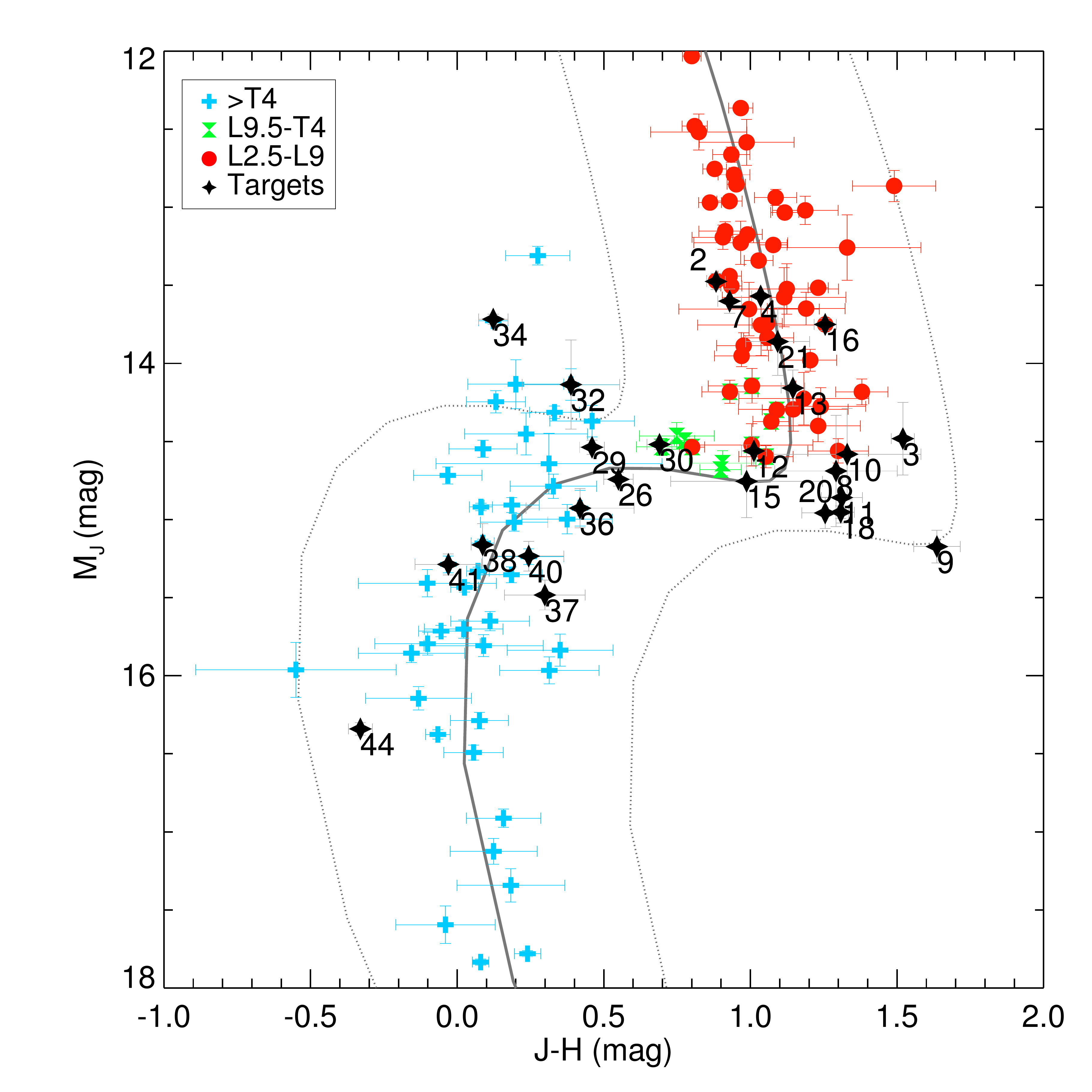}

\caption{\label{cmd_fig} 2MASS $J-H$ color versus $M_{J}$ 2MASS absolute magnitude diagram. The black stars correspond to L and T dwarfs of our sample with measured trigonometric parallaxes published in the literature. Red dots are L dwarfs, green dots are L-T transitions dwarfs and blue dots are T dwarfs with parallaxes published in \cite{Dupuy_Liu2012}. The solid grey lines represent the color-absolute magnitude relationships for brown dwarfs. The dotted grey lines represent the rms of that relation. {Objects lying outside the rms of the relation, are catalogued extremely red or blue, or overluminous}.}
\end{figure}

{We use near-infrared  color-magnitude diagrams (CMDs) for a simple, yet quantitative comparisons of the L and T brown dwarfs in this study to those published in \citet{Dupuy_Liu2012}, with the aim of identifying peculiar objects (extremely red or extremely blue dwarfs), or multiple systems. 
In Figure \ref{cmd_fig} we show the CMD plot, 2MASS $J-H$ color versus $J$-band absolute magnitude, for L and T brown dwarfs of Table \ref{all_data_hst} and objects from \cite{Dupuy_Liu2012} as a comparison. We calculated the absolute $J$-band magnitude using trigonometric parallaxes, {when} available.} Black stars represent objects from our sample listed in Table~\ref{all_data_hst} with trigonometric parallaxes available in the literature. {We do not include Y dwarfs in this CMD, as there are few other Y dwarfs for comparison}. Red dots represent L dwarfs, green dots represent L-T brown dwarfs, and blue dots represent T dwarfs with trigonometric parallaxes published in \cite{Dupuy_Liu2012}. The solid grey line represents the color-absolute magnitude relationship for brown dwarfs, and the dotted grey line represents the rms (root-mean-square) of that relation.

The object \object{2MASS J00470038+6803543} (hereafter W0047, L7, object 9)  stands out outside the rms of the color-absolute magnitude relation with red $J-H$ color index. Objects  \object{2MASS J17503293+1759042} (hereafter 2M1750+1759, object 32) and \object{2MASS J05591914-1404488} (hereafter 2M0559-1404, object 34) are overluminous with respect to the other L-T transition objects, as they are {above the rms of the color-magnitude relation for brown dwarfs \citep{Dupuy_Liu2012}}. {The cause of their overluminosity is unknown, as no multiplicity has been reported previously for these objects.}




\section{Low Surface Gravity Objects}\label{lowg_objects}

\subsection{Gravity Index Determination}\label{index_det}

\begin{figure}
\centering
\includegraphics[width=0.45\textwidth]{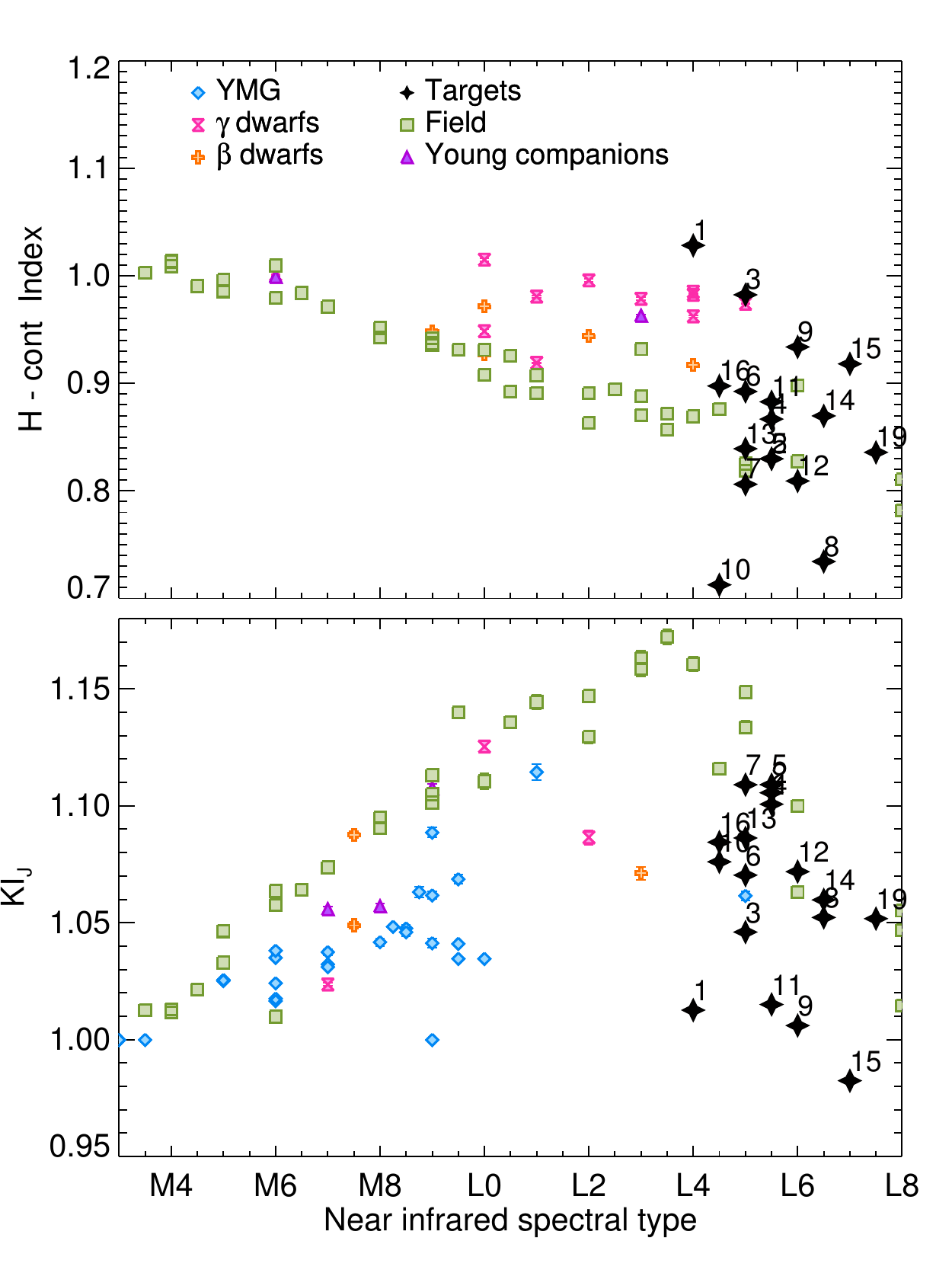}
\caption{\label{index_fig} Spectral types versus $H$-continuum  and $\mathrm{KI_{J}}$ indices from \cite{Allers2013}. For comparison, we show the value of these indices objects of the  field (green squares), young companions (purple triangles), $\beta$ dwarfs (intermediate surface gravity, orange crosses), $\gamma$ dwarfs (very low gravity, pink hourglass symbol) and young moving group members (blue diamonds). {Black stars belong to objects of the sample presented in this work with spectral types between L3 and L7. Numbers identify the objects from Table \ref{all_data_hst}}.}
\end{figure}

We use the low surface {gravity} indicators presented in \cite{Allers2013} that are applicable to our sample up to spectral type L7. We aim to {search for as-of-yet-unidentified low surface gravity objects and to confirm those that have been classified as low surface gravity objects previously}. Due to the spectral wavelength coverage of our HST/WFC3 spectra and their resolution, only the $H$- continuum and the $\mathrm{KI_{J}}$ indices are applicable to our spectra. The $H$-continuum measures the shape of the $H$-band, that has been found to be triangular for most of the very low-gravity brown dwarfs ($<$100~Myrs) with spectral types between M6 and L7. Intermediate and field gravity brown dwarfs show a "shoulder" at 1.57~$\mu$m indicative of the appearance of the $\mathrm{FeH}$ molecular absorption and the {collisionally-induced absorption by $\mathrm{H_{2}}$ molecules} \citep{Borysow1997,Allard2012a}. \cite{Allers2013} warned, though, that this index needs to be used in combination with others, as the $H$-band is also triangular for some objects that do not have low surface gravity. The $\mathrm{KI_{J}}$ index measures the alkali line doublet depth at 1.244 and 1.253~$\mu$m. These  have been found to be weaker for low surface gravity objects of spectral types between M5 to L7. The continuum, center, and bandwidth of these indices are described in Table~\ref{index}.  Their values are calculated using equation 1 from \cite{Allers2013}. Different values for the $\mathrm{KI_{J}}$ and  $H$-continuum indices correspond to different gravity scores: 0, 1 and 2, corresponding to field gravity (FLD-G), intermediate (INT-G), or low surface gravity objects (VL-G), respectively. The ranges of values of the $\mathrm{KI_{J}}$ and the $H$-continuum indices that correspond to different gravity scores are given in Table~9 of \cite{Allers2013}.  The  $\mathrm{KI_{J}}$ and the $H$-continuum indices and their gravity scores obtained for our objects are shown in Table~\ref{gs}. In Figure~\ref{index_fig} we show the spectral type versus the $\mathrm{KI_{J}}$ and the $H$-continuum indices for our sample (up to L7 spectral type) and for objects that belong to young moving groups, $\gamma$ and $\beta$ dwarfs\footnote{Optical classification for very low and intermediate surface gravity, respectively \citep{Cruz}.}, young companions \citep{Allers2013,Bonnefoy2014a}, and field brown dwarfs \citep{McLean,Cushing2005} for comparison.

\subsection{Results: gravity class determination}

Among the 16 objects in our sample with spectral types up to L7 for which the $\mathrm{KI_{J}}$ and the $H$-continuum indices could be measured, two have gravity scores consistent with very low surface gravities, namely: \object{CD-352722B} (object 1), and \object{2MASS J03552337+1133437} (object 3). Five objects had gravity scores consistent with intermediate gravities, namely: \object{2MASSI J0421072-630602} (object 6), \object{2MASS J00470038+6803543} (object 9),  \object{2MASS J01075242+0041563} (object 11), \object{PSO J318.5-22} (object 15), and \object{2MASSW J2224438-015852} (object 16). 

CD-352722B and 2MASS J03552337+1133437 had been reported before as low surface gravity objects (\citealt{Wahhaj} and \citealt{Faherty2013}, respectively), and they are also high-probability members of the AB-Doradus  moving group with an estimated age of $\sim$120~Myr \citep{Zuckerman2004}.

Object \object{2MASSI J0421072-630602} (object 6) was classified as intermediate-surface gravity (L5$\beta$) object by \cite{Cruz} in {optical wavelengths}. 

\object{2MASS J00470038+6803543} (object 9) is an extremely red L-dwarf, and a bona fide member of the AB Doradus  moving group, for which intermediate surface gravity characteristics have been previously reported \citep{Gizis2012, Allers2013}. 

PSO J318.5-22 is also an extremely red L-dwarf for which low surface gravity signatures have been found \cite{Liu2013} and is a bona fide member of the $\beta$-Pictoris  moving group (age = 23$\pm$3~Myr, \citealt{Zuckerman2001, Mamajek2014}). 

Object \object{2MASS J01075242+0041563} (object 11) does not clearly show low surface-gravity spectral characteristics \citep{Gagne2015}, and it was found to be a possible member of the Hyades association (age $\sim$625~Myr, \citealt{Bannister2007}). If confirmed, its age would not be consistent with its intermediate gravity classification (expected for objects with ages between 50 and 200~Myr, see \citealt{Allers2013}). 

Finally,  2MASSW J2224438-015852 (object 16) is an extremely red L4.5 dwarf, that was classified as a field dwarf by \cite{Liu_Dupuy_Allers2016} using the BANYAN II tool. \cite{Martin} classified it as a field gravity object using NIRSPEC/Keck~II spectra to obtain \cite{Allers2013} indices in the $J$-band.

\begin{table}
\begin{center}
\caption{\label{index} Spectral indices to segregate young brown dwarfs from \cite{Allers2013}.}
\begin{tabular}{lllll}
\hline
\hline
	Index & $\lambda_{line}$ ($\mu$m) & $\lambda_{cont1}$ ($\mu$m)& $\lambda_{cont2}$ ($\mu$m) & {Width} in $\lambda$   \\ 
    \hline
	$\mathrm{KI_{J}}$& 1.244 & 1.220 & 1.270  & 0.0166 \\ 
	$H$-cont  &1.560 & 1.470 & 1.670  &  0.0208\\ 

\hline
\end{tabular}
\end{center}
\end{table}

\begin{table*}
\begin{center}
\caption{\label{gs} Values obtained for the $H$-continuum and $\mathrm{KI_{J}}$ indeces for our objects with their corresponding gravity scores.}
\begin{tabular}{rlcccc}
\hline
\hline

	Object & Name & $H$-continuum &$\mathrm{KI_{J}}$ & Gravity score per index& Gravity class   \\ 
    \hline
    1   & CD-352722B &  1.028$\pm$0.001 &1.013$\pm$0.001& 2 2 & VL-G	\\
    2  & 2MASS J17502484-0016151 &0.829$\pm$0.001 &	1.106$\pm$0.001 & 0 0 & FLD-G \\
    3  & 2MASS J03552337+1133437 & 0.982$\pm$0.001 &	1.046$\pm$0.001 & 2 2 & VL-G  \\
    4   & 2MASS J18212815+1414010 & 0.867$\pm$0.001 & 1.101$\pm$0.001 & 0 0 & FLD-G\\
    5  & 2MASSW J1507476-162738  & 0.829$\pm$0.001 & 1.109$\pm$0.001& 0 0 & FLD-G 	\\
	6  &  2MASSI J0421072-630602 & 0.892$\pm$0.001 &1.070$\pm$0.001	& 1 1 & INT-G\\
    7  & 2MASS J05395200-0059019&0.806$\pm$0.001 &  1.109$\pm$0.001& 0 0 & FLD-G 	\\
     8  & 2MASSI J1711457+223204 & 0.734$\pm$0.001 & 1.052$\pm$0.001 & 0 -- & FLD-G 	\\
    9  & 2MASS J00470038+6803543&   0.934$\pm$0.001 &1.006$\pm$0.001 & 1 -- & INT-G	\\
    10  & LP261-75B & 0.713$\pm$0.004  &1.076$\pm$0.001 & 0 1 & FLD-G \\
     11 & 2MASS J01075242+0041563& 0.883$\pm$0.001 & 1.015$\pm$0.001 & 1 2 & INT-G	\\
    12  &2MASSW J1515008+484742 & 0.809$\pm$0.001 &1.072$\pm$0.001 & 0 -- & FLD-G	\\
    13  &2MASS J06244595-4521548&0.839$\pm$0.001 &  1.086$\pm$0.001 & 0 1 & FLD-G	\\
    14  &2MASSW J0801405+462850&0.869$\pm$0.001 &1.059$\pm$0.001 & 0 -- & FLD-G	\\
    15 & PSO J318.5-22& 0.918$\pm$0.002 & 0.982$\pm$0.001 & 1 --   & INT-G    \\
     16 & 2MASSW J2224438-015852 & 0.898$\pm$0.001   &  1.084$\pm$0.001 & 1 1 & INT-G	\\

    \hline
\end{tabular}
\end{center}
\end{table*}

\section{Candidates for composite atmospheres}\label{composite}

Our high-quality spectra are well-suited for exploring the diversity of ultracool atmospheres, including the identification of potentially composite (multi-component) spectra. An obvious source of such composite spectra are unresolved binaries with different spectral types. However, given the very high occurrence rate of heterogeneous cloud cover in brown dwarfs \citep[][]{Buenzli2014, Metchev2015}, it is expected that brown dwarfs with strong spectral heterogeneity will also contribute to the population of composite atmospheres.

\subsection{Search for composite atmosphere brown dwarfs}

To identify potential composite spectra candidates in our sample, we used the spectral indices from \cite{Burgasser2006, Burgasser2010} and \cite{Bardalez_Gagliuffi}. These indices examine peculiar spectral characteristics of spectroscopic L-plus-T and M-plus-T composite atmosphere brown dwarfs. L plus T composite spectra have bluer SEDs (spectral energy distributions) in the near-infrared than field objects of similar spectral type. In cases of L-plus-T spectroscopic binaries atomic and molecular features are blended, affecting the $\mathrm{H_{2}O}$ (1.15~$\mu$m) and $\mathrm{CH_{4}}$ (1.32~$\mu$m)  molecular features.
At 1.55~$\mu$m spectroscopic L+T binaries show larger flux from the T dwarf. \cite{Burgasser2006, Burgasser2010} and \cite{Bardalez_Gagliuffi} combine different pairs of indices, defining those that segregate brown dwarfs with composite spectra more efficiently. 

The criteria to select potential brown dwarf composite atmospheres, and the delimiters of the areas within the plots comparing different indices that segregate spectroscopic brown dwarfs with composite spectra are defined in Tables~\ref{spectral_indices}, \ref{criteria} and \ref{criteria2} of the Appendix. We do not measure indices that are outside the wavelength range of the HST/WFC3 G141 near-infrared spectra. Thus, only 13 out of 18 available  plots comparing indices from \cite{Burgasser2006, Burgasser2010} and \cite{Bardalez_Gagliuffi} are applicable to in our data (see Tables~\ref{criteria} and \ref{criteria2}). In Figures~\ref{index1}, \ref{index2} and \ref{index3} of the Appendix we show the comparison of the spectral indices listed in Tables~\ref{criteria} and \ref{criteria2} of the Appendix. To be considered a weak candidate to have a composite brown dwarf atmosphere,  \cite{Burgasser2006, Burgasser2010}, and \cite{Bardalez_Gagliuffi} established that an object needs to appear within the selection area of minimum four plots, and to be considered a strong candidate, it needs to appear at least in 8 plots. Although due to the wavelength coverage of the HST/WFC3 data we were not able to perform 5 of the index comparisons in \cite{Burgasser2006, Burgasser2010}, and \cite{Bardalez_Gagliuffi}, we conservatively use the same criteria to select weak and strong spectroscopic binaries.

\subsection{Results: composite atmosphere candidates}

We found that ten objects in our sample were selected by the indices as weak composite spectra candidates: 2M0310+16 (object 21), 2M0624-45 (object 13), 2MUCD 10802 (object 19), 2M0909+65 (object 25), 2M1039+32 (object 24), 2M1324+63 (object 27), 2M1515+48 (object 12), 2M1632+19 (object 20), 2M1711+22 (object 8), 2M1750-00 (object 2). In addition, we found 3  strong candidates to have composite spectra: 2M1507-16 (object 5), 2M1219+31 (object 22), and SIMP0136+09 (object 30). 

To confirm or reject the candidates, we compared our HST/WFC3 spectra to single template spectra from the SpeX Prism Spectral Library, and independently, to synthetic L plus T composite spectra created using single spectra from those libraries (see Table~\ref{all_candidates_hst}). To create the synthetic composite spectra, we scaled the fluxes of the components to 10~pc using the color-magnitude relation from \cite{Dupuy_Liu2012}, and coadded the two component fluxes.

Following \cite{Burgasser2006,Burgasser2010} and \cite{Bardalez_Gagliuffi}, we compared the goodness of the fit of the HST/WFC3 spectra to the single, and independently to synthetic L plus T composite spectra using a modified $\chi^{2}$ (G) using equation~\ref{eq_cushing} from \cite{Cushing2008}.

Finally, we tested if the fit to a composite template spectra was significantly better than the fit to a single template using one-sided F-test statistic. The distribution statistic ratio we used:

	\begin{equation}
	\eta_{SB} = \frac{min(G_{single}) \times df_{composite}}{min(G_{composite}) \times df_{single}}
	\label{F_test}
	\end{equation}
	{where min($G_{single}$) and min$(G_{composite})$ are the minimum $G$ for the best match to a single or a composite template}, and  $df_{composite}$ and $df_{single}$ are the degrees of freedom for the composite template spectra fit and the single template fit. The degrees of freedom are the number of data points used in the fit minus one, to account the scaling between our spectra and the template spectra. To rule out the null hypothesis, {meaning  that the candidate spectrum is not best described by a single template at the 99\% confidence level, we require $\eta_{SB} > 1.41$.} The F-statistic analysis rejected five of our candidates: 2M0310+16 (object 21), 2M0909+65 (object 25), 2M1750-00 (object 2), 2M1507-16 (object 5) and SIMP0136+09 (object 30). 
    
\begin{deluxetable*}{llllllc}
\tablecaption{{Candidates for composite spectra selected by \cite{Burgasser2006, Burgasser2010} and \cite{Bardalez_Gagliuffi} spectral indices.}  \label{all_candidates_hst}}
\tablewidth{700pt}
\tabletypesize{\scriptsize}
\tablehead{
\colhead{Num.} & \colhead{Name} &
\colhead{SpT} & \colhead{Single Component} & 
\colhead{Binary Components} & \colhead{$\eta_{SB}$} & \colhead{Variable?}
} 
\startdata			
    \hline
21	& 2M0310+16 & L8 & L8 (SD J121951.45+312849.4) & (L8) SD J085758.45+570851.4 +  T0 (SD J042348.57-041403.5) & 1.27  & Yes [1]   \\ 
{13}	& 2M0624-45  & L6.5 & L5 (2M J23512200+3010540)  &L3.5 (2M J2224438-015852) + T1 (SD J163239.34+415004.3) & {2.18} & Yes [1]    \\ 
{19}	& 2MUCD 10802&L7.5 & L7.5 (SD J115553.86+055957.5)& L7.5 (SD J115553.86+055957.5) + T0 (Gl337CD)& {1.43} & No [1]  \\ 
25  & 2M0909+65 & T1.5 & T2 (2M J11220826-3512363)  & L9.5 (SD J082030.12+103737.0) + T3.5 (SD J175032.96+175903.9) & 1.03 & No [1]     \\ 
{24}	& 2M1039+32 &  T1 & T1.5 (SD J090900.73+652527.2) & L8 (SD J121951.45+312849.4) + T4 (2M J2254188+312349) & {3.25} & Yes [1]  \\ 
{22}	& 2M1219+31 & L8 & L8 (2M J0328426+230205) & L9 (2M J0310599+164816) + T0 (Gl 337CD) & {2.07} & Yes [1] \\ 
{27}	& 2M1324+63 & T2 & T2 (SD J125453.90-012247.4) &L9 (SD J083008.12+482847.4) + T7.5 (2M J11145133-2618235) & {3.94} & Yes [1] \\ 
{12}	& 2M1515+48 & L6 & L9 (2M J0908380+503208) & L9 (2M J0908380+503208) + T0 (SD J152039.82+354619.8) & {1.69} & No [1]  \\ 
{20}	& 2M1632+19 & L8 & L8 (Gl584C) & L6(2M J0825196+211552)+T0(SD J204749.61-071818.3) & {2.71} & Yes [1]   \\
{8}	& 2M1711+22 & L5.0+T5.5 & T1 (SD J085834.42+325627.7) & L6(2M J0825196+211552)+T3(SD J102109-030420) & {2.30} & Binary \\ 
2	& 2M1750-00 & L5.5 & L5 (2M J18131803+5101246) & L4.5 (2M J15200224-4422419B) + T0.5 (SD J151643.01+305344.4) & 0.79  & Yes[1]   \\ 
5	& 2M1507-16 & L5.5 & L5 (2M J17461199+5034036)  & L5 (2M J10461875+4441149) + T0 (2M J0920122+351742) & 1.18 & Yes [2] \\ 
30	& SIMP0136+09 & T2.5 & T2 (2MASS J11220826-3512363)  & SD J213154.43-011939.3 (L9) + SD J092615.38+584720.9 (T4.5) & 0.95  & Yes [3] \\ 
\enddata
\tablecomments{[1] - \cite{Buenzli2014}, [2] - \cite{Yang2015}, [3] - \cite{Artigau2009}.}
\end{deluxetable*}

{Unresolved binaries should appear overluminous compared to single brown dwarfs. To test for evidence of overluminosity in our targets, we compared the absolute magnitudes, derived using the trigonometric parallaxes for our targets, with the spectro-photometric absolute magnitudes derived using the relation published by \cite{Dupuy_Liu2012} (see Table~\ref{variability_overluminosity}). To obtain spectro-photometric absolute magnitudes we used the spectral type of the principal component. We found that none of the sources with trigonometric parallaxes are overluminous. Actually, we find that some of them are slightly underluminous. This fact does not support the multiplicity hypothesis.}

\subsection{Rotational modulation and composite atmosphere candidates}    
    
     We also searched for published rotational modulation detections for our candidates as a potential marker for composite atmospheres. We found that nine of the thirteen composite atmosphere candidates have reported photometric variability due to cloud patterns (see Table~\ref{all_candidates_hst}). In fact, from the eight objects that satisfied the criteria for composite spectra candidates, six are known to have rotational modulations:  2M0624-45 (object 13), 2MUCD 10802 (object 19),  2M1039+32 (object 24), 2M1324+63 (object 27), 2M1219+31 (object 22), and 2M1632+19 (object 20). One is a confirmed binary: 2M1711+22 (object 8), and one has been reported as non-variable (2M1515+48, object 12). 
     
     These results suggest that \cite{Burgasser2006, Burgasser2010}, and \cite{Bardalez_Gagliuffi} spectral indices are biased towards L and T brown dwarfs that show photometric variability due to rotational modulations. Thus, these indices should also be useful to search for brown dwarfs candidates with heterogeneous cloud patterns in their atmospheres. In Table~\ref{variability_indexdet}, we list all objects with L4 to T4 spectral types in our sample for which the method presented in this Section are applicable. We specify which of them have reported rotational modulations in the literature, and we compare with those that have been found by the indices as composite spectra candidates. 
     
     We conclude that 21 out of the 32 objects listed in Table~\ref{variability_overluminosity} have  rotational modulations reported in the literature (see Table~\ref{variability_indexdet} for details); nonetheless, only nine are detected by the indices as candidates for composite spectra, with spectral types from L4 to T2 (spectral types from the literature). None of the low-gravity brown dwarfs found in Section~\ref{lowg_objects} with reported rotational modulations have been detected by indices. In addition, the indices have detected three other candidates to have composite spectra, but are not known to show rotational modulations. 
 
\begin{deluxetable*}{llcclll}
\tablecaption{{Photometric variability reported for final weak and strong candidates for composite spectra.}  \label{variability_overluminosity}}
\tablewidth{700pt}
\tabletypesize{\small}
\tablehead{
\colhead{Num.} & \colhead{Name} &
\colhead{SpT principal component} & \colhead{Variable?} & \colhead{$\mathrm{\pi_{Trig}}$ (mas)} & \colhead{${M_{J}}$ (mag)} & \colhead{${M_{J, SP}^{a}}$ (mag)}
} 
\startdata			
    \hline
13	& 2M0624-45  & L6.5 & Yes [1] & 86.21$\pm$4.46 & 14.16$\pm$0.12   & 14.11$\pm$0.40  \\ 
19	& 2MUCD 10802 & L7.5 & No [1] & &  &       \\ 
24	& 2M1039+32 &  T1  &  Yes [1]  & &    &     \\ 
22	& 2M1219+31 & L8 &  Yes [1]  &  &   &    \\ 
27	& 2M1324+63 & T2  & Yes [1]  &  &   &      \\ 
12	& 2M1515+48 & L6 &  No [4]   & 123.8$\pm$5.0  & 14.56$\pm$0.09   & 13.94$\pm$0.40   \\ 
20	& 2M1632+19 & L8 & Yes [2] & 65.19$\pm$2.16  & 14.96$\pm$0.10  &  14.51$\pm$0.40   \\
8	& 2M1711+22 &  L6.5 &  Binary [3] & 33.11$\pm$4.71 & 14.69$\pm$0.36     &  14.11$\pm$0.40   \\ 
\enddata
\tablecomments{[1] - \cite{Buenzli2014}, [2] - \cite{Metchev2015}, [3] - \cite{Burgasser2010}, [4] - \cite{Bardalez_Gagliuffi}, [5] - \cite{Artigau2009}, [6] - \cite{Yang2015}.}
\tablecomments{[a] - Absolute magnitude given by the empirical spectro-photometric relation by \cite{Dupuy_Liu2012}.}
\end{deluxetable*}

\begin{deluxetable*}{llccc}
\tablecaption{{L4--T4 dwarfs with reported rotational modulations and composite atmosphere candidates. \label{variability_indexdet}}}
\tablewidth{700pt}
\tabletypesize{\small}
\tablehead{
\colhead{Num.} & \colhead{Name} & \colhead{Variable?} & \colhead{Detected by indexes} & \colhead{Confirmed as candidate} 
} 
\startdata			
    \hline
	1 & CD-352722b & -- & No  & No  \\ 
	2 & 2MASS J17502484-0016151 & {Yes} [1] & {Yes} & No  \\ 
	3 & 2MASS J03552337+1133437 & -- & No &  No  \\     
	4 & 2MASS J18212815+1414010 & {Yes} [2] & {No} & No  \\    
	5 & 2MASSW J1507476-162738 & {Yes} [2] & {Yes} & No \\    
	6 & 2MASSI J0421072-630602 & No [1] & No & No \\ 
	7 & 2MASS J05395200-0059019 & No [1] & No & No \\ 
	8 & 2MASSI J1711457+223204 & {No, binary [1]} & {Yes} & Yes \\     
	9 & 2MASS J00470038+6803543 & {Yes} [3] & {No} & No  \\ 
	10 & LP261-75B  & {Yes} [4] &{No} & No  \\ 
	11 & 2MASS J01075242+0041563& {Yes} [5] & {No} & No \\     
	12 & 2MASSW J1515008+484742 & {No} [1] & {Yes} & No \\ 
	13 & 2MASS J06244595-4521548 & {Yes} [1]& {Yes} & Yes \\ 
	14 & 2MASSW J0801405+462850 & No [1] & No & No    \\ 
    15 & PSO J318.5-22 & {Yes} [6] & {No} & No  \\
  	16 & 2MASSW J2224438-015852 & No [7] &No & No  \\ 
	17 & Luh 16AB & {Yes} [8] & {No} & No   \\ 
	18 & 2MASSI J0825196+211552 & No [1] & No & No  \\  
	19 &2MUCD 10802 & {No} [1] & {Yes} & No   \\ 
	20 &2MASS J16322911+1904407 & {Yes} [1] & {Yes}  &Yes   \\ 
	21 &2MASSW J0310599+164816 & {Yes} [1] & {Yes} & No   \\ 
    22 &2MASS J12195156+3128497 & {Yes} [1] & {Yes} & Yes \\
	23 &SDSS J075840.33+324723.4 & Yes [13] & No & No  \\ 
	24 &2MASS J10393137+3256263 & {Yes} [1] & {Yes} & Yes \\ 
	25 &2MASS J09090085+6525275 & {No} [1] & {Yes} & No  \\ 
	26 &2MASS J21392676+0220226& {Yes} [9,14] & {No} & No \\     
	27 &2MASS J13243553+6358281 & {Yes} [1,14] & {Yes} & Yes \\ 
	28 &2MASS J16291840+0335371 & Yes [13] & No & No   \\ 
	29 &HN PEG B  & {Yes} [10] & {No} & No  \\ 
	30 &SIMP J013656.5+093347.3 & {Yes} [11,14] &{No} & No  \\ 
	31 &GU PSC B & {Yes} [12] & {No}  & No \\    
	32 &2MASS J17503293+1759042 & {Yes} [1] & {No} & No  \\ 
\enddata
\tablecomments{References: [1] - \cite{Buenzli2014}, [2] - \cite{Yang2015}, [3] - \cite{Lew2016}, [4] - \cite{Manjavacas2018}, [5] - Apai et al. in prep, [6] - \cite{Biller2018}, [7] - \cite{Metchev2015}, [8] - \cite{Buenzli2015}, [9] - \cite{Radigan2012}, [10] - \cite{Zhou2018}, [11] - \cite{Artigau2009}, [12] - \cite{Naud2017}, [13] - \cite{Radigan_Lafreniere2014}, [14] - \cite{Apai2017}.}
\end{deluxetable*}	

\section{{Methane  (1.2~$\mu$\MakeLowercase{m}) and Water (1.4~$\mu$\MakeLowercase{m}) Absorption Bands for Spectral Classification}\label{classification}}

\begin{figure}
\centering
\includegraphics[width=0.50\textwidth]{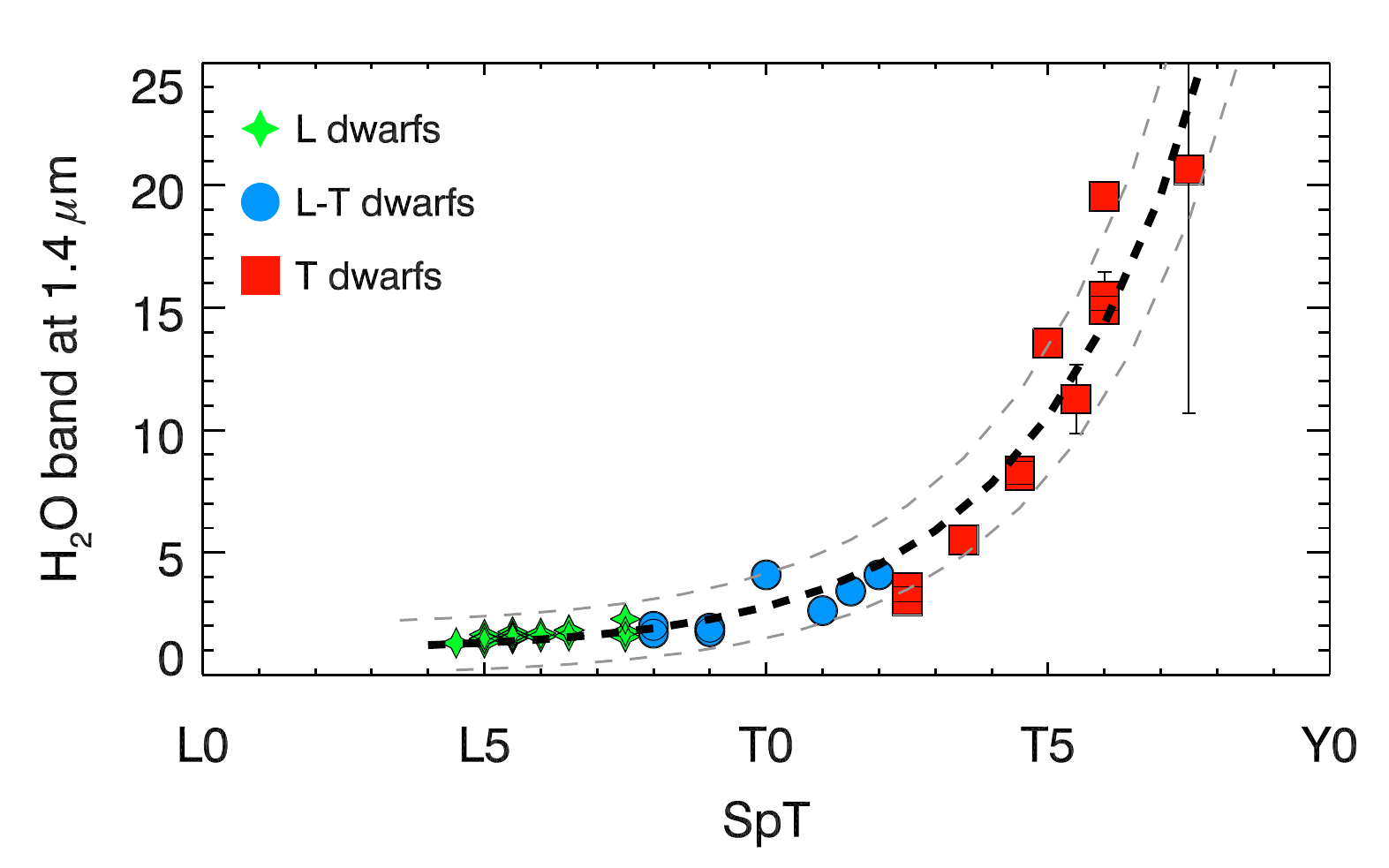}
\caption{Evolution of the depth of the $\mathrm{H_{2}O}$ band at 1.4~$\mu$m with near-infrared spectral types calculated using equation \ref{A_L2013}. The value of the depth of the $\mathrm{H_{2}O}$ band is dimensionless. \label{SpTJ_H_water}}
\end{figure}

\begin{figure}
\centering
\includegraphics[width=0.49\textwidth]{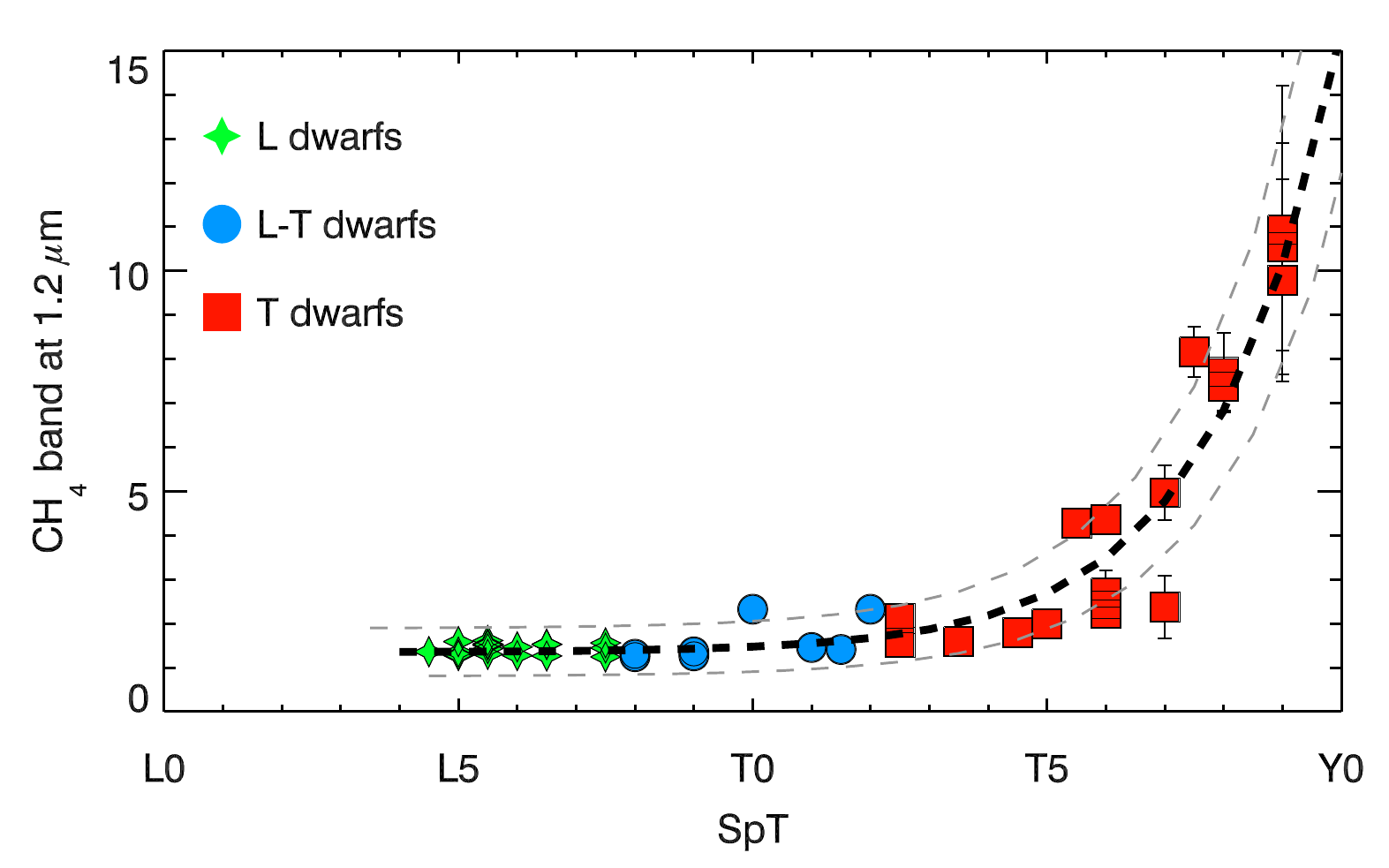}
\caption{Evolution of the depth of the $\mathrm{CH_{4}}$ band at 1.2~$\mu$m with near-infrared spectral types calculated using equation \ref{A_L2013}. The value of the $\mathrm{CH_{4}}$ band is dimensionless. \label{SpT_J_H_CH4}}
\end{figure}

The most prominent molecular absorption bands in the near-infrared spectra of brown dwarfs and substellar companions are $\mathrm{H_{2}O}$, $\mathrm{CO}$, and $\mathrm{CH_{4}}$. 
The absorption bands of the different molecules are controlled by the availability of C and O \citep{Marley2010}. Within the temperature range (approx. 1800~K to 600~K and below) of substellar objects in our sample -- corresponding to spectral types from L4 to Y1 -- carbon appears mainly in the form of CO in L dwarfs and as $\mathrm{CH_{4}}$ in T dwarfs \citep{Marley2015}. The equilibrium reaction that takes place is:
\begin{equation}
\label{reaction}
\mathrm{CO} + 3\mathrm{H_{2}} \leftrightarrow \mathrm{CH_{4}} + \mathrm{H_{2}O}
\end{equation}
At higher temperature (L dwarfs), the left side of the reaction is favored; thus there is an overabundance of CO, that implies an underabundance of $\mathrm{H_{2}O}$. At lower temperatures, below the L/T transition, the right hand-side of the reaction is favored, leading to higher abundances of  $\mathrm{CH_{4}}$ and $\mathrm{H_{2}O}$ \citep{Marley2015}. 
{The depth of some of these molecular bands in our HST/WFC3 spectra can serve to provide  a robust spectral classification of  substellar objects.}
In the HST/WFC3 1.10--1.69~$\mu$m spectra of brown dwarfs and substellar companions, we are able to measure the depth of the $\mathrm{CH_{4}}$ band at approximately 1.2~$\mu$m and the $\mathrm{H_{2}O}$ band at approximately 1.4~$\mu$m, and trace their {change in depth with  near-infrared spectral types} (see Figures \ref{SpTJ_H_water} and \ref{SpT_J_H_CH4}). We measured the depths of the $\mathrm{H_{2}O}$ and $\mathrm{CH_{4}}$ bands using equation 1 from \cite{Allers2013}:

\begin{equation}
\label{A_L2013}
\small
\mathrm{index} = \left(\frac{\lambda_{line}-\lambda_{cont1}}{\lambda_{cont2}-\lambda_{cont1}} F_{cont2} + \frac{\lambda_{cont2}-\lambda_{line}}{\lambda_{cont2}-\lambda_{cont1}} F_{cont1}\right)/F_{line}
\end{equation}

In Table~\ref{limits} we show the wavelengths in which the continuum, the center and the width of the bands are defined. The minimum value of this index is 1, implying that there is no absorption feature. In addition, we derive an exponential function to relate the near-infrared spectral types and the depths of the $\mathrm{CH_{4}}$,  and the $\mathrm{H_{2}O}$ bands.  To calculate the best-fit exponential function, we used the \texttt{IDL} function \texttt{COMFIT.PRO}, that fits an exponential equation of the form: $y = c_{0}*c_{1}^x + c_{2} $ using a gradient-expansion least-squares method \citep{marquardt1963}.  The exponential function was preferred over polynomial functions, as it provides a smaller reduced $\chi^2$. We did not include Y dwarfs in the fit due to their lower quality data (see Table~\ref{all_data_hst}), nor hot Jupiters, as their atmospheres are, in general, physically different.  To obtain the $\mathrm{CH_{4}}$ index, we discarded as well some companions (CD-352722 B, HN-Peg B, and GU PSC B), as they were outliers in the SpT vs. $\mathrm{CH_{4}}$ index plot, probably due to contamination of the star at the wavelength range in which the $\mathrm{CH_{4}}$ index is measured. {The best fits to exponential functions are displayed in Figures~\ref{SpTJ_H_water} and \ref{SpT_J_H_CH4} with a dashed thick black line. The grey dashed lines in those figures represent the standard deviation of the data points with respect the fitted function}. The values of the coefficients for each best fit exponential functions are displayed in Table~\ref{colour_age_table}. {In addition, in the former Table, we also show the function that provides spectral types of brown dwarfs given the value of the $\mathrm{H_{2}O}$ and $\mathrm{CH_{4}}$  bands}.  In Table~\ref{values_H2O_CH4}, we present the typical dimensionless values for the $\mathrm{H_{2}O}$ and $\mathrm{CH_{4}}$  bands calculated using the exponential functions indicated in Table~\ref{colour_age_table}. {The depths of the $\mathrm{H_{2}O}$ and $\mathrm{CH_{4}}$ bands can provide robust spectral classification for brown dwarfs with high-quality near-infrared spectra that includes the 1.4~$\mu$m water band. This is especially true for brown dwarfs with spectral types later than T2, for which the change of the band width is more abrupt with spectral type}.

\begin{table}
\begin{center}
\caption{Wavelength for the continuum and central wavelengths in which the depth of the $\mathrm{CH_{4}}$ and $\mathrm{H_{2}O}$ bands are measured. Wavelength units are $\mu$m. \label{limits}}
\begin{tabular}{lllll}
\hline
\hline
	Index & $\lambda_{line}$  & $\lambda_{cont1}$  & $\lambda_{cont2}$  & Band width   \\ 
    \hline
	$\mathrm{H_{2}O}$ & 1.40 & 1.31 & 1.47  & 0.08 \\ 
	$\mathrm{CH_{4}}$ & 1.18 & 1.10 & 1.30  &  0.12\\ 

\hline
\end{tabular}
\end{center}
\end{table}

\begin{table*}
	\caption{Exponential functions relating the depth of the $\mathrm{CH_{4}}$ band at 1.2~$\mu$m and $\mathrm{H_{2}O}$ band at 1.4~$\mu$m with near-infrared spectral types.\label{colour_age_table}}  
	\centering
	\begin{center}
		\begin{tabular}{llllll}
			\hline 
			
			 &    & \multicolumn{3}{l}{Exponential fit} \\		
			x & y 	        & $c_{0}$ & $c_{1}$ & $c_{2}$ &      $\chi^{2}_{red}$   \\
			\hline              		
  		NIR	SpT & $\mathrm{CH_{4}}$ & (1.13$\pm$1.09) $\times$ $\mathrm{10^{-3}}$  & 1.60$\pm$0.08 &  1.35$\pm$0.20 & 1.32 \\  
        NIR	SpT & $\mathrm{H_{2}O}$	& (6.51$\pm$1.90) $\times$ $\mathrm{10^{-2}}$ & 1.40$\pm$0.03 & (9.71$\pm$2.57)$\mathrm{10^{-1}}$  & 1.31 \\
			\hline		
		\end{tabular}
	\end{center}
		\tablecomments{The exponential function is defined as: $ y = c_{0}*c_{1}^x + c_{2}  $}
        \tablecomments{{To obtain NIR spectral types from the value of the index : x = $\log_{c_{1}} \left(\frac{y - c_{2}}{c_{0}}\right)$} }
\end{table*}

\begin{table}
\begin{center}
\caption{Typical dimensionless values for the $\mathrm{H_{2}O}$ and $\mathrm{CH_{4}}$  bands per spectral type calculated using the corresponding exponential function in Table~\ref{colour_age_table}. \label{values_H2O_CH4}}
\begin{tabular}{lll}
\hline
\hline
	
	SpT &  Value $\mathrm{CH_{4}}$ index& Value $\mathrm{H_{2}O}$   index     \\ 
    \hline
    L4 & 1.35$\pm$0.54 &	1.21$\pm$1.01		\\
    L5 & 1.36$\pm$0.54 &    1.31$\pm$1.01	\\
    L6 & 1.37$\pm$0.54 &	1.45$\pm$1.01		\\
    L7 & 1.38$\pm$0.54 &	1.64$\pm$1.01		\\
    L8 & 1.40$\pm$0.54 &	1.90$\pm$1.01		\\
    L9 & 1.43$\pm$0.54 &	2.27$\pm$1.01		\\
    T0 & 1.47$\pm$0.54 &	2.79$\pm$1.01		\\
    T1 & 1.55$\pm$0.54 &	3.50$\pm$1.01		\\
    T2 & 1.67$\pm$0.54 &	4.50$\pm$1.01		\\
    T3 & 1.87$\pm$0.54 &	5.89$\pm$1.01		\\
    T4 & 2.18$\pm$0.54 &	7.84$\pm$1.01		\\
    T5 & 2.68$\pm$0.54 &	10.54$\pm$1.01		\\
    T6 & 3.48$\pm$0.54 &	14.32$\pm$1.01		\\
    T7 & 4.77$\pm$0.54 &	19.60$\pm$1.01		\\
    T8 & 6.83$\pm$0.54 &	26.95$\pm$1.01		\\
    T9 & 10.13$\pm$054 &	37.20$\pm$1.01		\\

\hline
\end{tabular}
\end{center}
\end{table}

\section{{Comparison of brown dwarfs and hot Jupiters photometry and spectra}}\label{hot_jupiters}

{Color-magnitude diagrams have been traditionally used to directly compare the colors and absolute magnitudes of low-mass stars and brown dwarfs \citep[and references therein]{Burgasser2008,Faherty2012,Dupuy_Liu2012}, revealing that different parameters   influence the atmospheres of these objects. Beside effective temperatures, other secondary parameters that influence brown dwarfs colors include surface gravity, metallicity, dust sedimentation, and non-equilibrium chemistry. Brown dwarfs and hot Jupiters share similar effective temperatures and size ranges. Nevertheless, direct comparisons are usually challenging, as hot Jupiters orbit at close distance to their host stars, and they are, therefore, highly irradiated and difficult to observe directly. }

In this Section, we  compare the hot Jupiters HST/WFC3 {near-infrared} day-side emission photometry and spectra (see Table~\ref{all_hot_jupiters}) to {similar} photometry and spectra of field and young brown dwarfs, {to explore differences and similarities between these two classes of substellar objects}.

Before proceeding to the spectral comparison, we need to transform the relative flux density typically given for eclipse depth (ED) (a ratio between the flux densities of the planet and the host star , $F_{\lambda, \,planet}$/$F_{\lambda, \,star}$), to absolute (physical) flux density. If the flux of the planet is given in relative eclipse depth, we transform first those units to relative flux given in $F_{planet}$/$F_{star}$ by:

\begin{equation}\label{fplanet}
\frac{F_{\lambda, \,planet}}{F_{\lambda, \,planet}+F_{\lambda, \,star}} = ED  \rightarrow
\frac{F_{\lambda, \,planet}}{F_{\lambda, \,star}} = \frac{1}{1/ED -1} 
\end{equation}

Once the flux of the planet is given in $F_{\lambda, \,planet}$/$F_{\lambda, \,star}$, to transform to actual physical flux density units, we use a model spectrum for the temperature and surface gravity of the spectral type of each the parent stars given in Table~\ref{all_hot_jupiters}. We used the BT-Settl atmospheric models \citep{Allard2012a}, scaling  the model absolute flux by $(R/d)^2$, where \textit{R} is the radius of the star in m, and \textit{d} is the distance of the system to Earth. The star radii were obtained from Table 5 from \cite{Pecaut_Mamajek2013}. We used trigonometric distances available either from the Gaia  DR2 \citep{Gaia2018} or in the literature. Finally, we binned the stellar model spectra to match the corresponding HST/WFC3 spectra bins for each hot Jupiter, and obtained the physical flux density for each planet solving equation \ref{fplanet} as:

\begin{equation}
F_{\lambda, \,planet} =  \frac{F_{\lambda, \,star}}{1/ED - 1} 
\end{equation}

Once we transformed the units of hot Jupiter spectra to physical units ($F_{\lambda, \,planet}$ in erg $\mathrm{s^{-1}}$ $\mathrm{cm^{-2}}$ $\AA^{-1}$), we obtained HST/WFC3 photometric magnitudes in the $J$ (1.10-1.35 $\mu m$) and $H_{s}$-band (1.50-1.69 $\mu m$)  for each hot Jupiter and brown dwarf in our sample. The total flux in each band was calculated by integrating the flux densities in the relevant wavelength ranges. To obtain $J$ and $H_{s}$ HST/WFC3 magnitudes as:

\begin{equation}\label{STmag}
STmag_{J/H_{s}} = -2.5 \log F_{\lambda, \,planet} - ZP_{J/H_{s}}
\end{equation}

Where $F_{\lambda, \,planet}$ is given in  erg $\mathrm{s^{-1}}$ $\mathrm{cm^{-2}}$ $\buildrel _{\circ} \over {\mathrm{A}}$,  and \textit{ZP} is the zeropoint in the $J$ or $H_{s}$-band. We use the zero points for the {F125W ($ZP_{F125W}$ = 25.3293, in Vega magnitude) and the F160W filters ($ZP_{F160W}$ = 24.6949, in Vega magnitude)} that are centered at those bands\footnote{http://www.stsci.edu/hst/wfc3/ir\_phot\_zpt}.

\begin{figure*}
\centering
\label{CMD_HST}
\includegraphics[width=0.6\textwidth]{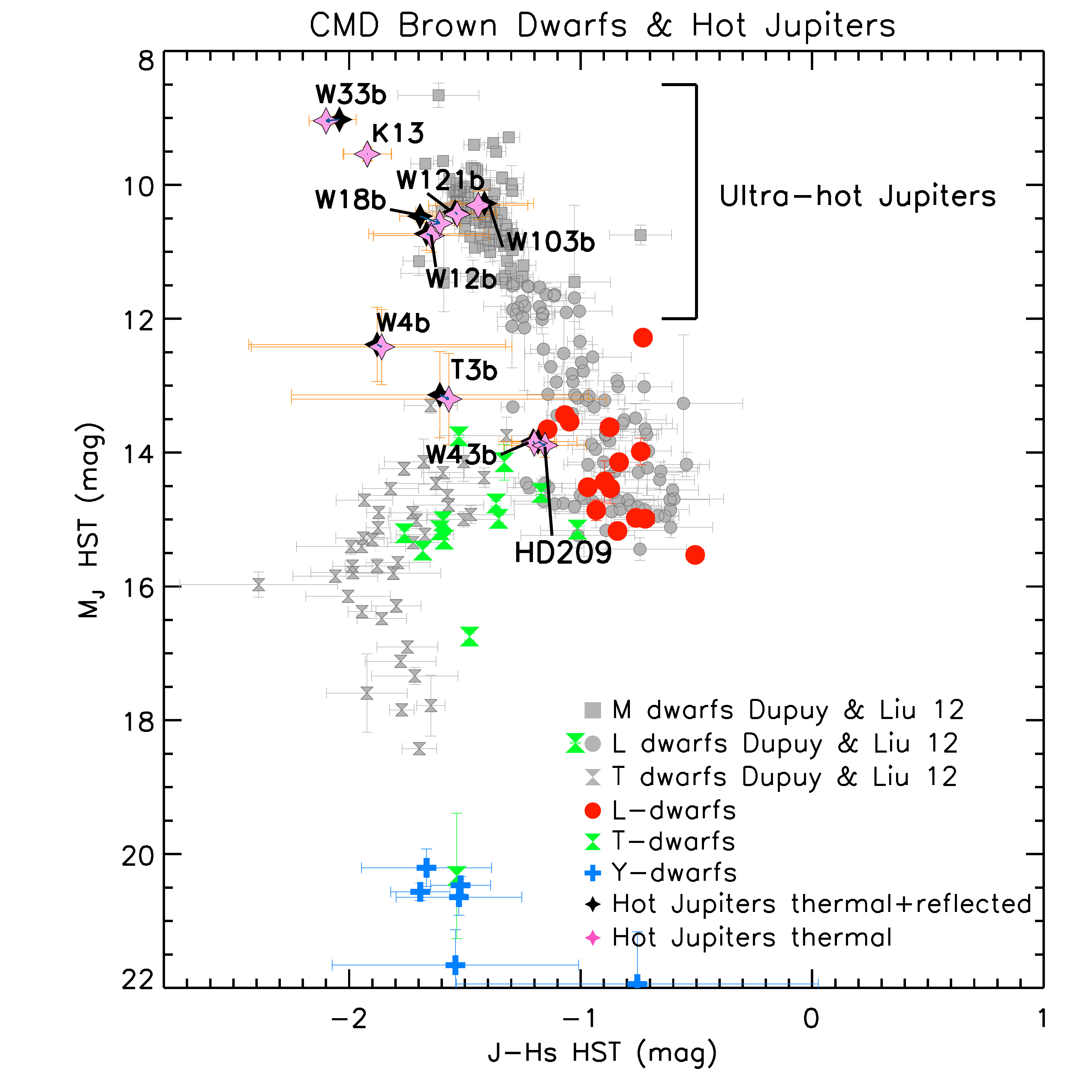}

\caption{\label{cmd_hot} CMD diagram showing brown dwarfs and hot Jupiters together. Red dots represent L dwarfs, green hour-glasses represent T dwarfs, and blue crosses represent Y dwarfs. hot Jupiters are shown as black stars, and hot Jupiter after removing the contribution of their albedo (thermal flux) are shown as pink stars.}
\end{figure*}

To compare the colors of highly-irradiated hot Jupiters and isolated brown dwarfs, we plot a $J$-$H_{s}$ HST color vs. $M_{J}$ HST magnitude  in a CMD diagram with all brown dwarfs with available trigonometric parallax and all hot Jupiters in Figure~\ref{CMD_HST}.   {For comparison, we include M-Y dwarfs presented in \cite{Dupuy_Liu2012} after transforming their 2MASS photometry to HST photometry using the polynomials presented in Section \ref{transf_mag} of the Appendix. Grey squares represent M dwarfs, grey dots are L dwarfs, and grey hourglass symbols correspond to T dwarfs. In addition, we add the targets from our sample with  trigonometric parallax: red dots are L dwarfs, green hourglass symbols are T dwarfs, blue crosses are Y dwarfs,  black stars are hot Jupiters, and pink stars are hot Jupiters  after removing the contribution from the reflected light of the host star}. The observed flux from the hot Jupiter is:

\begin{equation}
F_{\lambda, \,planet} = F_{\lambda, \,thermal} + F_{\lambda, \,reflected}  
\end{equation}

Where $\mathrm{F_{\lambda, \,planet}}$ is the observed hot Jupiter spectra, $F_{\lambda, \,thermal}$ is the thermal flux from the planet, and $\mathrm{F_{\lambda, \,reflected}}$ is the flux reflected from the star by the planet in the near-infrared, that depends on its albedo at those wavelengths.

\begin{equation}
F_{\lambda, \,thermal} = F_{\lambda, \,planet} - F_{\lambda, \,reflected}  
\end{equation}

Where,

\begin{equation}
F_{\lambda, \,reflected} = A \times \frac{F_{\lambda, \,star}}{4\pi a^{2}_{star-planet}} \times 4\pi R^{2}_{Planet}
\end{equation}

$A$ is the geometrical albedo in the near-infrared. The geometrical albedo is wavelength dependent, and varies depending on multiple factors, including the composition of the planetary atmosphere, particle sizes in its atmosphere, surface gravity, etc. \citep{Marley1999}. As it is non-trivial to determine the wavelength dependency of hot Jupiter geometrical albedo, we assume a maximum near-infrared constant albedo of 0.1 for the estimation of the reflected flux from the star, as predicted by \cite{Marley1999}. $F_{\lambda, \,star}$ is the flux density of the star given by the corresponding model spectra \citep{Allard2012a}, scaled to the star's distance from Earth. The scaling was done by multiplying the model spectrum by $(R_{star}/d_{star-\Earth})^2$, where $R_{star}$ is the radius of the star (obtained from \citealt{Pecaut_Mamajek2013}) and $d_{star-\Earth}$ is the distance between the Earth and the star (based on the trigonometric parallaxes of the host stars, see references in Table~\ref{all_hot_jupiters}). $a_{star-planet}$ is the  star-planet distance, available in the literature for all hot Jupiters  (see Table~\ref{all_hot_jupiters}). Finally, $R_{Planet}$ is the radius of each planet (Table~\ref{all_hot_jupiters}).

As seen in Figure~\ref{CMD_HST}, the contribution of the albedo-assumed reflected light to the observed  HST/WFC3 emission  spectra is almost negligible, and it does not significantly change the colors and/or absolute magnitudes of the hot Jupiters considered in this study.

\begin{figure*}
\centering
\includegraphics[width=0.45\textwidth]{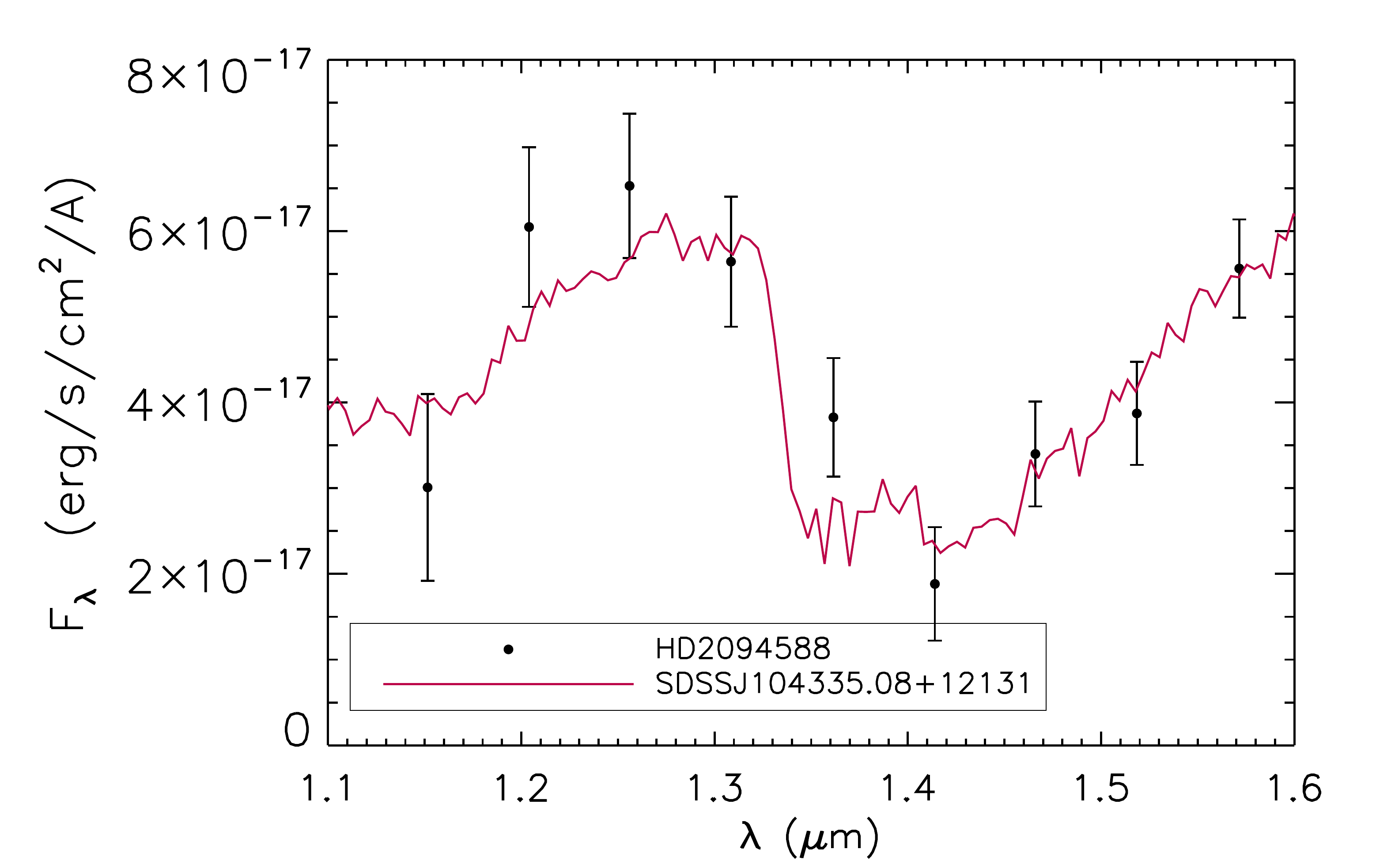}
\includegraphics[width=0.45\textwidth]{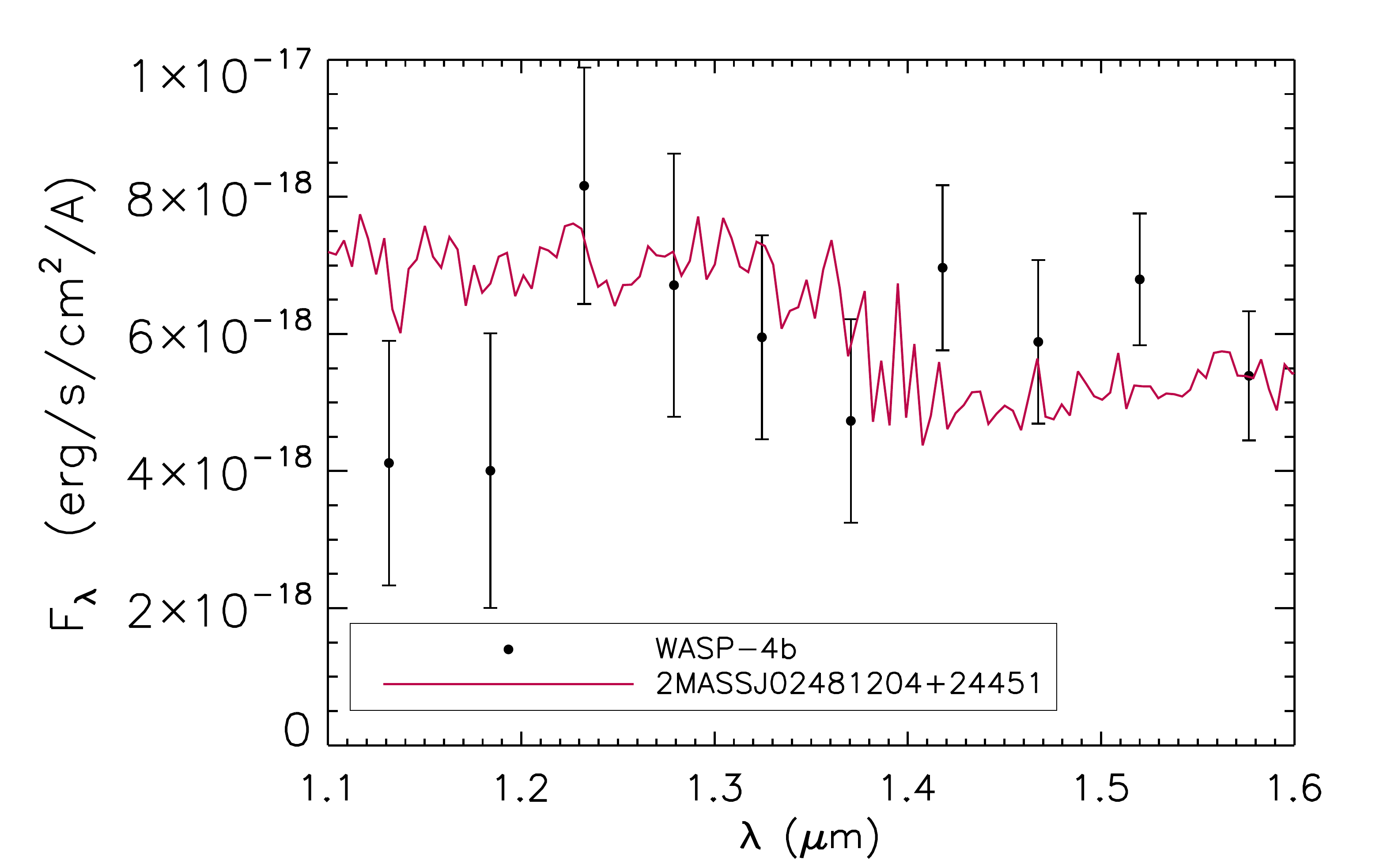}
\includegraphics[width=0.45\textwidth]{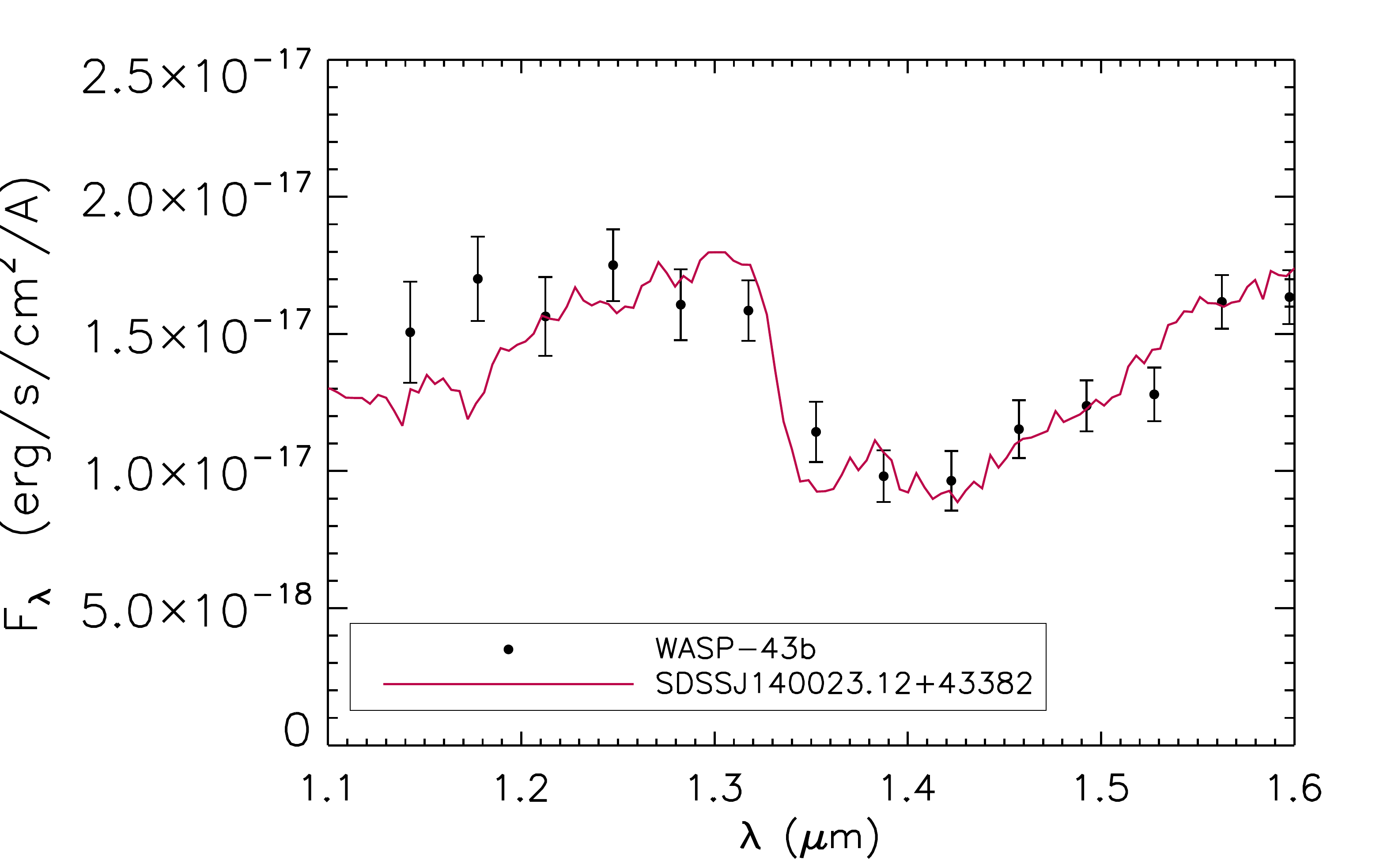}
\includegraphics[width=0.45\textwidth]{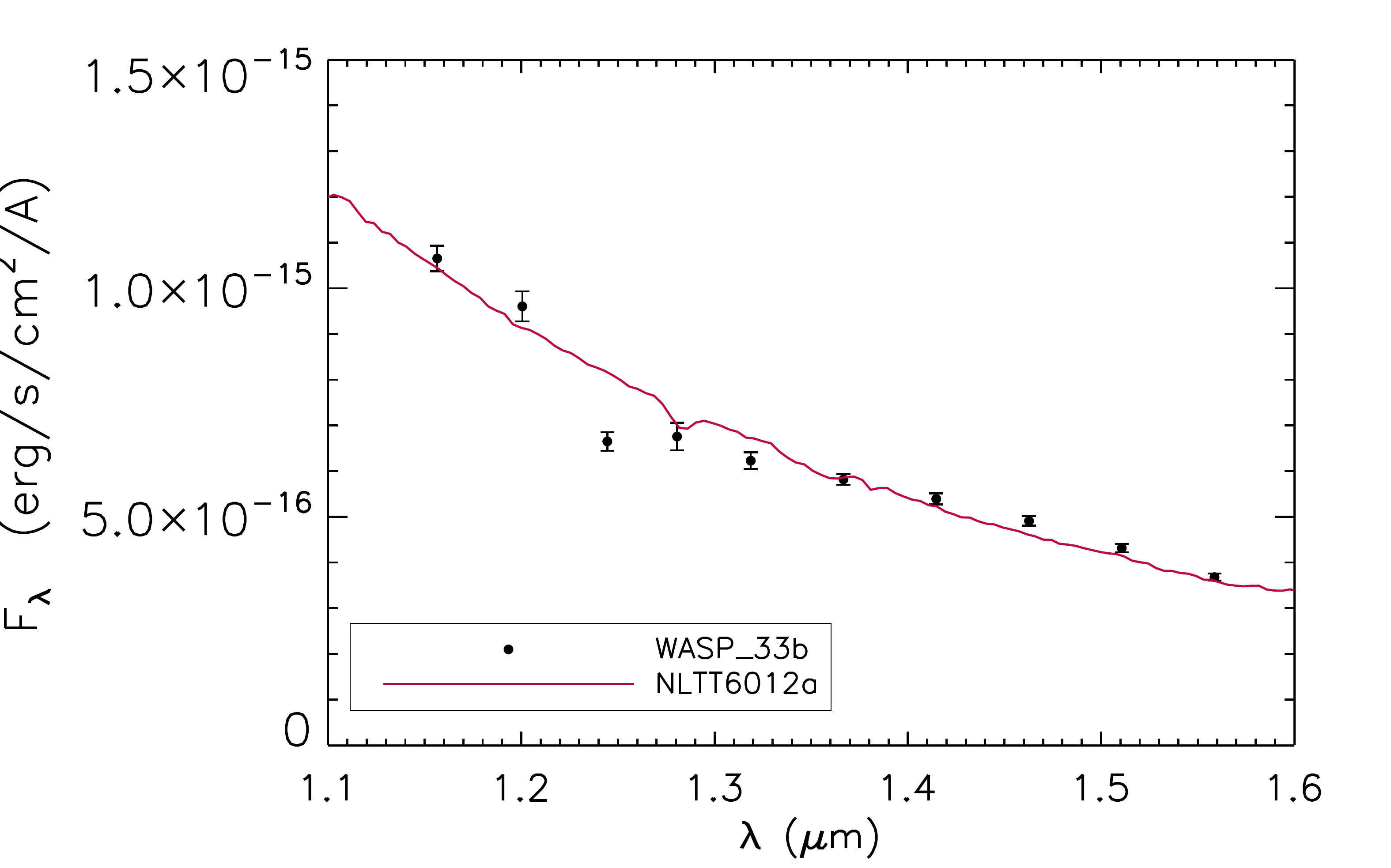}
\includegraphics[width=0.45\textwidth]{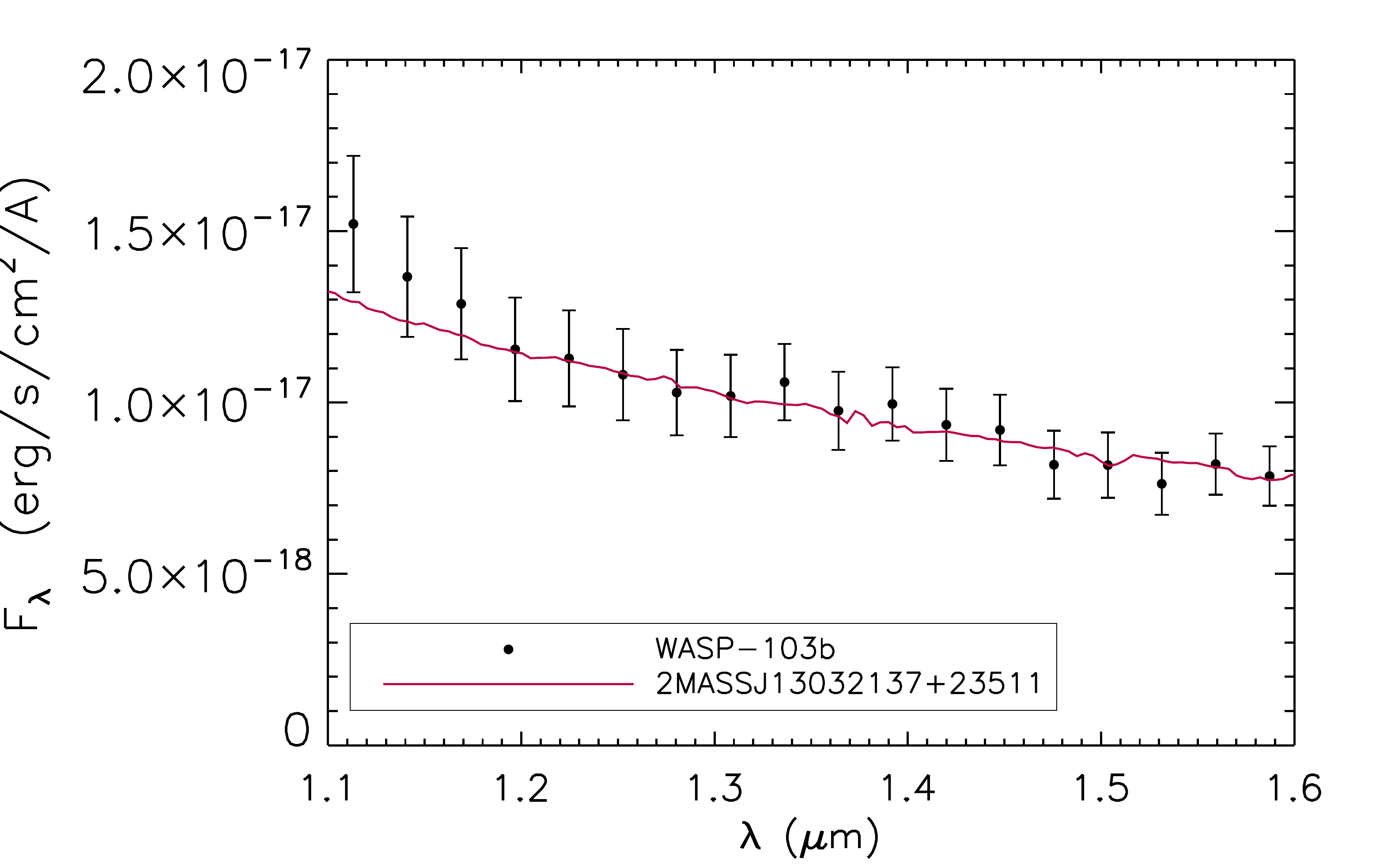}
\includegraphics[width=0.45\textwidth]{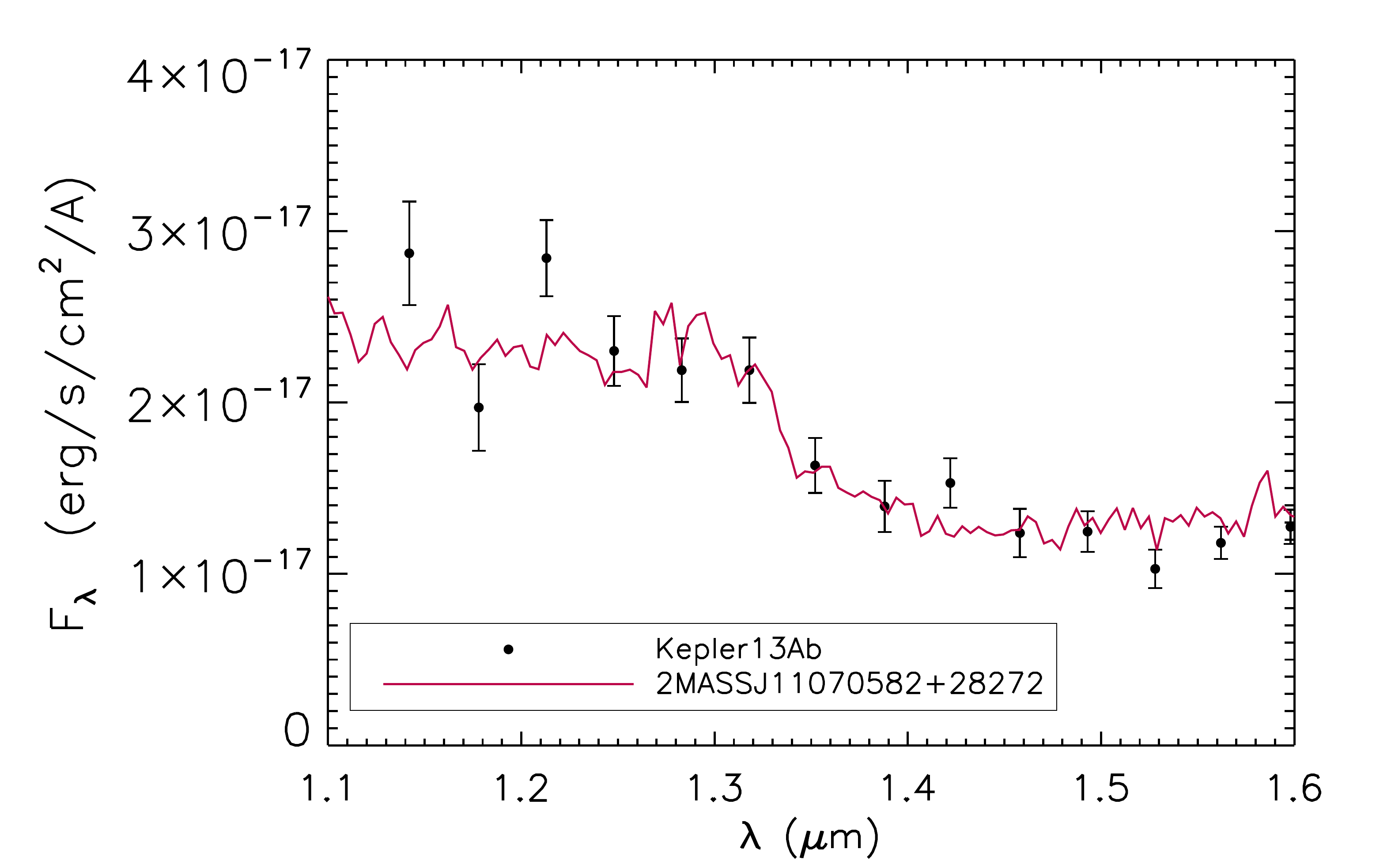}
\includegraphics[width=0.45\textwidth]{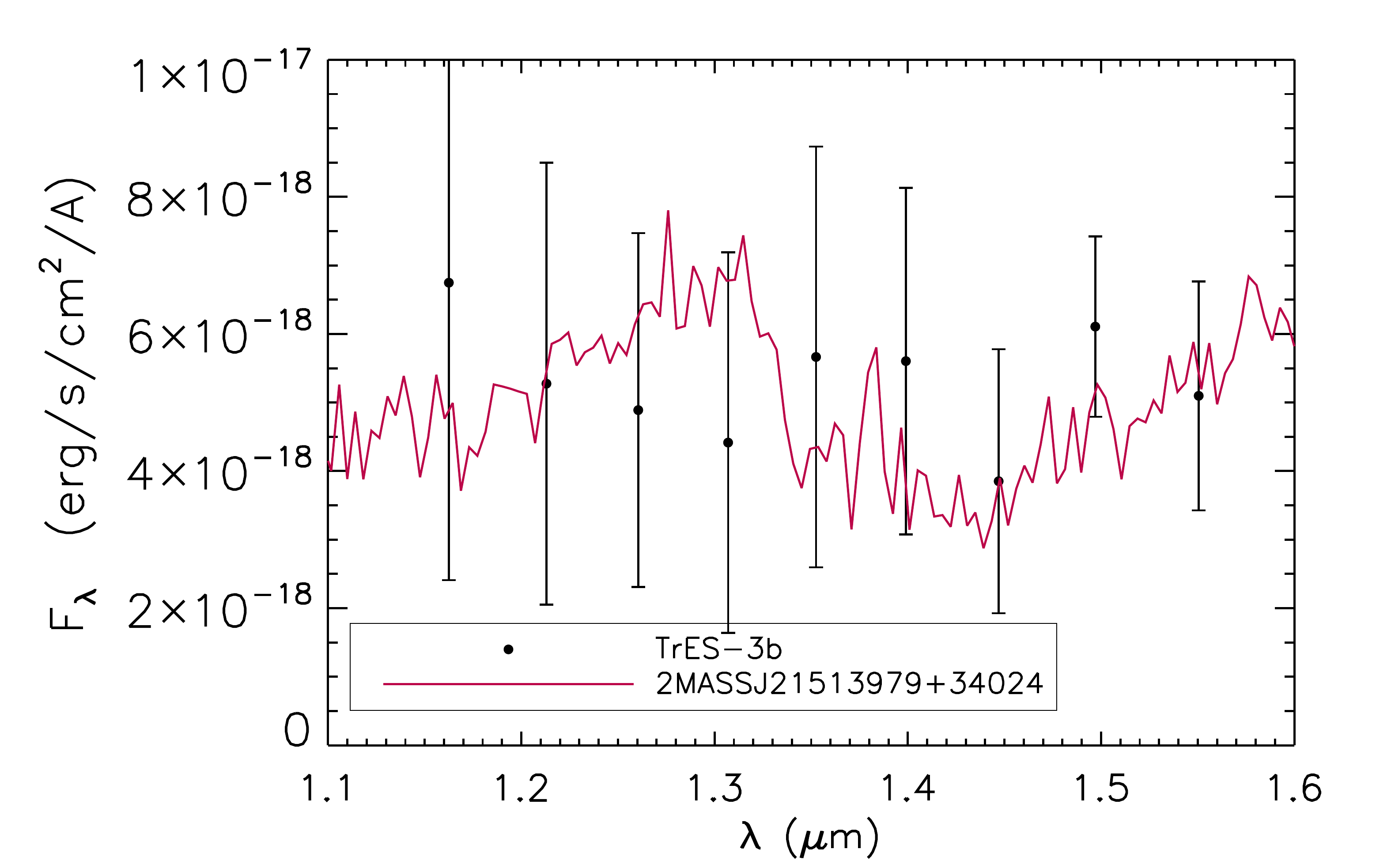}
\includegraphics[width=0.45\textwidth]{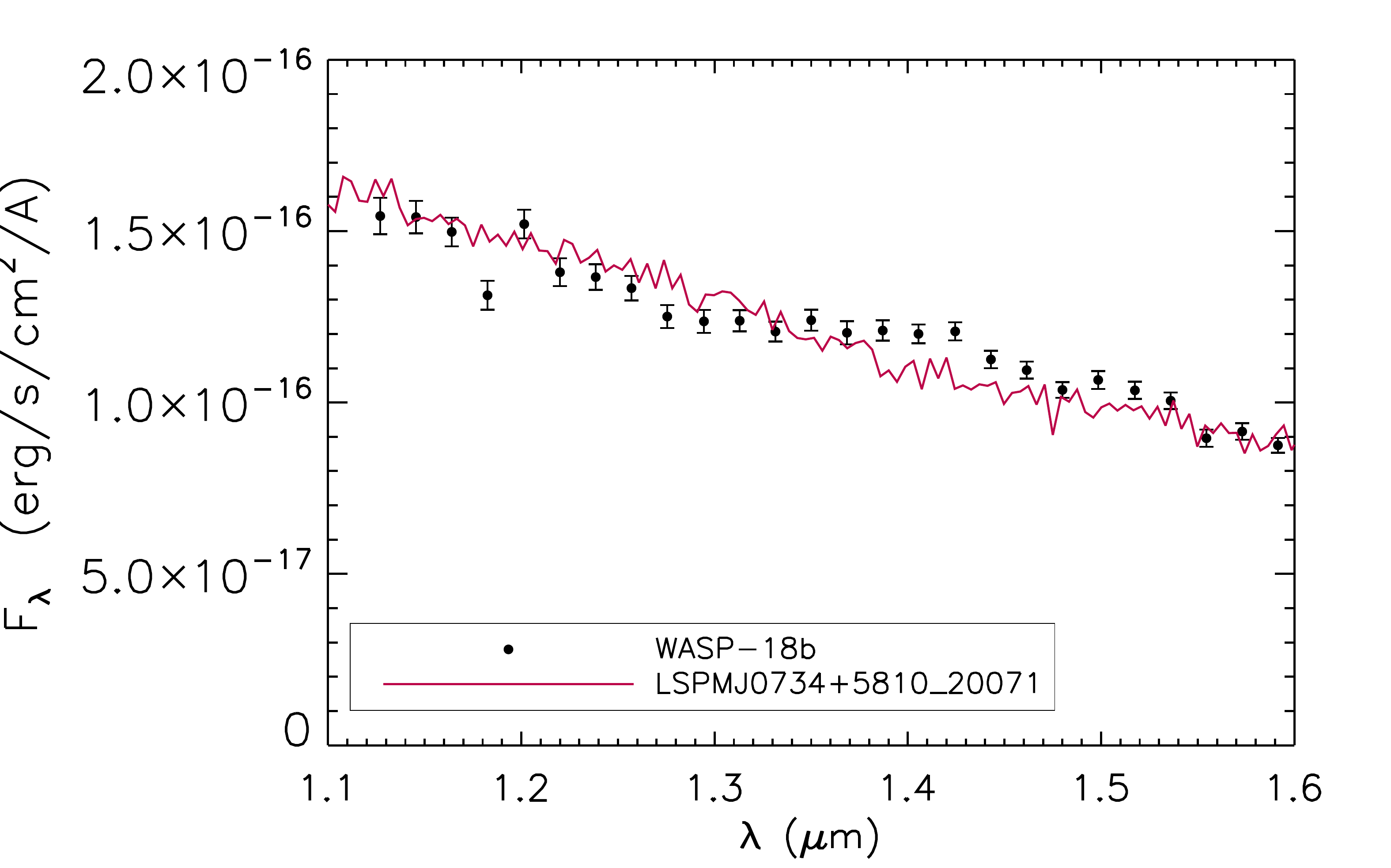}
\caption{\label{BD_best_match} Best matches of hot Jupiters to brown dwarf spectra and M-dwarfs.}
\end{figure*}

{A sub-group within hot Jupiters has been identified recently as \emph{ultra-hot Jupiters}. The six brightest hot Jupiters in our sample belong to this category (WASP-33B, Kepler-13Ab, WASP-18b, WASP-121b, WASP-103b, and WASP-12b). \citet{Lothringer2018} and \citet{Parmentier2018} among others proposed that under extreme irradiations strong molecular dissociations and H$^{-}$ opacity will significantly reduce or even eliminate the molecular absorption bands in the dayside emission spectra of hot Jupiters. In our spectral library, these \emph{ultra-hot Jupiters} all appear to lack the 1.4$\mu$m water absorption band  (Figure \ref{BD_best_match}), while they have consistent $J-H$ colors with the color sequence defined by M dwarfs (Figure \ref{CMD_HST}). This agrees with the $\mathrm{T_{eff}}$ of 2,500--3,000~K estimated by \cite{Haynes2015}, \cite{Beatty2017}, \cite{Sheppard2017}, \cite{Evans2017}, \cite{Cartier2017} and \cite{Stevenson2014}, respectively for the ultra-hot Jupiters in our sample. 
}  WASP-4b and TrES-3b have similar  $\mathrm{M_{J}}$ to those of early-type L dwarfs \citep{Dupuy_Liu2012}, with similar $\mathrm{T_{eff}}$ of $\sim$2,000~K \citep{Ranjan2014}. Finally, HD~209458B and WASP-43b have similar estimated $\mathrm{T_{eff}}$ of $\sim$1,500-1,700~K  \citep[respectively]{Line2016,Stevenson2014_Nat}, than mid-L dwarfs, and actually lie among other mid-L dwarfs in Figure~\ref{CMD_HST}.

Finally, we compared  the HST/WFC3 hot Jupiter emission spectra compiled in this work, to spectra collected in the SpeX Spectral Library. We chose the best-fits based on the value of their modified $\chi^{2}$ (G), as obtained using equation \ref{chi}, and visual inspection.
We found that {only three  of the ten hot Jupiters in our study had best matches to mid-L dwarfs: HD 209458B was matched to SDSS~J104335.08+12131 (L7),  WASP-43b was matched to SDSS~J140023.12+43382 (L7), and TrES-3b was matched to 2MASS J21513979+34024 (L7 peculiar).
The other seven hot Jupiters are best matched to M-dwarfs: WASP-33b, WASP-103b  and WASP-18b are best matched by  M3-type stars (NLTT 6012a, 2MASSJ13032137+23511, and LSPMJ0734+5810, respectively).  Kepler-13Ab is best matched by 2MASS J11070582+28272 (M7), {that is consistent with the result of \cite{Beatty2017}, who found a best match to an M8 brown dwarf}. WASP-4b best matched to 2MASS J02481204+24451 (M8). Finally, we did not find a best match for WASP-12b and WASP-121b (see Figure \ref{BD_best_match})}.

{These results are in general consistent with the positions of hot Jupiters within the CMD in Figure~\ref{CMD_HST}, and with the temperatures predicted for those objects by their respective authors. In addition, these results also agree with predictions made by the atmospheric models presented in \cite{Fortney2008}, in which they suggested that there are two classes of hot Jupiter day-side atmospheres analogous to the M- and L dwarfs spectral types, that they called \textit{pM} and \textit{pL}, respectively. The \textit{pM} class planets have hot stratospheres due to the high irradiation of their parent star ($\mathrm{T_{eff}>}$2000~K) with temperature inversion in their atmospheres, and molecular bands in emission. The models predict that the temperature differences between their day and night side are high due to radiative time constants at photospheric pressures are shorter than advective timescales. In contrast, the \textit{pL} class planets are less irradiated by their parent star. The incident flux from the parent star is absorbed  in the atmosphere, and redistributed easily, as there is no thermal inversion in their photospheres. Thus, they have cooler day sides and warmer night sides.  Their spectra are dominated by $\mathrm{H_{2}O}$, in the near-infrared, and by Na and K absorptions in the optical. }

{Our results are consistent with, and extend on, a previous study by \citet{Triaud2014a}, who compared in a CMD plot transiting planet dayside emission measured in two Spitzer/IRAC bands ([3.6] and [4.5] filters) to those of brown dwarfs. This comparison suggested overall similarity, with a few possible outliers. The Spitzer color-magnitude comparison is particularly sensitive to the presence/absence of methane absorption. 
In a follow-up work \citet{Triaud2014b} extended their study to near-infrared continuum bands, and to a larger sample using photometric distances to compare their objects with brown dwarfs in a CMD plot.  They found that, for a given luminosity, hot Jupiters' daysides show  larger ranges of colors than brown dwarfs, specially with decreasing intrinsic luminosity.
In contrast, our study suggests that brown dwarfs and hot jupiters tend to be similar in the shorter-wavelength continuum emission (J and H bands) bands and that, for many hot jupiters, brown dwarfs can provide surprisingly good spectral matches. }

\cite{Fortney2008} used these atmospheric models to predict the classes of several hot Jupiters, in which TrES-3b and HD 209458B were included. He found that TrES-3b should belong to the \textit{pM} class, as we consistently obtained. He predicted HD 209458B to be in the transition zone between both classes, but, as \cite{Line2016} also found, we conclude that this hot Jupiter matches better to an L-dwarf spectrum, thus, to the \textit{pL} class.

\section{Summary and Conclusions}\label{conclusions}

We present a very high-quality HST/WFC3 near-infrared spectral library of brown dwarfs (field and companion to stars), planetary-mass objects, and hot Jupiters, to enable quantitative comparative studies. In this paper, we provide an initial characterization and analysis of these HST/WFC3 near-infrared spectra:

\begin{enumerate}
\item In Section \ref{SpT} we derive near-infrared spectral types for the brown dwarfs and the substellar companions to stars uniformly, using as comparison the SpeX Spectral Library templates. We conclude that their spectral types are mostly consistent with the spectral types provided in the literature within $\pm$1.5 subspectral type. The only exception is for low surface-gravity objects, for which the differences found in spectral types are $\pm$3 spectral types. This is expected, as the SpeX Spectral Library templates are mostly composed of field gravity low-mass stars and brown dwarfs.

\item In Section \ref{cmd} we plot a $\mathrm{M_{J}}$ vs. $J-H$ color-magnitude diagram to compare our sample to other substellar objects, with the objective to identify brown dwarfs with peculiar colors/brightness, including red or blue objects, low-surface gravity objects, binaries, etc. We find that objects 32 (2M1750+1759, known binary), and 34 (2M0559-34, overluminous), are {overluminous in the color-magnitude diagram suggesting that they are potential multiple systems}. 

\item In Section \ref{lowg_objects} we obtain the $H$-continuum and $\mathrm{KI_{J}}$ near-infrared spectral indices from \cite{Allers2013} to search for potential L4 to L8 low-surface gravity substellar objects in our sample. We found two very low-gravity dwarfs:  CD$-$352722B (object 1) and 2M0355$+$1133 (object 3). In addition, we found five intermediate surface-gravity objects: 2M0421-6306 (object 6), W0047 (object 9), 2M0107+0041 (object 11), PSO J318.5-22 (object 15) and 2M2224-0158 (object 16).


\item In Section \ref{composite} we apply the method from \cite{Burgasser2006, Burgasser2010} and \cite{Bardalez_Gagliuffi} to search for candidates for composite spectra in our sample. Their spectral indices selected 13 composite spectra candidates, from which eight were selected from the F-statistic analysis described in Section~\ref{composite}. None of these eight objects are overluminous, as would be expected for binary or multiple brown dwarfs systems. In addition, we found that five of the eight selected objects have been reported in the literature as photometrically variable. Thus, this method might be useful to find potential variable late L and early T dwarfs. Nevertheless, we also found that not all objects in our sample with reported photometric variability have been detected by  \cite{Burgasser2006, Burgasser2010}, and \cite{Bardalez_Gagliuffi} method. The indices itself detected  9 out of 19 variables and 3 no variable objects in our sample, with spectral types between L4 and T2. 

\item In Section \ref{classification} we measure the depths of the water band at $\sim$1.4~$\mu$m and the methane band at $\sim$1.2~$\mu$m for brown dwarfs and substellar companions to stars. We derive a relation between their near-infrared spectral types and the depths of those bands, providing a tool for spectral classification of other substellar objects. 

\item In Section \ref{hot_jupiters} we compare the emission spectra of the dayside of hot Jupiters to the spectra of brown dwarfs and substellar companions to stars in our sample and low-mass stars and brown dwarfs from the SpeX spectral library. We found best matches to either L or M-dwarfs for eight out of the ten hot Jupiters of our sample. In addition, we plot a color-magnitude diagram using $J$ and $H_{short}$ HST bands for all our sample. The hottest hot Jupiters, WASP-33B, Kepler-13Ab, WASP-18b, WASP-121b, WASP-103b, and WASP-12b have similar $\mathrm{M_{J}}$ magnitudes to mid-M dwarfs \citep{Dupuy_Liu2012}, which agrees with the $\mathrm{T_{eff}}$ of 2,500--3,000~K estimated by their respective authors. WASP-4b and TrES-3b have similar  $\mathrm{M_{J}}$ that early L dwarfs \citep{Dupuy_Liu2012}, with similar $\mathrm{T_{eff}}$ of $\sim$2000~K. Finally, HD~209458B and WASP-43b have similar estimated $\mathrm{T_{eff}}$ of $\sim$1500-1700~K  \citep[respectively]{Line2016,Stevenson2014_Nat}, than mid-L dwarfs, and actually lie among other mid-L dwarfs in the CMD diagram.

\end{enumerate}

{The HST/WFC3 near-infrared spectra presented in this work will be available as csv files that will include wavelength, flux and uncertainty in flux. The spectral library will be available in a machine-readable format at the High-level Science Products website at the MAST archive under the Cloud Atlas program's page \doi{doi:10.17909/t9-asft-6k38}.}

\acknowledgments

We thank our anonymous referee for his/her useful comments that helped to improve our paper. Based on observations made with the NASA/ESA Hubble
Space Telescope, obtained at the Space Telescope Institute, which is operated by AURA, Inc., under NASA contract NAS 5-26555, under  GO-13241, GO-14241, GO-12550, GO-13176, GO-12550, GO-13299, GO-13280, GO-13281, GO-12314, GO-14051, GO-12217, GO-13178, GO-12970, GO-12230,  GO-13467, GO-12495, GO-14050, GO-13467, GO-12181, GO-13308, and GO-14767. This publication makes use of data
products from the Two Micron All Sky Survey, which is a joint
project of the University of Massachusetts and the Infrared
Processing and Analysis Center/California Institute of Technology,
funded by the National Aeronautics and Space
Administration and the National Science Foundation. This work makes use of the SpeX  Prism Spectral Library.
This work has made use of data from the European Space Agency (ESA) mission
{\it Gaia} (\url{https://www.cosmos.esa.int/gaia}), processed by the {\it Gaia}
Data Processing and Analysis Consortium (DPAC,
\url{https://www.cosmos.esa.int/web/gaia/dpac/consortium}). Funding for the DPAC
has been provided by national institutions, in particular, the institutions
participating in the {\it Gaia} Multilateral Agreement.
We have done use of the Matplotlib python library \citep{Hunter_2007}.

\appendix

\setcounter{table}{0}
\renewcommand{\thetable}{A\arabic{table}}

\setcounter{figure}{0}
\renewcommand{\thefigure}{A\arabic{figure}}

\begin{longrotatetable}
\begin{deluxetable*}{llllll}
\tablecaption{Log of the sample of L, L-T, T and Y dwarfs with HST/WFC3 spectroscopy.\label{quality_spectra}}
\tablewidth{700pt}
\tabletypesize{\scriptsize}
\tablehead{
\colhead{Num.} & \colhead{Name} & 
\colhead{Obs. dates} & \colhead{Num. orbits$\times$visits} &\colhead{$\mathrm{T_{exp}}$ single expo. (s)} & 
\colhead{Num. single expo./orbit} 
} 
\startdata
	1 & CD-352722b & 2015 Sept 07 & 2x1 & 29.6 &  98  \\ 
	2 & 2MASS J17502484-0016151 & 2012 Jun 15 & 1x1 & 22.6 & 64   \\ 
	3 & 2MASS J03552337+1133437 & 2015 Oct 06 & 2x1 &  89.6 &  46  \\     
	4 & 2MASS J18212815+1414010 & 2013 Jun 09 \& Jun 27 & 3x2 & 112.0 & 19  \\    
	5 & 2MASSW J1507476-162738 & 2013 Apr 30 \& May 12 & 4x2 &  67.3 & 30  \\    
	6 & 2MASSI J0421072-630602 & 2012 Mar 20 & 1x1 & 112.0 & 19\\ 
	7 & 2MASS J05395200-0059019 &2012 Mar 01 & 1x1 & 45.0 & 37 \\ 
	8 & 2MASSI J1711457+223204 & 2012 Aug 01 & 1x1 & 223.7 & 9 \\     
	9 & 2MASS J00470038+6803543 & 2016 Jun 06/07 & 6x1 & 201.3 & 65  \\ 
	10 & LP261-75B & 2016 Dec 20 & 6x1 & 201.4 & 66 \\ 
	11 & 2MASS J01075242+0041563& 2017 Jan 02 & 6x1 & 201.4 &  22 \\     
	12 & 2MASSW J1515008+484742 & 2012 Feb 25 & 1x1 & 45.0& 36 \\ 
	13 & 2MASS J06244595-4521548 & 2012 May 08 & 1x1 & 45.0 & 38  \\ 
	14 & 2MASSW J0801405+462850 & 2011 Nov 10 & 1x1 & 223.7 & 11   \\ 
    15 & PSO J318.5-22 & 2016 Sept 08 & 5x1 & 278.0 & 9 \\
  	16 & 2MASSW J2224438-015852 & 2015 Sept 09 & 2x1 & 89.6 &  26 \\ 
	17 & Luh 16AB &2013 Nov 8 & 5x1 & 76.2 & 100    \\ 
	18 & 2MASSI J0825196+211552 & 2012 May 09 & 1x1 & 112.0 & 21 \\  
	19 &2MUCD 10802 & 2011 Dec 09 & 1x1 & 45.0 & 40  \\ 
	20 &2MASS J16322911+1904407 & 2012 Aug 11 & 1x1  & 223.7 & 9 \\ 
	21 &2MASSW J0310599+164816 & 2012 Aug 25 & 1x1 & 223.0 & 10 \\ 
	22 &2MASS J12195156+3128497 & 2012 Jun 18 & 1x1 & 223.7 & 9\\
	23 &SDSS J075840.33+324723.4 & 2014 Apr 12 & 5x1 &  112.0 & 22  \\ 
	24 &2MASS J10393137+3256263 & 2012 May 08 & 1x1 & 223.7 & 11 \\ 
	25 &2MASS J09090085+6525275 & 2012 Aug 21 & 1x1 & 223.7 & 10   \\ 
	26 &2MASS J21392676+0220226& 2010 Oct 21 & 6x1 & 22.3 & 11\\     
	27 &2MASS J13243553+6358281 & 2012 Feb 25 & 1x1 & 112.0 & 21  \\ 
    28 &2MASS J16291840+0335371 & 2015 June 06  & 4x1 & 112.0  &  21  \\ 
	29 &HNPEGB  & 2017 May 16 & 6x1 & 201.4 & 65  \\ 
	30 &SIMP J013656.5+093347.3 & 2013 Sep 28 \& Oct 7 & 4x2 & 112.0 & 19  \\ 
	31 &GUPSCB & 2018 Jan 08 & 6x1 & 201.4 & 43  \\    
	32 &2MASS J17503293+1759042 & 2012 Oct 05 & 1x1 & 223.7 & 10   \\ 
	33 &2MASS J00001354+2554180 & 2012 Sept 13 & 1x1 & 45.0 & 40 \\     
	34 &2MASS J05591914-1404488 & 2011 Oct 16 & 1x1 & 22.6 & 62 \\ 
	35 &2MASSI J2339101+135230 & 2012 Aug 21 & 1x1 & 223.7 & 11 \\ 
	36 &2MASS J11101001+0116130 & 2016 Feb 10 & 2x1 & 201.4 & 21  \\     
	37 &2MASS J22282889-4310262& 2013 Jul 20 \& 27 & 4x2 & 201.3 & 39 \\  
	38 &2MASS J08173001-6155158 & 2011 Oct 09 & 1x1 & 22.6 & 67   \\ 
	39 &S Ori J053814.5-024512   & 2010 Sept 05 & 1x1 &602.7 &  4 \\ 
	40 &2MASSI J0243137-245329 & 2011 Dec 31 & 1x1 & 112.0 & 19   \\ 
	41 &2MASS J16241436+0029158 & 2012 Jul 13 & 1x1 & 112.0 & 17 \\ 
	42 &CFBDSIR2149-0403 & 2015 Sept 09 \& Nov 18 & 4x2 & 290.7 & 16 \\ 
    43 &S Ori J053814.5-024512  & 2010 Oct 06 & 1x1 & 602.7 & 4  \\ 
	44 &ROSS458C & 2018 Jan 05 \& 06 & 7x1 & 201.4 &  21  \\  
	45 &WISEA J032504.52–504403.0  & 2013 Aug 04 & 1x1  & 403.0 & 4   \\ 
	46 &WISEA J033515.07+431044.7  & 2013 Aug 30  &  1x1 & 453.0 &  4  \\ 
	47 &WISEA J040443.50–642030.0  & 2013 Apr 09 & 1x1 & 453.0  & 4  \\    
	48 &WISEA J221216.27–693121.6 & 2013 Sept 11 & 1x1  & 453.0 & 4  \\ 
	49 &WISEA J094306.00+360723.3 & 2013 Feb 20 & 1x1 & 503.0 & 4 \\     
	50 &WISEA J154214.00+223005.2 & 2012 Mar 04& 1x1 & 503.0 & 4  \\ 
	51 &WISEA J035934.07–540154.8 & 2011 Aug 10 & 1x1 & 553.0 & 4 \\ 
	52 &WISEA J041022.75+150247.9  & 2012 Sept 01 & 1x1 & 503.0 & 4 \\     
	53 &WISEA J073444.03–715743.8 & 2013 May 20 & 1x1 & 453.0 &4  \\ 
	54 &WISEA J120604.25+840110.5 & 2013 July 15 & 1x1 & 453.0 & 4   \\ 
	55 &WISE J154151.65–225024.9 & 2013 May 09 & 1x1& 453.0 &4  \\ 
	56 &WISEA J173835.52+273258.8  & 2011 May 12 & 1x1 & 503.0 & 4 \\     
	57 &WISEA J205628.88+145953.6 & 2011 Sept 04 & 1x1 & 503.0 & 4 \\ 
	58 &WISEA J222055.34–362817.5 & 2013 Jun 20 & 1x1 & 1103.0 & 4 \\ 
	59 &WISEA J220905.75+271143.6 & 2012 Sept 15 & 1x1 & 503.0 & 4  \\ 
	60 &WISEA J082507.37+280548.2 & 2014 Jan 17 & 1x1 & 2406.0 & 3 \\ 
	61 &WISEA J140518.32+553421.3 & 2011 Mar 14 & 1x1 & 553.0 & 4  \\     
	62 &WISEA J163940.84–684739.4 & 2013 Oct 29 & 1x1 & 602.9 & 4  \\ 
	63 &WISEA J053516.87–750024.6  & 2011 Sept 27   & 1x2 & 553.0 & 4  \\ 
    63 &WISEA J053516.87–750024.6  & 2012 Sept 17   & 1x2 & 553.0 & 4  \\ 
    63 &WISEA J053516.87–750024.6  & 2013 Sept 27 & 1x1 & 1269.0 & 6  \\ 
	64 &WISEA J035000.31–565830.5  & 2011 Aug 13  & 1x1  & 553.0 & 4  \\ 
	65 &WISEA J064723.24–623235.4  & 2013 May 13 \& 2013 Dec 29 &1x2 & 1203.0 & 6 \\
	66 &WISEA J235402.79+024014.1 & 2013 Sept 22 & 1x1 & 806.0 & 4 \\ 
    67 & WASP-18b &	2014 Apr-Jun \& Aug	&	6x4	&	73.74	&  8? \\
    68 &WASP-33b   & 2012 Nov 25 \& 2013 Jan 14 & 5x2  &  51.7 & 119 \\ 
    69 &WASP-12b   & 2011 Apr 12 & 5x1 &7.35  &  188 \\ 
    70 &WASP-121b    & 2016 Nov 10 & 5x2  & 103 &  16 \\ 
    71 &WASP-43b    & 2013 Nov 09 \& Dec 5  & 14x2 & 103.129 & 19 \\
    72 &WASP-103b   & 2015 Jun 17 \& 17 & 5x2 & 81.089 &  12\\
    73 &TrES-3b   & 2011 March 02 & 4x1 & 36.02 & 219 \\
    74 &Kepler-13Ab    & 2014 Apr 28 \& Oct 13 & 5x2 & 7.6 & 101 \\ 
    75 & HD 209458B   & Sept-Dec 2014 & 5x5 &14.971  & 43  \\ 
	76 &WASP-4b   & 2010 Nov 25 & 5x1 & 36.02 & 268 \\ 
\enddata
\end{deluxetable*}
\end{longrotatetable}


\section{{Dataset description and Data reduction}}\label{sec:data_red}

{In the following we summarize the key steps and references for the different data reductions  performed on the spectra compiled in this work.}

\subsection{Brown Dwarf and Low-mass Companion Spectra}\label{BD_data}


\subsubsection{{Time-resolved spectra}}
\label{UAdata}

{We present the datasets with time-resolved spectroscopy taken for several of the brown dwarfs compiled in this work. }


{The \cite{Apai2013} dataset consists of the first two brown dwarfs (SIMP J013656+093347 and 2MASS J21392676+0220226) observed in time-resolved  observations, and obtained in the GO-12314 program (PI Apai). Each objects was observed in six consecutive HST orbits.  \cite{Apai2013} provides a detailed summary of the reduction procedure.}
In this spectral library, we present the median  of the time-resolved spectra for each object.

The \citet{Buenzli2012} study presented near-infrared, time-resolved, six orbit-long  spectroscopy of a single target (\object{2M2228}, object 37) and was reduced with a method identical to that in \cite{Apai2013}. These data were also taken in program GO-12314 (PI: Apai), and the target was also a known variable brown dwarf. The observations showed spectral variability with pressure-dependent phase shifts \citet{Buenzli2012}. We took the median of the time-resolved spectra for our library.

The \citet{Buenzli2014} sample consists of 22 brown dwarfs with spectral types between L5 and T6. These data were collected in an HST SNAP program (PO 12550, PI Apai).  Basic reduction followed the same steps as for the previous programs. The ramp effect was corrected using an analytical function fitted to the flux of a non-variable star as in \cite{Apai2013}, in addition to removing the first 180~s of each time series where the scatter in the ramp effect is substantial.

The \citet{Buenzli2015} study presented spatially and temporally resolved spectroscopy for the \object{Luhman 16 A and B} binary brown dwarf components. Reduction followed the steps described in \cite{Apai2013,Buenzli2014}. We present the combined Luhman 16 A and B spectra in this paper.

The time-domain programs described above focused on relative variations and did not correct for wavelength-dependent aperture losses, which is not relevant in the related studies. However, for our purposes these corrections are necessary. We performed a uniform aperture correction on all sources from the above studies to correct for flux loss due to the finite width of the spectral extraction windows. We corrected for the missing flux per wavelength on the basis of measured wavelength-dependent flux losses, performing  a bi-linear interpolation in wavelength and aperture width of the values of the aperture corrections tabulated in Table 6 from \cite{Hartig2009}. 


{Within the \textit{Cloud Atlas} HST treasury program (HST GO 14241),  time-resolved spectroscopy observations  for eight L4 to T7 high- and low-surface gravity brown dwarfs were obtained}. The data were collected between September 2015 and September 2018. The {\em Cloud Atlas} program uses time-resolved spectroscopy to probe the spatial distribution and properties of condensate clouds. A publication in preparation (Apai et al., in prep.) will provide an overview of the program and its key results from the time-resolved spectroscopy. Results for three objects have already been published (\object{WISE0047}: \citealt{Lew2016}; \object{LP261-75B}: \citealt{Manjavacas2018}; \object{HN Peg B}: \citealt{Zhou2018}), while other papers are in preparation (\object{S0107}: Apai et al. 2018, in prep.).  Here we present the spectral results based on time-averaged spectra for all objects. We performed the {data reduction using very similar methods as described above for previous studies \citep[and references therein]{Apai2013,Buenzli2012,Buenzli2014,Buenzli2015}}. 
An important difference, however, is the use of a significantly improved WFC3 ramp correction method. \cite{Zhou2017}  identified charge trapping and delayed release as the cause of the ``ramp effect'' and developed a solid-state physics-based model capable of reliably correcting this effect in a wide variety of WFC3 data. Most of the \textit{Cloud Atlas} datasets published use the ramp effect correction by \citet{Zhou2017}.
{The uncertainty level for our spectra after the data reduction is 0.1--0.3\% per spectral bin, {measured} using the reduced individual spectra. These uncertainties are due to photon noise, errors in the sky subtraction, and the read-out noise. 
    {Finally, we performed aperture corrections following the same procedure as for the other spectra mentioned previously in this Section}.
    
  
{Within the HST  GO 13299 and 14051 (P.I. Radigan), time-resolved near-infrared spectra observations were obtained with HST/WFC3 to study the rotational modulations of two unusually blue L dwarfs. The  objective of this project was making spectrally and spatially resolved maps of these objects. These objects are SDSS~J075840.33+324723.4 (object 23), and 2MASS~J16291840+0335371 (object 28).
SDSS J075840.33+324723.4 was observed during five consecutive orbits, and 2MASS J16291840+0335371 was observed during four consecutive orbits.
The data reduction was performed using a similar procedure as for the \textit{Cloud Atlas} treasury program data. In this paper, we present the median combined spectra of all the time-resolved near-infrared spectra taken during the consecutive orbits in which these objects were observed.
The uncertainty level for these spectra after median combine all time-resolved spectra is $\sim$0.03\%  at 1.25~$\mu$m. These uncertainties are due to photon noise, errors in the sky subtraction, and the read-out noise.}

   {Finally, \cite{Biller2018} presents time-resolved spectroscopy of the red L7 dwarf, PSO~318-22}. A difference with the previous studies is that the ramp correction was corrected using four background stars in the field of view (2-3 times brighter than the target). They median combined and normalized the while light curves of the background stars to produce a calibration curve. Then they divided the target's light curve by the calibration curve to eliminate the ramp effect and other systematics, following a similar approach as done by previous ground studies \citep{Biller2015, Radigan2014}.

	\subsubsection{{Single spectra}}

{S Ori 70 and S Ori 73 \citep{Pena_ramirez2015}} are T7$\pm$0.5 and  T4.5$\pm$0.5 dwarfs, respectively.}
    S Ori 70 and S Ori 73 were observed with HST/WFC3 (PI Lucas, HST-GO-12217). {Details on the data reduction can be found in \cite{Pena_ramirez2015}.}

    
    

	
{In addition, we include 22  T8 to Y2 brown dwarf spectra presented and analyzed in \cite{Schneider2015}}. 
    The observations were carried out within the P.I. Kirkpatrick, programs 12330, 13178 and P.I. Cushing, programs HST-GO-12544 and HST-GO-12970. 
    As G141 with which the observations were performed is slitless grism, the source spectra are sometimes contaminated by photons from nearby sources. To address this problem, \cite{Schneider2015} developed a source extraction routine to define source apertures and background regions on the individual images. After the best aperture is defined, aperture corrections and flux calibrations are performed following \cite{Kuntschner2011}. For objects with multiple visits, the images have been median combined to produce a final spectroscopic image. Finally, spectra are extracted as indicated above. The published spectra are time-averaged spectra.     
	
	\subsection{Hot Jupiter emission spectra}
	
    {In this Section, we summarize the different reduction methods performed by the respective authors in which hot Jupiter's emission spectra were published.}
    
  \subsubsection{WASP-18b }
    
{The emission spectrum of WASP-18b  was presented in \cite{Sheppard2017}}. Observations of three individual eclipse events were obtained during three epochs as part of the program GO 13467.  At a forth epoch observations were obtained with two eclipses within an orbital phase curve. Grism observations were taken in spatial scan mode with forward-reverse cadence \citep{Dressel}. {Further details in the data reduction are found in \cite{Sheppard2017}}. Finally, a forward scan and a reverse scan light curve were obtained and analyzed separately. To correct non-astrophysical effects, the systematic trends were removed using parametric marginalization \citep{Wakeford2016}, and then further detrending was performed by the subtraction of scaled band-integrated residuals from wavelength bins \citep{Haynes2015}. The wavelength bins of the spectrum are given in Table 1 of \cite{Sheppard2017}.

    \subsubsection{WASP-33b }
    
    {The emission spectrum of WASP-33b was first published by \cite{Haynes2015}}. WASP-33b is orbiting a $\delta$-Scuti star \citep{Herrero2011}, and its modulations were model with sine functions. 
    {To produce the 2D spectral frames from the files provided by the standard pipeline}, a top hat mask was applied in the spatial direction of each read of a width of 20 pixels tall \citep{Herrero2011}; then, subsequent reads were subtracted and added to differenced frames to create one scanned image \citep{Deming2013}.  To correct bad pixels, the method by \cite{Mandell2013} was used within the combined spectral frames, and combine the images into 1D spectra. To perform the wavelength and flat-field wavelength dependent calibrations, the coefficients from \cite{Wilkins2014} were used. 
    
    \subsubsection{WASP-12b }
    
    {The emission spectrum of WASP-12b was published in \cite{Stevenson2014}}. The observations of WASP-12b were taken in five consecutive  in staring mode. Further details on the observations can be found in \cite{Swain2013}.
   Data were reduced using the standard HST pipeline as explained in detailed in \cite{Stevenson2014}.
    
    To trace the first order spectra, the direct image was located using a two-dimensional Gaussian and then use Table 1 in \cite{Kuntschner2009} to provide a direct-to-dispersed image offset. The wavelength calibration is performed using the coefficients provided in Table 5 from \cite{Kuntschner2009}. The flat field was modeled using the standard calibration flat files. The spectral extraction was performed within a box of 150$\times$150 pixels centered on the spectrum. The spectral extraction was performed along 40 pixels in the spatial direction, and the remaining pixels in the box were used for the background subtraction, generating eleven light curves.
    
     \subsubsection{WASP-121b}
    
    {The emission spectrum of WASP-121b was first presented in  \cite{Evans2017}}. 
    {Data reduction performed using the HST/WFC3 standard pipeline. Details explained in detailed in \cite{Evans2017}}. The target flux was extracted taking the difference between successive non-destructive reads. The background was measured as a median count of a box of 110 columns along the dispersion axis and 20 rows along the cross-dispersion axis. To remove flux contributions from nearby stars and cosmic ray hits, all pixels above and below 35 pixels from the center of the spectrum along the cross-dispersion axis were set to zero. Finally, all frames were added together. The spectrum was then extracted by summing the flux within a rectangular aperture across the dispersion axis with apertures from 100 to 200 pixels.
    The data taken during the first HST orbit were discarded due to a strong ramp effect, as well as the first exposure of the remaining HST orbits.
    
    \subsubsection{WASP-43b }
    
    {The emission spectrum of WASP-43b was first presented in  \cite{Stevenson2014_Nat}}. The observations were performed during 13-14 HST orbits on each primary transit or secondary eclipse visit, each of them consisting of four orbits.   Due to the ramp effect, the first orbit of each visit was removed from the analysis. For the rest, the ramp was fitted with an exponential ramp model. {Further details on the data reduction can be found in \cite{Stevenson2014_Nat}}.
        
    \subsubsection{WASP-103b }
    
    {The emission spectrum of WASP-103b was first presented in  \cite{Cartier2017}}. WASP-103b was observed with ten \textit{HST} orbits in two visits. The first orbit of both visits was  discarded.
    
    To remove the background, images were created using sequential pairs of up-the-ramp readouts within each exposure.  For those subframes, a conservative mask was used to determine the background region and measure the sky background level, assuming that is spatially flat and uniform due to the short exposure times. This background was subtracted from all subframes. In addition, a smaller mask was defined \citep{Deming2013, Knutson2014a} and  all pixels outside of the mask were zeroed. This helps to reduce noise and exclude cosmic rays (CRs) in the background area when later combining all subframes to determine the flux for each exposure.
Special flat fields were created for the data reduction using the determined centroids in the spectral direction (X) and scan direction (Y) direct image frame, assuming that every column has the same wavelength \citep{Cartier2017}. Finally, to remove additional cosmic rays and bad pixels, a moving median filter was applied. The final extracted spectrum was binned to 22 wavelength channels.
    
    Instrumental effects and systematics due to the ramp correction were removed using Gaussian Processes (GP) regression \citep{Rasmussen_Williams2006} that does not need to pre-specify a parametric model. To find the best fit light curve eclipse model  GP regression was used \citep{Cartier2017}.
    
    \subsubsection{TrES-3b and WASP-4b }
    
    {The emission spectra of TrES-3b and WASP-4b  were first presented in  \cite{Ranjan2014}}. The observations were carried out during four consecutive orbits during the eclipse of TrES-3b, and five consecutive orbits during the eclipse of WASP-4b. The first orbit of each observation was discarded to avoid the most prominent ramp effect systematics.
    
    {The details on the data reduction can be found in  \cite{Ranjan2014}}.  The background subtraction was performed by choosing a fixed area on the detector, matching the wavelength range of the spectrum free of object flux in the individual 2D images. These background columns are scaled to match the spectral extraction aperture.  Finally, the extracted spectra were binned in wavelength to enhance the signal-to-noise per resolution element (see the bin's wavelength ranges in \citealt{Ranjan2014} for each object). 
       
    \subsubsection{Kepler-13Ab }
    {{The emission spectra of Kepler-13Ab was first presented in  \cite{Beatty2017}}. The Kepler-13Ab system is composed by three stars: the planet host, Kepler-13A, and the unresolved binary Kepler-13BC, with the two components separated by 1."15 \citep{Shporer2014}. }
    
        {The observations were carried out during two visits were composed of a total of five HST orbits}. The planet host star is in a close binary system. The data reduction includes primary subtraction.
    All details of the data reduction can be found in \cite{Beatty2017}. The cosmic-rays hits were removed separately in an area around the stellar spectra, and the area dominated by the sky background. Finally, the background was subtracted from each exposure by defining two background regions across the bottom and top of each of these images.  
    
    {To perform a spectral extraction of Kepler-13Ab, the contribution of Kepler-13BC needs to be subtracted first. Using the WAYNE simulator \citep{Varley2017}, the artificial 2D spectra of Kepler-13BC was created and subtracted to create an undiluted 2D spectrum of Kepler-13A. Finally, to perform the light-curve extraction, the spectral trace of Kepler-13A was fitted with a Gaussian profile along the detector columns. Then the columns along the detector were summed using an extraction aperture with a half width of 4.5 pixels, to generate a 1D spectrum of Kepler-13A.}
    
    {The wavelength calibrations was done using the direct image taken at the beginning of each of the visits. The X- and Y- location of both Kepler-13A and Kepeler-13BC where determined on the detector subarray, and then \cite{Kuntschner2009} wavelength calibration method was implemented to calculate a wavelength solution for each star. The Paschen-$\beta$ line visible at 1.282~$\mu$m was used to verified the accuracy of the wavelength calibration.}

    \subsubsection{HD 209458B}
    
{    {The emission spectra of HD 209458B  was first presented in  \cite{Line2016}}. HD~209458b was observed as part of the GO 13467 HST treasury program. It was observed during secondary eclipse over five visits, each with five HST orbits. The first orbit of each visit were excluded from the analysis to minimize the impact of the ramp effect on the dataset. A direct image was taken at the beginning of each orbit to aid the wavelength calibration.}
   
  {All details about the data reduction can be found in \cite{Line2016}}. To extract the 1D spectra, and optimal extraction was used \citep{Horne1986} with a extraction window of 110 pixel rows centered on the spectra and flanked by additional 110 pixels rows for background extraction. The spectral of all frames were combined. Finally, the combined spectra was divided into 10 spectroscopic bins.

\section{Transformation between 2MASS to HST magnitudes}\label{transf_mag}

\begin{figure*}
\centering
\includegraphics[width=0.45\textwidth]{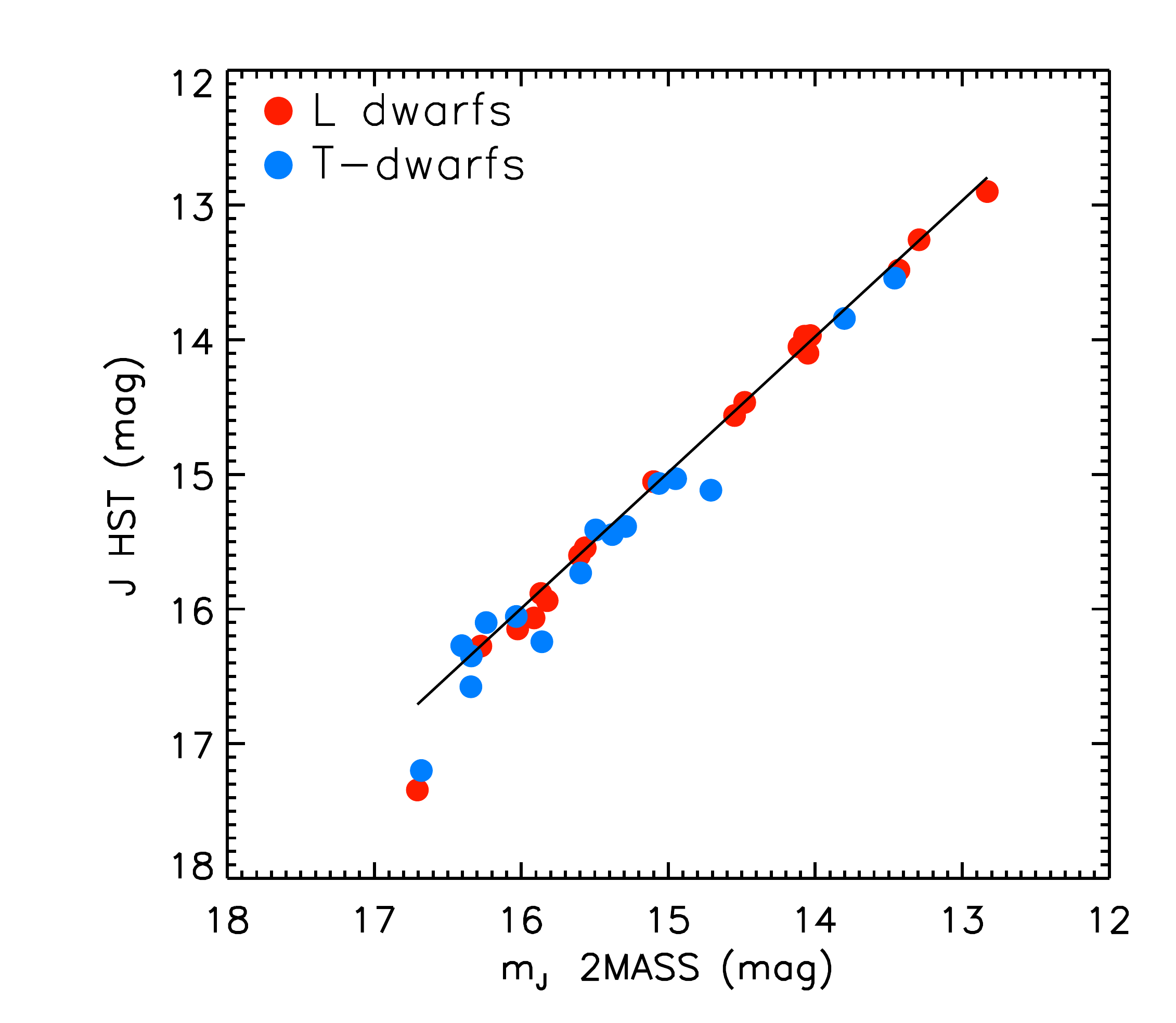}
\includegraphics[width=0.45\textwidth]{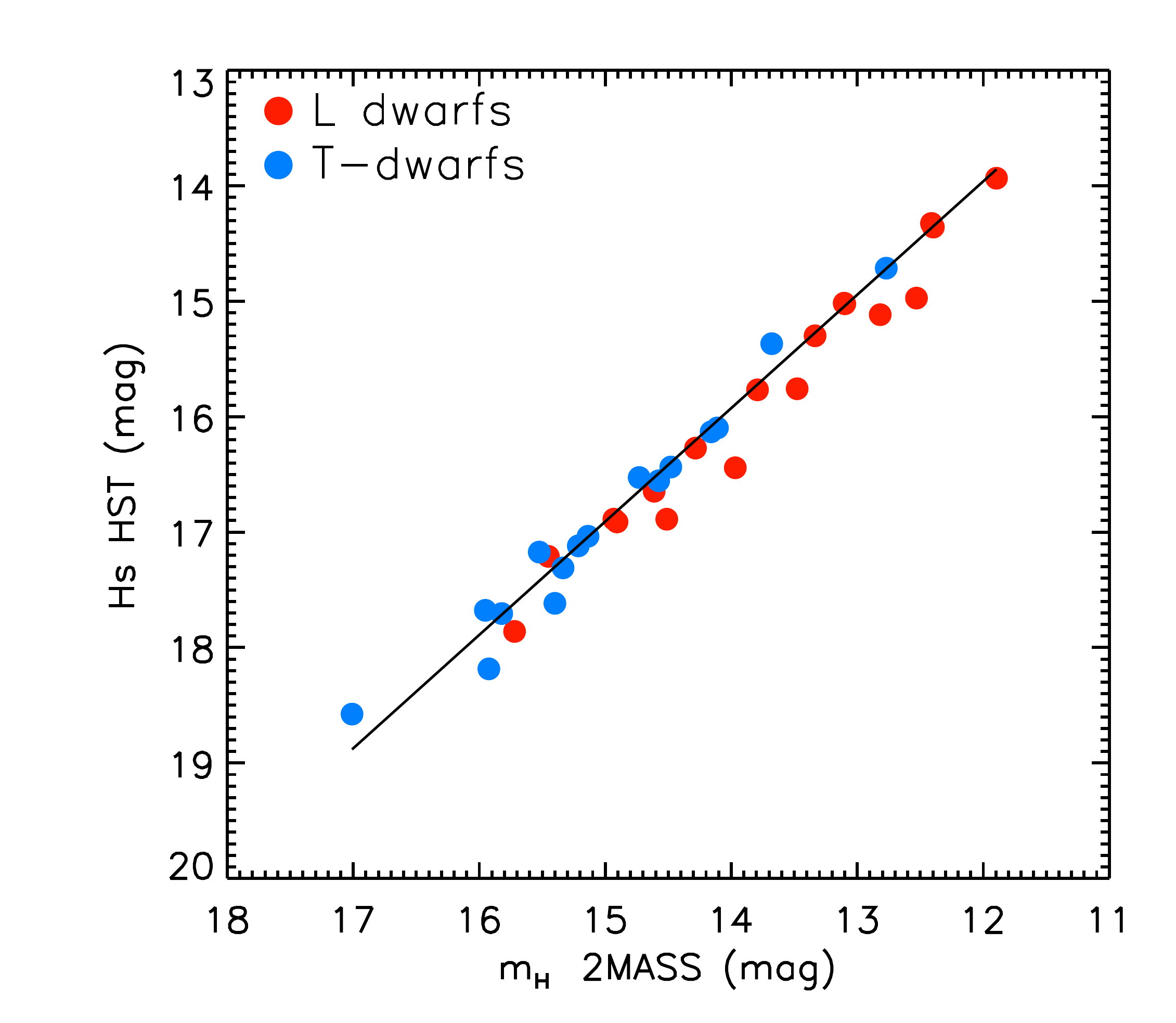}
\caption{Relation between $J$ and $H$ 2MASS magnitudes, and $J$ and $Hs$ HST magnitudes for L and T dwarfs. L dwarfs are shown as red points, and T dwarfs are shown as blue points. Photometric uncertainties are smaller than the symbols. \label{phot_rel}}
\end{figure*}

We derive empirical relations to transform L and T brown dwarf magnitudes from the 2MASS to the HST photometric system. To obtain the HST/WFC3 near-infrared magnitudes, we follow the same procedure than in Section~\ref{hot_jupiters}, equation~\ref{STmag}. In Figure \ref{phot_rel} we show the relation between the $J$ and $H$-band 2MASS magnitudes, and the $J$ and $H_{s}$-bands in the HST photometric system. We do not include the T9-T9.5 dwarfs with high phometric uncertainties. Finally, we calculate a linear relationship between both photometric systems for the $J$ and the $H$-band independently. The coefficients for both relations are presented in Table~\ref{photometric_trasf}.

\begin{table}
	\caption{Linear functions relating the $J$ and $H$-band 2MASS magnitudes and the $J$ and $Hs$ respectively for L and T brown dwarfs. \label{photometric_trasf}}  
	\centering
	\begin{center}
		\begin{tabular}{llll}
			\hline 
			
			 &    & \multicolumn{2}{l}{Linear fit} \\		
			x & y 	        & $c_{0}$ & $c_{1}$   \\
			\hline              		
  		$J$ 2MASS &$J$ HST & -0.15535$\pm$0.00254   & 1.00938$\pm$0.00017 \\  
        $H$ 2MASS  & $Hs$ HST	& 2.17362$\pm$0.00206 &  0.98229$\pm$0.00014 \\
			\hline		
		\end{tabular}
	\end{center}
		\tablecomments{ The linear function is defined as: $ y = c_{0} + c_{1} x  $}
\end{table}

			\begin{table*}
				\caption{Spectral indices to select L plus T brown dwarf binary candidates. \label{spectral_indices}}
				\centering
				\small
				
				\begin{center}
					\begin{tabular}{l l l l l}
						\hline
						\hline
						Index & Numerator Range$^{a}$ & Denominator Range$^{a}$& Feature & Reference \\
						
						\hline
						
						$\mathrm{H_{2}O}$-J & 1.140-1.165 & 1.260-1.285 & 1.150~$\mu$m $\mathrm{H_{2}O}$ &1 \\
						$\mathrm{CH_{4}}$-J & 1.315-1340 & 1.260-1285 & 1.320~$\mu$m $\mathrm{CH_{4}}$ &1 \\
						
						$\mathrm{H_{2}O}$-H & 1.480-1.520 & 1.560-1.600 & 1.400~$\mu$m $\mathrm{H_{2}O}$ &1 \\
						$\mathrm{CH_{4}}$-H & 1.635-1.675 & 1.560-1.600 & 1.650~$\mu$m $\mathrm{CH_{4}}$ &1 \\
						
						
						$H$-dip      & 1.610-1.640 & 1.560-1.590 + 1.660-1.690$^{b}$ &1.650~$\mu$m $\mathrm{CH_{4}}$& 2 \\
						
						$J$-slope    & 1.27-1.30  & 1.30-1.33  & 1.28~$\mu$m flux peak shape & 4 \\
						
						$J$-curve    & 1.04-1.07+1.26-1.29$^{c}$ & 1.14-1.17 & Curvature across J-band & 4\\
						$H$-bump     & 1.54-1.57  & 1.66-1.69  & Slope across H-band peak & 4\\
						
						Derived NIR SpT  &             &            & near-infrared spectral type$^{d}$&1\\
						
						\hline
					\end{tabular}
				\tablecomments{a: Wavelength range in nm over which flux density is integrated; b: denominator is the sum of the flux in the two wavelength ranges; c: numerator is the sum of the two ranges; d: near-infrared  spectral type derived using comparison to SpeX spectra.
References: 1 -- \cite{Burgasser2006}; 2 -- \cite{Burgasser2010}; 3 -- \cite{Burgasser2002}; 4 --\cite{Bardalez_Gagliuffi}.}
				\end{center}

									
			\end{table*}
			
			
			\begin{table*}
				\caption{Index criteria for the selection of potential brown dwarf binary systems \label{criteria}}
				\centering
				\begin{center}
					\begin{tabular}{l l l }
						\hline
						\hline
						Abscissa &Ordinate & Inflection Points \\
						
						\hline
						
						$\mathrm{H_{2}O}$-H & H-dip               & (0.5,0.49),(0.875,0.49) \\
						Spex SpT & $\mathrm{H_{2}O}$-J/$\mathrm{H_{2}O}$-H &  (L8.5,0.925),(T1.5,0.925),(T3,0.85) \\
						
						\hline
					\end{tabular}
				\end{center}
				
			\end{table*}

			\begin{table*}
				\caption{Delimiters for selection regions of potential brown dwarf binary systems\label{criteria2}}
				\centering
				
				\begin{center}
					\begin{tabular}{l l l}
						\hline
						\hline
						Abscissa & Ordinate & Limits \\
						
						\hline
						
						SpT & $\mathrm{CH_{4}}$-H &  Best fit curve: y = $-4.3x 10^{-4}x^2$+0.0253x + 0.7178       \\
						$\mathrm{H_{2}O}$-J & $\mathrm{CH_{4}}$-H & Intersection of: -0.08x+1.09 and x = 0.90         \\
						$\mathrm{H_{2}O}$-J & H-bump & Intersection of: y = 0.16x+0.806 and x = 0.90              \\
						$\mathrm{CH_{4}}$-J  & $\mathrm{CH_{4}}$-H & Intersection of: y = -0.56x + 1.41 and y = 1.04      \\
						$\mathrm{CH_{4}}$-J & H-bump & Intersection of: y = 1.00x + 0.24, x = 0.74 and y = 0.91 \\
						$\mathrm{CH_{4}}$-H & J-slope & Intersection of: y = 1.250x -0.207, x = 1.03 and y = 1.03 \\
						$\mathrm{CH_{4}}$-H & J-curve & Best fit curve: y = 1.245$x^{2}$ - 1.565x + 2.224\\
						$\mathrm{CH_{4}}$-H  & H-bump & Best fit curve: y = 1.36$x^{2}$ - 4.26x + 3.89 \\
						J-slope & H-dip & Intersection of y = 0.20x + 0.27 and x = 1.03 \\
						J-slope & H-bump & Intersection of: y = -2.75x + 3.84 and y = 0.91 \\
						J-curve & H-bump & Best fit curve: y = 0.269$x^{2}$ - 1.326 + 2.479 \\

						\hline
					\end{tabular}
				\end{center}
				
			\end{table*}

\begin{figure*}
\centering
\includegraphics[width=0.45\textwidth]{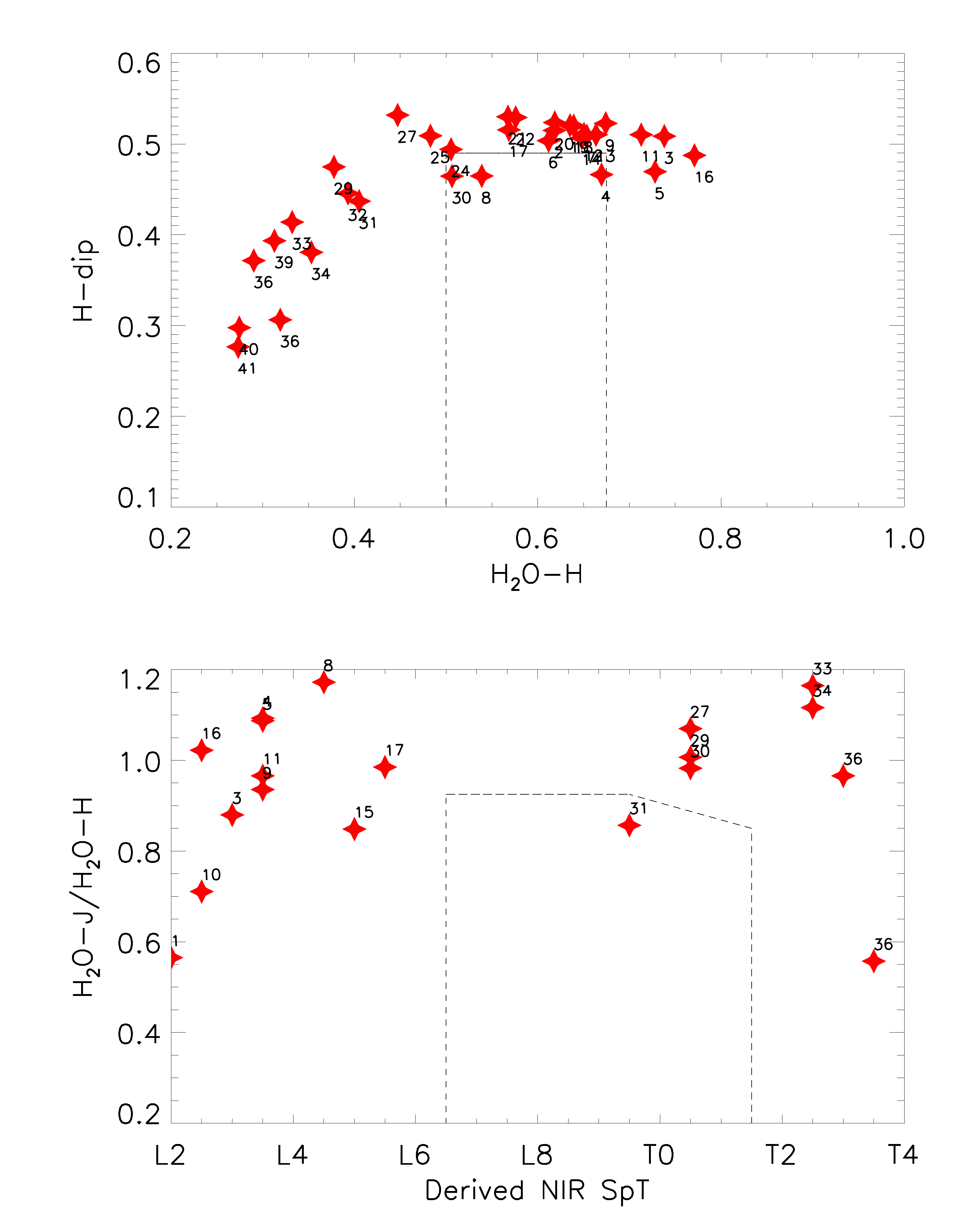}
\caption{Spectral index selection. Numbers correspond to our objects. The boxes shown with dashed lines mark the areas where the selection criteria of Table \ref{criteria} and \ref{criteria2} are valid. \label{index1}}
\end{figure*}

\begin{figure*}
\centering
\includegraphics[width=0.9\textwidth]{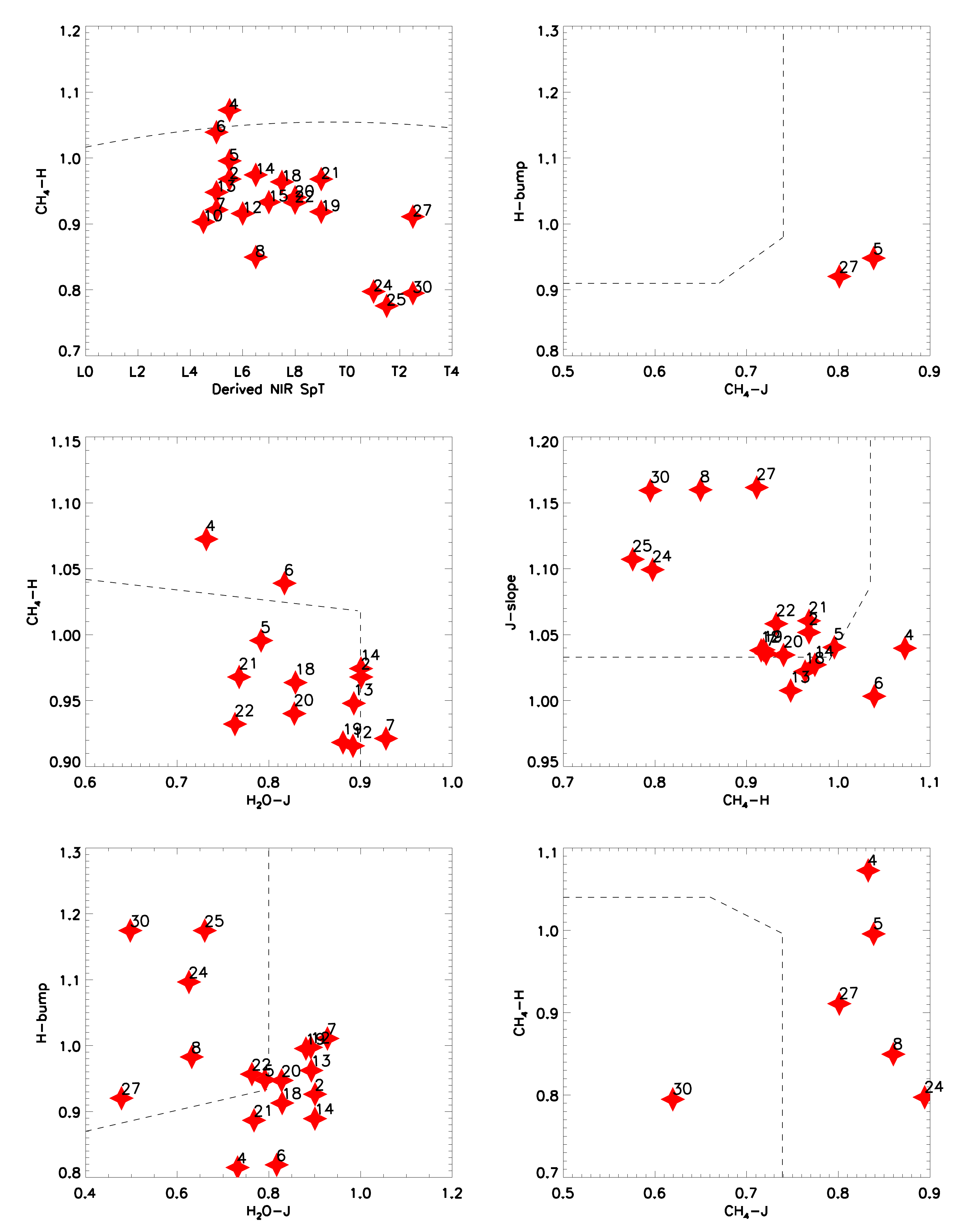}
\caption{Spectral index selection. Numbers correspond to our objects. The boxes shown with dashed lines mark the areas where the selection criteria of Table \ref{criteria} and \ref{criteria2} are valid.\label{index2}}
\end{figure*}

\begin{figure*}
\centering
\includegraphics[width=0.9\textwidth]{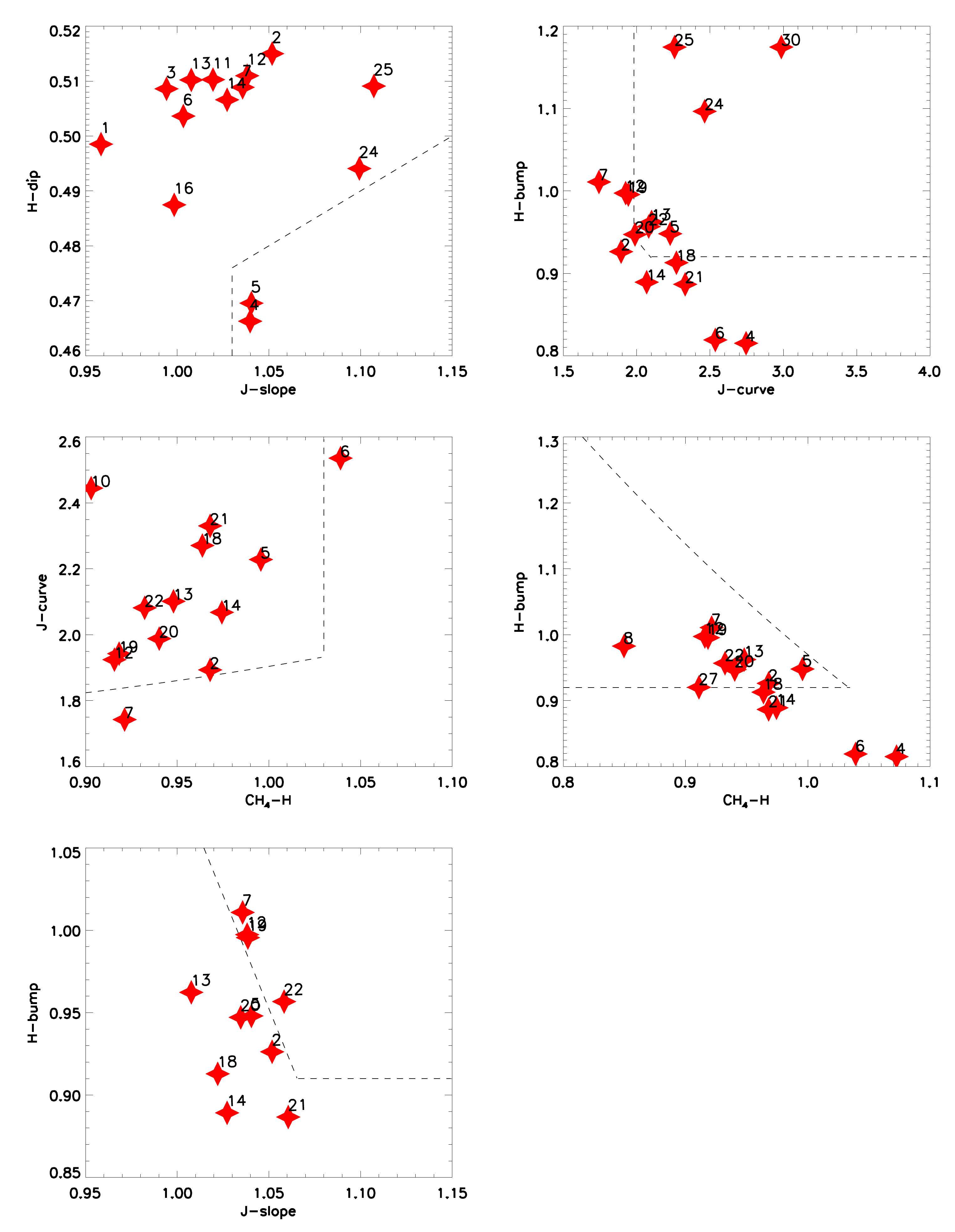}
\caption{Spectral index selection. Numbers correspond to our objects. The boxes shown with dashed lines mark the areas where the selection criteria of Table \ref{criteria} and \ref{criteria2} are valid.\label{index3}}
\end{figure*}

\bibliographystyle{apj}
\bibliography{HST_library}

\begin{thebibliography}{}
\expandafter\ifx\csname natexlab\endcsname\relax\def\natexlab#1{#1}\fi

\bibitem[{{Allard} {et~al.}(2012){Allard}, {Homeier}, \&
  {Freytag}}]{Allard2012a}
{Allard}, F., {Homeier}, D., \& {Freytag}, B. 2012, Royal Society of London
  Philosophical Transactions Series A, 370, 2765

\bibitem[{{Aller} {et~al.}(2016){Aller}, {Liu}, {Magnier}, {Best}, {Kotson},
  {Burgett}, {Chambers}, {Hodapp}, {Flewelling}, {Kaiser}, {Metcalf}, {Tonry},
  {Wainscoat}, \& {Waters}}]{Aller2016}
{Aller}, K.~M., {Liu}, M.~C., {Magnier}, E.~A., {et~al.} 2016, \apj, 821,
  doi:10.3847/0004-637X/821/2/120

\bibitem[{{Allers} {et~al.}(2016){Allers}, {Gallimore}, {Liu}, \&
  {Dupuy}}]{Allers2016}
{Allers}, K.~N., {Gallimore}, J.~F., {Liu}, M.~C., \& {Dupuy}, T.~J. 2016,
  \apj, 819, 133

\bibitem[{{Allers} \& {Liu}(2013)}]{Allers2013}
{Allers}, K.~N., \& {Liu}, M.~C. 2013, \apj, 772, 79

\bibitem[{{Andrei} {et~al.}(2011){Andrei}, {Smart}, {Penna}, {d'Avila},
  {Bucciarelli}, {Camargo}, {Crosta}, {Dapr{\`a}}, {Goldman}, {Jones},
  {Lattanzi}, {Nicastro}, {Pinfield}, {da Silva Neto}, \&
  {Teixeira}}]{Andrei2011}
{Andrei}, A.~H., {Smart}, R.~L., {Penna}, J.~L., {et~al.} 2011, \aj, 141, 54

\bibitem[{{Apai} {et~al.}(2013){Apai}, {Radigan}, {Buenzli}, {Burrows}, {Reid},
  \& {Jayawardhana}}]{Apai2013}
{Apai}, D., {Radigan}, J., {Buenzli}, E., {et~al.} 2013, \apj, 768, 121

\bibitem[{{Apai} {et~al.}(2017){Apai}, {Karalidi}, {Marley}, {Yang}, {Flateau},
  {Metchev}, {Cowan}, {Buenzli}, {Burgasser}, {Radigan}, {Artigau}, \&
  {Lowrance}}]{Apai2017}
{Apai}, D., {Karalidi}, T., {Marley}, M.~S., {et~al.} 2017, Science, 357, 683

\bibitem[{{Artigau} {et~al.}(2009){Artigau}, {Bouchard}, {Doyon}, \&
  {Lafreni{\`e}re}}]{Artigau2009}
{Artigau}, {\'E}., {Bouchard}, S., {Doyon}, R., \& {Lafreni{\`e}re}, D. 2009,
  \apj, 701, 1534

\bibitem[{{Bannister} \& {Jameson}(2007)}]{Bannister2007}
{Bannister}, N.~P., \& {Jameson}, R.~F. 2007, \mnras, 378, L24

\bibitem[{{Bardalez Gagliuffi} {et~al.}(2014){Bardalez Gagliuffi}, {Burgasser},
  {Gelino}, {Looper}, {Nicholls}, {Schmidt}, {Cruz}, {West}, {Gizis}, \&
  {Metchev}}]{Bardalez_Gagliuffi}
{Bardalez Gagliuffi}, D.~C., {Burgasser}, A.~J., {Gelino}, C.~R., {et~al.}
  2014, \apj, 794, 143

\bibitem[{{Beatty} {et~al.}(2017){Beatty}, {Madhusudhan}, {Tsiaras}, {Zhao},
  {Gilliland}, {Knutson}, {Shporer}, \& {Wright}}]{Beatty2017}
{Beatty}, T.~G., {Madhusudhan}, N., {Tsiaras}, A., {et~al.} 2017, \aj, 154, 158

\bibitem[{{Bedin} {et~al.}(2017){Bedin}, {Pourbaix}, {Apai}, {Burgasser},
  {Buenzli}, {Boffin}, \& {Libralato}}]{Bedin2017}
{Bedin}, L.~R., {Pourbaix}, D., {Apai}, D., {et~al.} 2017, \mnras, 470, 1140

\bibitem[{{Benneke} \& {Seager}(2012)}]{Benneke2012}
{Benneke}, B., \& {Seager}, S. 2012, \apj, 753, 100

\bibitem[{{Biller} {et~al.}(2015){Biller}, {Vos}, {Bonavita}, {Buenzli},
  {Baxter}, {Crossfield}, {Allers}, {Liu}, {Bonnefoy}, {Deacon}, {Brandner},
  {Schlieder}, {Dupuy}, {Kopytova}, {Manjavacas}, {Allard}, {Homeier}, \&
  {Henning}}]{Biller2015}
{Biller}, B.~A., {Vos}, J., {Bonavita}, M., {et~al.} 2015, \apjl, 813, L23

\bibitem[{{Biller} {et~al.}(2018){Biller}, {Vos}, {Buenzli}, {Allers},
  {Bonnefoy}, {Charnay}, {B{\'e}zard}, {Allard}, {Homeier}, {Bonavita},
  {Brandner}, {Crossfield}, {Dupuy}, {Henning}, {Kopytova}, {Liu},
  {Manjavacas}, \& {Schlieder}}]{Biller2018}
{Biller}, B.~A., {Vos}, J., {Buenzli}, E., {et~al.} 2018, \aj, 155, 95

\bibitem[{{Bonnefoy} {et~al.}(2014){Bonnefoy}, {Chauvin}, {Lagrange}, {Rojo},
  {Allard}, {Pinte}, {Dumas}, \& {Homeier}}]{Bonnefoy2014a}
{Bonnefoy}, M., {Chauvin}, G., {Lagrange}, A.-M., {et~al.} 2014, \aap, 562,
  A127

\bibitem[{{Borysow} {et~al.}(1997){Borysow}, {Jorgensen}, \&
  {Zheng}}]{Borysow1997}
{Borysow}, A., {Jorgensen}, U.~G., \& {Zheng}, C. 1997, \aap, 324, 185

\bibitem[{{Buenzli} {et~al.}(2014){Buenzli}, {Apai}, {Radigan}, {Reid}, \&
  {Flateau}}]{Buenzli2014}
{Buenzli}, E., {Apai}, D., {Radigan}, J., {Reid}, I.~N., \& {Flateau}, D. 2014,
  \apj, 782, 77

\bibitem[{{Buenzli} {et~al.}(2015){Buenzli}, {Saumon}, {Marley}, {Apai},
  {Radigan}, {Bedin}, {Reid}, \& {Morley}}]{Buenzli2015}
{Buenzli}, E., {Saumon}, D., {Marley}, M.~S., {et~al.} 2015, \apj, 798, 127

\bibitem[{{Buenzli} {et~al.}(2012){Buenzli}, {Apai}, {Morley}, {Flateau},
  {Showman}, {Burrows}, {Marley}, {Lewis}, \& {Reid}}]{Buenzli2012}
{Buenzli}, E., {Apai}, D., {Morley}, C.~V., {et~al.} 2012, \apjl, 760, L31

\bibitem[{{Burgasser} {et~al.}(2010){Burgasser}, {Cruz}, {Cushing}, {Gelino},
  {Looper}, {Faherty}, {Kirkpatrick}, \& {Reid}}]{Burgasser2010}
{Burgasser}, A.~J., {Cruz}, K.~L., {Cushing}, M., {et~al.} 2010, \apj, 710,
  1142

\bibitem[{{Burgasser} {et~al.}(2006{\natexlab{a}}){Burgasser}, {Geballe},
  {Leggett}, {Kirkpatrick}, \& {Golimowski}}]{2006ApJ...637.1067B}
{Burgasser}, A.~J., {Geballe}, T.~R., {Leggett}, S.~K., {Kirkpatrick}, J.~D.,
  \& {Golimowski}, D.~A. 2006{\natexlab{a}}, \apj, 637, 1067

\bibitem[{{Burgasser} {et~al.}(2006{\natexlab{b}}){Burgasser}, {Geballe},
  {Leggett}, {Kirkpatrick}, \& {Golimowski}}]{Burgasser2006}
---. 2006{\natexlab{b}}, \apj, 637, 1067

\bibitem[{{Burgasser} {et~al.}(2008){Burgasser}, {Looper}, {Kirkpatrick},
  {Cruz}, \& {Swift}}]{Burgasser2008}
{Burgasser}, A.~J., {Looper}, D.~L., {Kirkpatrick}, J.~D., {Cruz}, K.~L., \&
  {Swift}, B.~J. 2008, \apj, 674, 451

\bibitem[{{Burgasser} {et~al.}(2002){Burgasser}, {Kirkpatrick}, {Brown},
  {Reid}, {Burrows}, {Liebert}, {Matthews}, {Gizis}, {Dahn}, {Monet}, {Cutri},
  \& {Skrutskie}}]{Burgasser2002}
{Burgasser}, A.~J., {Kirkpatrick}, J.~D., {Brown}, M.~E., {et~al.} 2002, \apj,
  564, 421

\bibitem[{{Burningham} {et~al.}(2011){Burningham}, {Leggett}, {Homeier},
  {Saumon}, {Lucas}, {Pinfield}, {Tinney}, {Allard}, {Marley}, {Jones},
  {Murray}, {Ishii}, {Day-Jones}, {Gomes}, \& {Zhang}}]{Burningham2011}
{Burningham}, B., {Leggett}, S.~K., {Homeier}, D., {et~al.} 2011, \mnras, 414,
  3590

\bibitem[{{Cartier} {et~al.}(2017){Cartier}, {Beatty}, {Zhao}, {Line}, {Ngo},
  {Mawet}, {Stassun}, {Wright}, {Kreidberg}, {Fortney}, \&
  {Knutson}}]{Cartier2017}
{Cartier}, K.~M.~S., {Beatty}, T.~G., {Zhao}, M., {et~al.} 2017, \aj, 153, 34

\bibitem[{{Cruz} {et~al.}(2009){Cruz}, {Kirkpatrick}, \& {Burgasser}}]{Cruz}
{Cruz}, K.~L., {Kirkpatrick}, J.~D., \& {Burgasser}, A.~J. 2009, \aj, 137, 3345

\bibitem[{{Cushing} {et~al.}(2005){Cushing}, {Rayner}, \&
  {Vacca}}]{Cushing2005}
{Cushing}, M.~C., {Rayner}, J.~T., \& {Vacca}, W.~D. 2005, \apj, 623, 1115

\bibitem[{{Cushing} {et~al.}(2008){Cushing}, {Marley}, {Saumon}, {Kelly},
  {Vacca}, {Rayner}, {Freedman}, {Lodders}, \& {Roellig}}]{Cushing2008}
{Cushing}, M.~C., {Marley}, M.~S., {Saumon}, D., {et~al.} 2008, \apj, 678, 1372

\bibitem[{{Cutri} {et~al.}(2003){Cutri}, {Skrutskie}, {van Dyk}, {Beichman},
  {Carpenter}, {Chester}, {Cambresy}, {Evans}, {Fowler}, {Gizis}, {Howard},
  {Huchra}, {Jarrett}, {Kopan}, {Kirkpatrick}, {Light}, {Marsh}, {McCallon},
  {Schneider}, {Stiening}, {Sykes}, {Weinberg}, {Wheaton}, {Wheelock}, \&
  {Zacarias}}]{Cutri2003}
{Cutri}, R.~M., {Skrutskie}, M.~F., {van Dyk}, S., {et~al.} 2003, VizieR Online
  Data Catalog, 2246, 0

\bibitem[{{Dahn} {et~al.}(2002){Dahn}, {Harris}, {Vrba}, {Guetter}, {Canzian},
  {Henden}, {Levine}, {Luginbuhl}, {Monet}, {Monet}, {Pier}, {Stone}, {Walker},
  {Burgasser}, {Gizis}, {Kirkpatrick}, {Liebert}, \& {Reid}}]{Dahn2002}
{Dahn}, C.~C., {Harris}, H.~C., {Vrba}, F.~J., {et~al.} 2002, \aj, 124, 1170

\bibitem[{{Delorme} {et~al.}(2017){Delorme}, {Dupuy}, {Gagn{\'e}}, {Reyl{\'e}},
  {Forveille}, {Liu}, {Artigau}, {Albert}, {Delfosse}, {Allard}, {Homeier},
  {Malo}, {Morley}, {Naud}, \& {Bonnefoy}}]{Delorme2017}
{Delorme}, P., {Dupuy}, T., {Gagn{\'e}}, J., {et~al.} 2017, \aap, 602,
  doi:10.1051/0004-6361/201629633

\bibitem[{{Deming} {et~al.}(2013){Deming}, {Wilkins}, {McCullough}, {Burrows},
  {Fortney}, {Agol}, {Dobbs-Dixon}, {Madhusudhan}, {Crouzet}, {Desert},
  {Gilliland}, {Haynes}, {Knutson}, {Line}, {Magic}, {Mandell}, {Ranjan},
  {Charbonneau}, {Clampin}, {Seager}, \& {Showman}}]{Deming2013}
{Deming}, D., {Wilkins}, A., {McCullough}, P., {et~al.} 2013, \apj, 774, 95

\bibitem[{{Dressel}(2018)}]{Dressel}
{Dressel}, L. 2018, Baltimore: STScI

\bibitem[{{Dupuy} \& {Liu}(2012)}]{Dupuy_Liu2012}
{Dupuy}, T.~J., \& {Liu}, M.~C. 2012, \apjs, 201, 19

\bibitem[{{Evans} {et~al.}(2017){Evans}, {Sing}, {Kataria}, {Goyal}, {Nikolov},
  {Wakeford}, {Deming}, {Marley}, {Amundsen}, {Ballester}, {Barstow},
  {Ben-Jaffel}, {Bourrier}, {Buchhave}, {Cohen}, {Ehrenreich}, {Garc{\'{\i}}a
  Mu{\~n}oz}, {Henry}, {Knutson}, {Lavvas}, {Lecavelier Des Etangs}, {Lewis},
  {L{\'o}pez-Morales}, {Mandell}, {Sanz-Forcada}, {Tremblin}, \&
  {Lupu}}]{Evans2017}
{Evans}, T.~M., {Sing}, D.~K., {Kataria}, T., {et~al.} 2017, \nat, 548, 58

\bibitem[{{Faherty} {et~al.}(2013){Faherty}, {Rice}, {Cruz}, {Mamajek}, \&
  {N{\'u}{\~n}ez}}]{Faherty2013}
{Faherty}, J.~K., {Rice}, E.~L., {Cruz}, K.~L., {Mamajek}, E.~E., \&
  {N{\'u}{\~n}ez}, A. 2013, \aj, 145, 2

\bibitem[{{Faherty} {et~al.}(2012){Faherty}, {Burgasser}, {Walter}, {Van der
  Bliek}, {Shara}, {Cruz}, {West}, {Vrba}, \&
  {Anglada-Escud{\'e}}}]{Faherty2012}
{Faherty}, J.~K., {Burgasser}, A.~J., {Walter}, F.~M., {et~al.} 2012, \apj,
  752, 56

\bibitem[{{Fisher} \& {Heng}(2018)}]{FisherHeng2018}
{Fisher}, C., \& {Heng}, K. 2018, ArXiv e-prints, arXiv:1809.06894

\bibitem[{{Fortney} {et~al.}(2008){Fortney}, {Lodders}, {Marley}, \&
  {Freedman}}]{Fortney2008}
{Fortney}, J.~J., {Lodders}, K., {Marley}, M.~S., \& {Freedman}, R.~S. 2008,
  \apj, 678, 1419

\bibitem[{{Gagn{\'e}} {et~al.}(2015{\natexlab{a}}){Gagn{\'e}}, {Burgasser},
  {Faherty}, {Lafreni{\'e}re}, {Doyon}, {Filippazzo}, {Bowsher}, \&
  {Nicholls}}]{Gagne2015_s1110}
{Gagn{\'e}}, J., {Burgasser}, A.~J., {Faherty}, J.~K., {et~al.}
  2015{\natexlab{a}}, \apjl, 808, L20

\bibitem[{{Gagn{\'e}} {et~al.}(2015{\natexlab{b}}){Gagn{\'e}}, {Faherty},
  {Cruz}, {Lafreni{\'e}re}, {Doyon}, {Malo}, {Burgasser}, {Naud}, {Artigau},
  {Bouchard}, {Gizis}, \& {Albert}}]{Gagne2015}
{Gagn{\'e}}, J., {Faherty}, J.~K., {Cruz}, K.~L., {et~al.} 2015{\natexlab{b}},
  \apjs, 219, 33

\bibitem[{{Gagn{\'e}} {et~al.}(2017){Gagn{\'e}}, {Faherty}, {Burgasser},
  {Artigau}, {Bouchard}, {Albert}, {Lafreni{\`e}re}, {Doyon}, \& {Bardalez
  Gagliuffi}}]{Gagne2017}
{Gagn{\'e}}, J., {Faherty}, J.~K., {Burgasser}, A.~J., {et~al.} 2017, \apj,
  841, doi:10.3847/2041-8213/aa70e2

\bibitem[{{Gaia Collaboration} {et~al.}(2018){Gaia Collaboration}, {Brown},
  {Vallenari}, {Prusti}, {de Bruijne}, {Babusiaux}, \&
  {Bailer-Jones}}]{Gaia2018}
{Gaia Collaboration}, {Brown}, A.~G.~A., {Vallenari}, A., {et~al.} 2018, ArXiv
  e-prints, arXiv:1804.09365

\bibitem[{{Gandhi} \& {Madhusudhan}(2017)}]{Gandhi2017}
{Gandhi}, S., \& {Madhusudhan}, N. 2017, \mnras, 472, 2334

\bibitem[{{Gizis} {et~al.}(2015){Gizis}, {Allers}, {Liu}, {Harris}, {Faherty},
  {Burgasser}, \& {Kirkpatrick}}]{Gizis2015}
{Gizis}, J.~E., {Allers}, K.~N., {Liu}, M.~C., {et~al.} 2015, \apj, 799, 203

\bibitem[{{Gizis} {et~al.}(2012){Gizis}, {Faherty}, {Liu}, {Castro}, {Shaw},
  {Vrba}, {Harris}, {Aller}, \& {Deacon}}]{Gizis2012}
{Gizis}, J.~E., {Faherty}, J.~K., {Liu}, M.~C., {et~al.} 2012, \aj, 144, 94

\bibitem[{{Hartig}(2009)}]{Hartig2009}
{Hartig}, G.~F. 2009, {WFC3 SMOV Programs 11437/9: IR On-orbit PSF Evaluation},
  Tech. rep.

\bibitem[{{Haynes} {et~al.}(2015){Haynes}, {Mandell}, {Madhusudhan}, {Deming},
  \& {Knutson}}]{Haynes2015}
{Haynes}, K., {Mandell}, A.~M., {Madhusudhan}, N., {Deming}, D., \& {Knutson},
  H. 2015, \apj, 806, 146

\bibitem[{{Herrero} {et~al.}(2011){Herrero}, {Morales}, {Ribas}, \&
  {Naves}}]{Herrero2011}
{Herrero}, E., {Morales}, J.~C., {Ribas}, I., \& {Naves}, R. 2011, \aap, 526,
  L10

\bibitem[{{Horne}(1986)}]{Horne1986}
{Horne}, K. 1986, \pasp, 98, 609

\bibitem[{Hunter(2007)}]{Hunter_2007}
Hunter, J.~D. 2007, Computing In Science \& Engineering, 9, 90

\bibitem[{{Kirkpatrick}(2005)}]{Kirkpatrick}
{Kirkpatrick}, J.~D. 2005, \araa, 43, 195

\bibitem[{{Kirkpatrick} {et~al.}(1999){Kirkpatrick}, {Reid}, {Liebert},
  {Cutri}, {Nelson}, {Beichman}, {Dahn}, {Monet}, {Gizis}, \&
  {Skrutskie}}]{Kirkpatrick1999}
{Kirkpatrick}, J.~D., {Reid}, I.~N., {Liebert}, J., {et~al.} 1999, \apj, 519,
  802

\bibitem[{{Kirkpatrick} {et~al.}(2000){Kirkpatrick}, {Reid}, {Liebert},
  {Gizis}, {Burgasser}, {Monet}, {Dahn}, {Nelson}, \&
  {Williams}}]{Kirkpatrick_2000}
---. 2000, \aj, 120, 447

\bibitem[{{Kirkpatrick} {et~al.}(2011){Kirkpatrick}, {Cushing}, {Gelino},
  {Griffith}, {Skrutskie}, {Marsh}, {Wright}, {Mainzer}, {Eisenhardt},
  {McLean}, {Thompson}, {Bauer}, {Benford}, {Bridge}, {Lake}, {Petty},
  {Stanford}, {Tsai}, {Bailey}, {Beichman}, {Bloom}, {Bochanski}, {Burgasser},
  {Capak}, {Cruz}, {Hinz}, {Kartaltepe}, {Knox}, {Manohar}, {Masters},
  {Morales-Calder{\'o}n}, {Prato}, {Rodigas}, {Salvato}, {Schurr}, {Scoville},
  {Simcoe}, {Stapelfeldt}, {Stern}, {Stock}, \& {Vacca}}]{Kirkpatrick2011}
{Kirkpatrick}, J.~D., {Cushing}, M.~C., {Gelino}, C.~R., {et~al.} 2011, \apjs,
  197, 19

\bibitem[{{Knutson} {et~al.}(2014){Knutson}, {Benneke}, {Deming}, \&
  {Homeier}}]{Knutson2014a}
{Knutson}, H.~A., {Benneke}, B., {Deming}, D., \& {Homeier}, D. 2014, \nat,
  505, 66

\bibitem[{{Kuntschner} {et~al.}(2009){Kuntschner}, {Kuemmel}, {Walsh}, \&
  {Bushouse}}]{Kuntschner2009}
{Kuntschner}, H., {Kuemmel}, M., {Walsh}, J., \& {Bushouse}, H. 2009, Space
  Telescope European Coordinating Facility Newsletter, 47, 4

\bibitem[{{Kuntschner} {et~al.}(2011){Kuntschner}, {K{\"u}mmel}, {Walsh}, \&
  {Bushouse}}]{Kuntschner2011}
{Kuntschner}, H., {K{\"u}mmel}, M., {Walsh}, J.~R., \& {Bushouse}, H. 2011,
  {Revised Flux Calibration of the WFC3 G102 and G141 grisms}, Tech. rep.

\bibitem[{{Lavie} {et~al.}(2017){Lavie}, {Mendon{\c c}a}, {Mordasini}, {Malik},
  {Bonnefoy}, {Demory}, {Oreshenko}, {Grimm}, {Ehrenreich}, \&
  {Heng}}]{Lavie2017}
{Lavie}, B., {Mendon{\c c}a}, J.~M., {Mordasini}, C., {et~al.} 2017, \aj, 154,
  91

\bibitem[{{Lee} {et~al.}(2014){Lee}, {Irwin}, {Fletcher}, {Heng}, \&
  {Barstow}}]{Lee2014}
{Lee}, J.-M., {Irwin}, P.~G.~J., {Fletcher}, L.~N., {Heng}, K., \& {Barstow},
  J.~K. 2014, \apj, 789, 14

\bibitem[{{Leggett} {et~al.}(2017){Leggett}, {Tremblin}, {Esplin}, {Luhman}, \&
  {Morley}}]{Leggett2017}
{Leggett}, S.~K., {Tremblin}, P., {Esplin}, T.~L., {Luhman}, K.~L., \&
  {Morley}, C.~V. 2017, \apj, 842, 118

\bibitem[{{Leggett} {et~al.}(2000){Leggett}, {Geballe}, {Fan}, {Schneider},
  {Gunn}, {Lupton}, {Knapp}, {Strauss}, {McDaniel}, {Golimowski}, {Henry},
  {Peng}, {Tsvetanov}, {Uomoto}, {Zheng}, {Hill}, {Ramsey}, {Anderson},
  {Annis}, {Bahcall}, {Brinkmann}, {Chen}, {Csabai}, {Fukugita}, {Hennessy},
  {Hindsley}, {Ivezi{\'c}}, {Lamb}, {Munn}, {Pier}, {Schlegel}, {Smith},
  {Stoughton}, {Thakar}, \& {York}}]{Leggett2000}
{Leggett}, S.~K., {Geballe}, T.~R., {Fan}, X., {et~al.} 2000, \apjl, 536, L35

\bibitem[{{Lew} {et~al.}(2016){Lew}, {Apai}, {Zhou}, {Schneider}, {Burgasser},
  {Karalidi}, {Yang}, {Marley}, {Cowan}, {Bedin}, {Metchev}, {Radigan}, \&
  {Lowrance}}]{Lew2016}
{Lew}, B.~W.~P., {Apai}, D., {Zhou}, Y., {et~al.} 2016, \apjl, 829, L32

\bibitem[{{Line} {et~al.}(2015){Line}, {Teske}, {Burningham}, {Fortney}, \&
  {Marley}}]{Line2015}
{Line}, M.~R., {Teske}, J., {Burningham}, B., {Fortney}, J.~J., \& {Marley},
  M.~S. 2015, \apj, 807, 183

\bibitem[{{Line} {et~al.}(2013){Line}, {Wolf}, {Zhang}, {Knutson}, {Kammer},
  {Ellison}, {Deroo}, {Crisp}, \& {Yung}}]{Line2013}
{Line}, M.~R., {Wolf}, A.~S., {Zhang}, X., {et~al.} 2013, \apj, 775, 137

\bibitem[{{Line} {et~al.}(2016){Line}, {Stevenson}, {Bean}, {Desert},
  {Fortney}, {Kreidberg}, {Madhusudhan}, {Showman}, \&
  {Diamond-Lowe}}]{Line2016}
{Line}, M.~R., {Stevenson}, K.~B., {Bean}, J., {et~al.} 2016, \aj, 152, 203

\bibitem[{{Line} {et~al.}(2017){Line}, {Marley}, {Liu}, {Burningham}, {Morley},
  {Hinkel}, {Teske}, {Fortney}, {Freedman}, \& {Lupu}}]{Line2017}
{Line}, M.~R., {Marley}, M.~S., {Liu}, M.~C., {et~al.} 2017, \apj, 848, 83

\bibitem[{{Liu} {et~al.}(2016){Liu}, {Dupuy}, \&
  {Allers}}]{Liu_Dupuy_Allers2016}
{Liu}, M.~C., {Dupuy}, T.~J., \& {Allers}, K.~N. 2016, \apj, 833, 96

\bibitem[{{Liu} {et~al.}(2013){Liu}, {Magnier}, {Deacon}, {Allers}, {Dupuy},
  {Kotson}, {Aller}, {Burgett}, {Chambers}, {Draper}, {Hodapp}, {Jedicke},
  {Kaiser}, {Kudritzki}, {Metcalfe}, {Morgan}, {Price}, {Tonry}, \&
  {Wainscoat}}]{Liu2013}
{Liu}, M.~C., {Magnier}, E.~A., {Deacon}, N.~R., {et~al.} 2013, \apjl, 777, L20

\bibitem[{{Lothringer} {et~al.}(2018){Lothringer}, {Benneke}, {Crossfield},
  {Henry}, {Morley}, {Dragomir}, {Barman}, {Knutson}, {Kempton}, {Fortney},
  {McCullough}, \& {Howard}}]{Lothringer2018}
{Lothringer}, J.~D., {Benneke}, B., {Crossfield}, I.~J.~M., {et~al.} 2018, \aj,
  155, 66

\bibitem[{{Luhman} \& {Esplin}(2016)}]{Luhman_Esplin2016}
{Luhman}, K.~L., \& {Esplin}, T.~L. 2016, \aj, 152, 78

\bibitem[{{MacKenty} {et~al.}(2010){MacKenty}, {Kimble}, {O'Connell}, \&
  {Townsend}}]{MacKenty2010}
{MacKenty}, J.~W., {Kimble}, R.~A., {O'Connell}, R.~W., \& {Townsend}, J.~A.
  2010, in \procspie, Vol. 7731, Space Telescopes and Instrumentation 2010:
  Optical, Infrared, and Millimeter Wave, 77310Z

\bibitem[{{Madhusudhan} {et~al.}(2016){Madhusudhan}, {Apai}, \&
  {Gandhi}}]{Madhusudhan2016}
{Madhusudhan}, N., {Apai}, D., \& {Gandhi}, S. 2016, ArXiv e-prints,
  arXiv:1612.03174

\bibitem[{{Madhusudhan} \& {Seager}(2009)}]{Madhusudhan2009}
{Madhusudhan}, N., \& {Seager}, S. 2009, \apj, 707, 24

\bibitem[{{Mamajek} \& {Bell}(2014)}]{Mamajek2014}
{Mamajek}, E.~E., \& {Bell}, C.~P.~M. 2014, \mnras, 445, 2169

\bibitem[{{Mandell} {et~al.}(2013){Mandell}, {Haynes}, {Sinukoff},
  {Madhusudhan}, {Burrows}, \& {Deming}}]{Mandell2013}
{Mandell}, A.~M., {Haynes}, K., {Sinukoff}, E., {et~al.} 2013, \apj, 779, 128

\bibitem[{{Manjavacas} {et~al.}(2018){Manjavacas}, {Apai}, {Zhou}, {Karalidi},
  {Lew}, {Schneider}, {Cowan}, {Metchev}, {Miles-P{\'a}ez}, {Burgasser},
  {Radigan}, {Bedin}, {Lowrance}, \& {Marley}}]{Manjavacas2018}
{Manjavacas}, E., {Apai}, D., {Zhou}, Y., {et~al.} 2018, \aj, 155, 11

\bibitem[{{Marley} {et~al.}(1999){Marley}, {Gelino}, {Stephens}, {Lunine}, \&
  {Freedman}}]{Marley1999}
{Marley}, M.~S., {Gelino}, C., {Stephens}, D., {Lunine}, J.~I., \& {Freedman},
  R. 1999, \apj, 513, 879

\bibitem[{{Marley} \& {Robinson}(2015)}]{Marley2015}
{Marley}, M.~S., \& {Robinson}, T.~D. 2015, \araa, 53, 279

\bibitem[{{Marley} {et~al.}(2010){Marley}, {Saumon}, \&
  {Goldblatt}}]{Marley2010}
{Marley}, M.~S., {Saumon}, D., \& {Goldblatt}, C. 2010, \apjl, 723, L117

\bibitem[{{Marocco} {et~al.}(2013){Marocco}, {Andrei}, {Smart}, {Jones},
  {Pinfield}, {Day-Jones}, {Clarke}, {Sozzetti}, {Lucas}, {Bucciarelli}, \&
  {Penna}}]{Marocco2013}
{Marocco}, F., {Andrei}, A.~H., {Smart}, R.~L., {et~al.} 2013, \aj, 146, 161

\bibitem[{Marquardt(1963)}]{marquardt1963}
Marquardt, D.~W. 1963, SIAM Journal on Applied Mathematics, 11, 431

\bibitem[{{Marsh} {et~al.}(2013){Marsh}, {Wright}, {Kirkpatrick}, {Gelino},
  {Cushing}, {Griffith}, {Skrutskie}, \& {Eisenhardt}}]{Marsh2013}
{Marsh}, K.~A., {Wright}, E.~L., {Kirkpatrick}, J.~D., {et~al.} 2013, \apj,
  762, 119

\bibitem[{{Martin} {et~al.}(2018){Martin}, {Kirkpatrick}, {Beichman}, {Smart},
  {Faherty}, {Gelino}, {Cushing}, {Schneider}, {Wright}, {Lowrance}, {Ingalls},
  {Tinney}, {McLean}, {Logsdon}, \& {Lebreton}}]{Martin2018}
{Martin}, E.~C., {Kirkpatrick}, J.~D., {Beichman}, C.~A., {et~al.} 2018, ArXiv
  e-prints, arXiv:1809.06479

\bibitem[{{Martin} {et~al.}(1996){Martin}, {Rebolo}, \&
  {Zapatero-Osorio}}]{Martin}
{Martin}, E.~L., {Rebolo}, R., \& {Zapatero-Osorio}, M.~R. 1996, \apj, 469, 706

\bibitem[{{McLean} {et~al.}(2003){McLean}, {McGovern}, {Burgasser},
  {Kirkpatrick}, {Prato}, \& {Kim}}]{McLean}
{McLean}, I.~S., {McGovern}, M.~R., {Burgasser}, A.~J., {et~al.} 2003, \apj,
  596, 561

\bibitem[{{Metchev} {et~al.}(2015){Metchev}, {Heinze}, {Apai}, {Flateau},
  {Radigan}, {Burgasser}, {Marley}, {Artigau}, {Plavchan}, \&
  {Goldman}}]{Metchev2015}
{Metchev}, S.~A., {Heinze}, A., {Apai}, D., {et~al.} 2015, \apj, 799, 154

\bibitem[{{Naud} {et~al.}(2017){Naud}, {Artigau}, {Rowe}, {Doyon}, {Malo},
  {Albert}, {Gagn{\'e}}, \& {Bouchard}}]{Naud2017}
{Naud}, M.-E., {Artigau}, {\'E}., {Rowe}, J.~F., {et~al.} 2017, \aj, 154, 138

\bibitem[{Parmentier {et~al.}(2018)Parmentier, Line, Bean, Mansfield,
  Kreidberg, Lupu, Visscher, Desert, Fortney, Deleuil, Arcangeli, Showman, \&
  Marley}]{Parmentier2018}
Parmentier, V., Line, M.~R., Bean, J.~L., {et~al.} 2018, A{\&}A,
  arXiv:1805.00096

\bibitem[{{Pe{\~n}a Ram{\'{\i}}rez} {et~al.}(2015){Pe{\~n}a Ram{\'{\i}}rez},
  {Zapatero Osorio}, \& {B{\'e}jar}}]{Pena_ramirez2015}
{Pe{\~n}a Ram{\'{\i}}rez}, K., {Zapatero Osorio}, M.~R., \& {B{\'e}jar},
  V.~J.~S. 2015, \aap, 574, A118

\bibitem[{{Pecaut} \& {Mamajek}(2013)}]{Pecaut_Mamajek2013}
{Pecaut}, M.~J., \& {Mamajek}, E.~E. 2013, \apjs, 208, 9

\bibitem[{{Pinhas} {et~al.}(2018){Pinhas}, {Rackham}, {Madhusudhan}, \&
  {Apai}}]{Pinhas2018}
{Pinhas}, A., {Rackham}, B.~V., {Madhusudhan}, N., \& {Apai}, D. 2018, \mnras,
  480, 5314

\bibitem[{{Radigan}(2014)}]{Radigan2014}
{Radigan}, J. 2014, \apj, 797, 120

\bibitem[{{Radigan} {et~al.}(2012){Radigan}, {Jayawardhana}, {Lafreni{\`e}re},
  {Artigau}, {Marley}, \& {Saumon}}]{Radigan2012}
{Radigan}, J., {Jayawardhana}, R., {Lafreni{\`e}re}, D., {et~al.} 2012, \apj,
  750, 105

\bibitem[{{Radigan} {et~al.}(2014){Radigan}, {Lafreni{\`e}re}, {Jayawardhana},
  \& {Artigau}}]{Radigan_Lafreniere2014}
{Radigan}, J., {Lafreni{\`e}re}, D., {Jayawardhana}, R., \& {Artigau}, E. 2014,
  \apj, 793, 75

\bibitem[{{Rajan} {et~al.}(2017){Rajan}, {Rameau}, {De Rosa}, {Marley},
  {Graham}, {Macintosh}, {Marois}, {Morley}, {Patience}, {Pueyo}, {Saumon},
  {Ward-Duong}, {Ammons}, {Arriaga}, {Bailey}, {Barman}, {Bulger}, {Burrows},
  {Chilcote}, {Cotten}, {Czekala}, {Doyon}, {Duch{\^e}ne}, {Esposito},
  {Fitzgerald}, {Follette}, {Fortney}, {Goodsell}, {Greenbaum}, {Hibon},
  {Hung}, {Ingraham}, {Johnson-Groh}, {Kalas}, {Konopacky}, {Lafreni{\`e}re},
  {Larkin}, {Maire}, {Marchis}, {Metchev}, {Millar-Blanchaer}, {Morzinski},
  {Nielsen}, {Oppenheimer}, {Palmer}, {Patel}, {Perrin}, {Poyneer},
  {Rantakyr{\"o}}, {Ruffio}, {Savransky}, {Schneider}, {Sivaramakrishnan},
  {Song}, {Soummer}, {Thomas}, {Vasisht}, {Wallace}, {Wang}, {Wiktorowicz}, \&
  {Wolff}}]{Rajan2017}
{Rajan}, A., {Rameau}, J., {De Rosa}, R.~J., {et~al.} 2017, \aj, 154, 10

\bibitem[{{Ranjan} {et~al.}(2014){Ranjan}, {Charbonneau}, {D{\'e}sert},
  {Madhusudhan}, {Deming}, {Wilkins}, \& {Mandell}}]{Ranjan2014}
{Ranjan}, S., {Charbonneau}, D., {D{\'e}sert}, J.-M., {et~al.} 2014, \apj, 785,
  148

\bibitem[{{Rasmussen} \& {Williams}(2006)}]{Rasmussen_Williams2006}
{Rasmussen}, C.~E., \& {Williams}, C.~K.~I. 2006, {Gaussian Processes for
  Machine Learning}

\bibitem[{{Sahlmann} {et~al.}(2016){Sahlmann}, {Lazorenko}, {Bouy},
  {Mart{\'{\i}}n}, {Queloz}, {S{\'e}gransan}, \& {Zapatero
  Osorio}}]{Sahlmann2016}
{Sahlmann}, J., {Lazorenko}, P.~F., {Bouy}, H., {et~al.} 2016, \mnras, 455, 357

\bibitem[{{Samland} {et~al.}(2017){Samland}, {Molli{\`e}re}, {Bonnefoy},
  {Maire}, {Cantalloube}, {Cheetham}, {Mesa}, {Gratton}, {Biller}, {Wahhaj},
  {Bouwman}, {Brandner}, {Melnick}, {Carson}, {Janson}, {Henning}, {Homeier},
  {Mordasini}, {Langlois}, {Quanz}, {van Boekel}, {Zurlo}, {Schlieder},
  {Avenhaus}, {Beuzit}, {Boccaletti}, {Bonavita}, {Chauvin}, {Claudi}, {Cudel},
  {Desidera}, {Feldt}, {Fusco}, {Galicher}, {Kopytova}, {Lagrange}, {Le
  Coroller}, {Martinez}, {Moeller-Nilsson}, {Mouillet}, {Mugnier}, {Perrot},
  {Sevin}, {Sissa}, {Vigan}, \& {Weber}}]{Samland2017}
{Samland}, M., {Molli{\`e}re}, P., {Bonnefoy}, M., {et~al.} 2017, \aap, 603,
  A57

\bibitem[{{Schneider} {et~al.}(2015){Schneider}, {Cushing}, {Kirkpatrick},
  {Gelino}, {Mace}, {Wright}, {Eisenhardt}, {Skrutskie}, {Griffith}, \&
  {Marsh}}]{Schneider2015}
{Schneider}, A.~C., {Cushing}, M.~C., {Kirkpatrick}, J.~D., {et~al.} 2015,
  \apj, 804, 92

\bibitem[{{Sheppard} {et~al.}(2017){Sheppard}, {Mandell}, {Tamburo}, {Gandhi},
  {Pinhas}, {Madhusudhan}, \& {Deming}}]{Sheppard2017}
{Sheppard}, K.~B., {Mandell}, A.~M., {Tamburo}, P., {et~al.} 2017, \apjl, 850,
  L32

\bibitem[{{Shporer} {et~al.}(2014){Shporer}, {O'Rourke}, {Knutson},
  {Szab{\'o}}, {Zhao}, {Burrows}, {Fortney}, {Agol}, {Cowan}, {Desert},
  {Howard}, {Isaacson}, {Lewis}, {Showman}, \& {Todorov}}]{Shporer2014}
{Shporer}, A., {O'Rourke}, J.~G., {Knutson}, H.~A., {et~al.} 2014, \apj, 788,
  92

\bibitem[{{Stevenson} {et~al.}(2014{\natexlab{a}}){Stevenson}, {Bean},
  {Madhusudhan}, \& {Harrington}}]{Stevenson2014}
{Stevenson}, K.~B., {Bean}, J.~L., {Madhusudhan}, N., \& {Harrington}, J.
  2014{\natexlab{a}}, \apj, 791, 36

\bibitem[{{Stevenson} {et~al.}(2014{\natexlab{b}}){Stevenson}, {D{\'e}sert},
  {Line}, {Bean}, {Fortney}, {Showman}, {Kataria}, {Kreidberg}, {McCullough},
  {Henry}, {Charbonneau}, {Burrows}, {Seager}, {Madhusudhan}, {Williamson}, \&
  {Homeier}}]{Stevenson2014_Nat}
{Stevenson}, K.~B., {D{\'e}sert}, J.-M., {Line}, M.~R., {et~al.}
  2014{\natexlab{b}}, Science, 346, 838

\bibitem[{{Swain} {et~al.}(2013){Swain}, {Deroo}, {Tinetti}, {Hollis},
  {Tessenyi}, {Line}, {Kawahara}, {Fujii}, {Showman}, \&
  {Yurchenko}}]{Swain2013}
{Swain}, M., {Deroo}, P., {Tinetti}, G., {et~al.} 2013, \icarus, 225, 432

\bibitem[{{Tinney} {et~al.}(2003){Tinney}, {Burgasser}, \&
  {Kirkpatrick}}]{Tinney2003}
{Tinney}, C.~G., {Burgasser}, A.~J., \& {Kirkpatrick}, J.~D. 2003, \aj, 126,
  975

\bibitem[{{Tinney} {et~al.}(2014){Tinney}, {Faherty}, {Kirkpatrick}, {Cushing},
  {Morley}, \& {Wright}}]{Tinney2014}
{Tinney}, C.~G., {Faherty}, J.~K., {Kirkpatrick}, J.~D., {et~al.} 2014, \apj,
  796, 39

\bibitem[{{Todorov} {et~al.}(2016){Todorov}, {Line}, {Pineda}, {Meyer},
  {Quanz}, {Hinkley}, \& {Fortney}}]{Todorov2016}
{Todorov}, K.~O., {Line}, M.~R., {Pineda}, J.~E., {et~al.} 2016, \apj, 823, 14

\bibitem[{{Triaud}(2014)}]{Triaud2014a}
{Triaud}, A.~H.~M.~J. 2014, \mnras, 439, L61

\bibitem[{{Triaud} {et~al.}(2014){Triaud}, {Lanotte}, {Smalley}, \&
  {Gillon}}]{Triaud2014b}
{Triaud}, A.~H.~M.~J., {Lanotte}, A.~A., {Smalley}, B., \& {Gillon}, M. 2014,
  \mnras, 444, 711

\bibitem[{{Varley} {et~al.}(2017){Varley}, {Tsiaras}, \&
  {Karpouzas}}]{Varley2017}
{Varley}, R., {Tsiaras}, A., \& {Karpouzas}, K. 2017, \apjs, 231, 13

\bibitem[{{Vrba} {et~al.}(2004){Vrba}, {Henden}, {Luginbuhl}, {Guetter},
  {Munn}, {Canzian}, {Burgasser}, {Kirkpatrick}, {Fan}, {Geballe},
  {Golimowski}, {Knapp}, {Leggett}, {Schneider}, \& {Brinkmann}}]{Vrba2004}
{Vrba}, F.~J., {Henden}, A.~A., {Luginbuhl}, C.~B., {et~al.} 2004, \aj, 127,
  2948

\bibitem[{{Wahhaj} {et~al.}(2011){Wahhaj}, {Liu}, {Biller}, {Clarke},
  {Nielsen}, {Close}, {Hayward}, {Mamajek}, {Cushing}, {Dupuy}, {Tecza},
  {Thatte}, {Chun}, {Ftaclas}, {Hartung}, {Reid}, {Shkolnik}, {Alencar},
  {Artymowicz}, {Boss}, {de Gouveia Dal Pino}, {Gregorio-Hetem}, {Ida},
  {Kuchner}, {Lin}, \& {Toomey}}]{Wahhaj}
{Wahhaj}, Z., {Liu}, M.~C., {Biller}, B.~A., {et~al.} 2011, \apj, 729, 139

\bibitem[{{Wakeford} {et~al.}(2016){Wakeford}, {Sing}, {Evans}, {Deming}, \&
  {Mandell}}]{Wakeford2016}
{Wakeford}, H.~R., {Sing}, D.~K., {Evans}, T., {Deming}, D., \& {Mandell}, A.
  2016, \apj, 819, 10

\bibitem[{{Wang} {et~al.}(2018){Wang}, {Smart}, {Shao}, {Jones}, {Marocco},
  {Luo}, {Burgasser}, {Zhong}, \& {Du}}]{Wang2018}
{Wang}, Y., {Smart}, R.~L., {Shao}, Z., {et~al.} 2018, \pasp, 130, 064402

\bibitem[{{Weinberger} {et~al.}(2013){Weinberger}, {Anglada-Escud{\'e}}, \&
  {Boss}}]{Weinberger}
{Weinberger}, A.~J., {Anglada-Escud{\'e}}, G., \& {Boss}, A.~P. 2013, \apj,
  762, 118

\bibitem[{{Wilkins} {et~al.}(2014){Wilkins}, {Deming}, {Madhusudhan},
  {Burrows}, {Knutson}, {McCullough}, \& {Ranjan}}]{Wilkins2014}
{Wilkins}, A.~N., {Deming}, D., {Madhusudhan}, N., {et~al.} 2014, \apj, 783,
  113

\bibitem[{{Yang} {et~al.}(2015){Yang}, {Apai}, {Marley}, {Saumon}, {Morley},
  {Buenzli}, {Artigau}, {Radigan}, {Metchev}, {Burgasser}, {Mohanty},
  {Lowrance}, {Showman}, {Karalidi}, {Flateau}, \& {Heinze}}]{Yang2015}
{Yang}, H., {Apai}, D., {Marley}, M.~S., {et~al.} 2015, \apjl, 798, L13

\bibitem[{{Yang} {et~al.}(2016){Yang}, {Apai}, {Marley}, {Karalidi}, {Flateau},
  {Showman}, {Metchev}, {Buenzli}, {Radigan}, {Artigau}, {Lowrance}, \&
  {Burgasser}}]{Yang2016}
---. 2016, \apj, 826, 8

\bibitem[{{Zhou} {et~al.}(2017){Zhou}, {Apai}, {Lew}, \&
  {Schneider}}]{Zhou2017}
{Zhou}, Y., {Apai}, D., {Lew}, B.~W.~P., \& {Schneider}, G. 2017, \aj, 153, 243

\bibitem[{{Zhou} {et~al.}(2018){Zhou}, {Apai}, {Metchev}, {Lew}, {Schneider},
  {Marley}, {Karalidi}, {Manjavacas}, {Bedin}, {Cowan}, {Miles-P{\'a}ez},
  {Lowrance}, {Radigan}, \& {Burgasser}}]{Zhou2018}
{Zhou}, Y., {Apai}, D., {Metchev}, S., {et~al.} 2018, \aj, 155, 132

\bibitem[{{Zuckerman} {et~al.}(2004){Zuckerman}, {Song}, \&
  {Bessell}}]{Zuckerman2004}
{Zuckerman}, B., {Song}, I., \& {Bessell}, M.~S. 2004, \apjl, 613, L65

\bibitem[{{Zuckerman} {et~al.}(2001){Zuckerman}, {Song}, {Bessell}, \&
  {Webb}}]{Zuckerman2001}
{Zuckerman}, B., {Song}, I., {Bessell}, M.~S., \& {Webb}, R.~A. 2001, \apjl,
  562, L87

\end{thebibliography}
\end{document}